%% file: main.tex
\documentclass[11pt,3p,doublespacing]{elsarticle}
\journal{.}
\usepackage{times}
\emergencystretch=\maxdimen
\biboptions{sort&compress}
\usepackage{hyperref}
 \hypersetup{
 	plainpages = false, 
	bookmarksopen = true,
	bookmarksnumbered = true,
	breaklinks = true,
	linktocpage,
	colorlinks = true,
	linkcolor = purple,
	urlcolor  = black,
	citecolor = purple,
	}
\usepackage{multicol}
\usepackage{multirow}
\usepackage{imakeidx}
\makeindex
\usepackage{nomencl}
\makenomenclature
\usepackage[xindy]{glossaries}
\makeglossaries
\usepackage{graphics}
\usepackage{graphicx}
\usepackage{epsf,psfrag}
\usepackage{epstopdf}
\usepackage[para]{threeparttable}
\usepackage{subfigure}
\usepackage{epsfig}
\usepackage{pdflscape}
\usepackage{rotating}
\usepackage{lscape}
\usepackage{float}
\usepackage{stfloats}
\usepackage{textgreek}
\usepackage{longtable}
\usepackage{lscape}
\usepackage{capt-of}
\usepackage{comment}
\usepackage{tablefootnote}
\usepackage{footnote}
\usepackage{caption}
\usepackage[autopunct=true]{csquotes}
\usepackage{rotating}
\usepackage{txfonts}
\usepackage[normalem]{ulem}
\usepackage{resizegather}
\usepackage{siunitx}
\usepackage{setspace}
\usepackage{color,soul}
\usepackage[usenames,dvipsnames]{xcolor}
\usepackage{amsfonts,amsmath,amssymb}
\usepackage{latexsym,array}
\usepackage{textcomp}

\newcommand{\RomanNumeralCaps}[1]
\linenumbers
\makesavenoteenv{tabular}
\makesavenoteenv{table}
\usepackage[left,displaymath, mathlines]{lineno}
\DeclareMathAlphabet{\mathpzc}{OT1}{pzc}{m}{it}
\def\fig{Figure~}
\def\figs{Figures~}
\def\eqn{Eq.~}
\def\eqns{Eqs.~}
\def\tab{Table~}

\def\micro{\textmu}

\providecommand\p{{\partial}}

\newcommand{\percent}{{\%}}
\newcommand\mm[1]{$#1$}  

\newcommand\sur{S_{\text{r}}}
\newcommand\sut{S_{\text{t}}}

\newcommand{\myvec}[1]{\mathbf{#1}}     
\def\tsc#1{\csdef{#1}{\textsc{\lowercase{#1}}\xspace}}
\tsc{WGM}
\tsc{QE}
\tsc{EP}
\tsc{PMS}
\tsc{BEC}
\tsc{DE}
\newcommand{\del}[1]{{\textcolor{orange}{\expandafter\sout\expandafter{#1}}}}

\newcommand{\rev}[1]{\textcolor{black}{#1}}     
     
\begin{document}

%
%
\setcounter{page}{1}
\begin{frontmatter} 
%
%
%
%
%
\title{Electroviscous effects in electrolyte liquid flow through an oppositely-charged contraction-expansion  microfluidic slit device}
\author[labela]{Jitendra {Dhakar}}
\author[labela]{Ram Prakash {Bharti}\corref{coradd}}
\emailauthor{rpbharti@iitr.ac.in}{R.P. Bharti}
\address[labela]{Complex Fluid Dynamics and Microfluidics (CFDM) Lab, Department of Chemical Engineering,  Indian Institute of Technology Roorkee, Roorkee - 247667, Uttarakhand, India}
%
%
\cortext[coradd]{\textit{Corresponding author. }}
%
\begin{abstract}
\fontsize{11}{15pt}\selectfont
%
\noindent 
Electrokinetic flows in microchannels with opposite charge asymmetry, i.e., unequal and contrasting surface charges on opposing channel walls, significantly influence microfluidic hydrodynamics and can be exploited for enhanced control of mass transfer, mixing, and heat transport in practical microfluidic applications. In this study, the electroviscous flow of a liquid electrolyte through an oppositely charged non-uniform microslit is investigated numerically. The flow governing equations, Poisson, Nernst-Planck, and Navier-Stokes (P-NP-NS) equations, are solved using the finite element method (FEM) to determine the coupled electrokinetic flow fields for a wide range of dimensionless parameters: Reynolds number ($Re = 10^{-2}$), Schmidt number ($\mathit{Sc} = 10^3$), inverse Debye length ($2 \le K \le 20$), top wall surface charge density ($4 \le S_\text{t} \le 16$), surface charge density ratio ($-2 \le S_\text{r} \le 0$), and contraction ratio ($d_\text{c} = 0.25$). For completeness and comparative analysis, results for like-charged microfluidic devices, where both walls carry surface charges of the same sign ($0 \le S_r \le 2$), are also included.
The results reveal that the maximum enhancement in total electrical potential ($|\Delta U|$) and pressure drop ($|\Delta P|$) is achieved as 296.82\% at $K = 20$, $S_\text{t} = 4$, $-2 \le S_\text{r} \le -1.25$, and 
14.57\% at $S_\text{t} = 16$, $S_\text{r} = 0$, $2 \le K \le 20$, respectively. The electroviscous correction factor $Y$ (apparent-to-physical viscosity ratio) exhibits maximum increases of 14.02\% at $K = 2$, $S_\text{t} = 16$; 11.81\% at $K = 2$, $S_\text{r} = 0$; and 14.57\% at $S_\text{r} = 0$, $S_\text{t} = 16$, when $S_\text{r}$ is increased (from $-0.75$ to $0$), $S_\text{t}$ is increased (from $4$ to $16$), and $K$ is decreased (from $20$ to $2$), respectively. The overall maximum increment in $Y$ is obtained as 15.13\% at $K = 2$, $S_\text{r} = 0$, $S_t = 16$, relative to the non-electroviscous flow (non-EVF) case ($S_\text{k} = 0$ or $K = \infty$).
These findings demonstrate that opposite charge asymmetry, while weaker than that of similar asymmetry, strongly influences electroviscous flow in non-uniform microchannel geometries. The present numerical results provide useful insights for the design and optimization of microchips in engineering and biomedical applications. Furthermore, this charge asymmetry offers a powerful mechanism to regulate electrokinetic transport in microfluidic devices, with direct implications for developing efficient practical systems.
\end{abstract}
\begin{keyword}
\fontsize{11}{16pt}\selectfont
%
electroviscous effects; opposite charge asymmetry; pressure-driven microfluidics; sustainable microfluidic systems; enhanced mixing and mass transfer; heat and mass transfer optimization
\end{keyword}
\end{frontmatter}
%
\section{Introduction}\label{sec:intro}
\noindent
Over the past decades, microfluidics has emerged as a promising technology with widespread applications in biomedical and engineering fields, including drug delivery systems \citep{vladisavljevic2013industrial,nguyen2013design}, DNA analysis \citep{foudeh2012microfluidic,bruijns2016microfluidic,li2021microfluidic}, micro heat exchangers \citep{xue2001performance,stogiannis2015efficacy,bahiraei2017efficacy,pan2019numerical}, medical diagnostic sensors \citep{wu2018lab}, and electronic chip cooling \citep{imran2018numerical,abdulqadur2019performance,tan2019heat,zhuang2020optimization}. It is well established that electrolyte liquids can be transported effectively in microfluidic devices. Unlike conventional macro-scale channel flows, transport in micron-sized channels exhibits unique characteristics, where surface-related forces (electrical, magnetic, and interfacial tension) play a dominant role in flow behavior. Consequently, a comprehensive understanding of electroviscous (EV) effects is essential for accurately accounting for electrical forces in microscale flow analysis, thereby enabling the design of efficient and reliable microdevices for diverse microfluidic applications.  In particular, investigating EV effects under opposite charge asymmetry, where microchannel walls carry unequal and contrasting surface charges, is of great significance, as such conditions profoundly alter hydrodynamic behavior and offer new opportunities for regulating electrokinetic transport in non-uniform microfluidic systems.

\begin{figure}[!hb]
	\centering
	\includegraphics[width=1\linewidth]{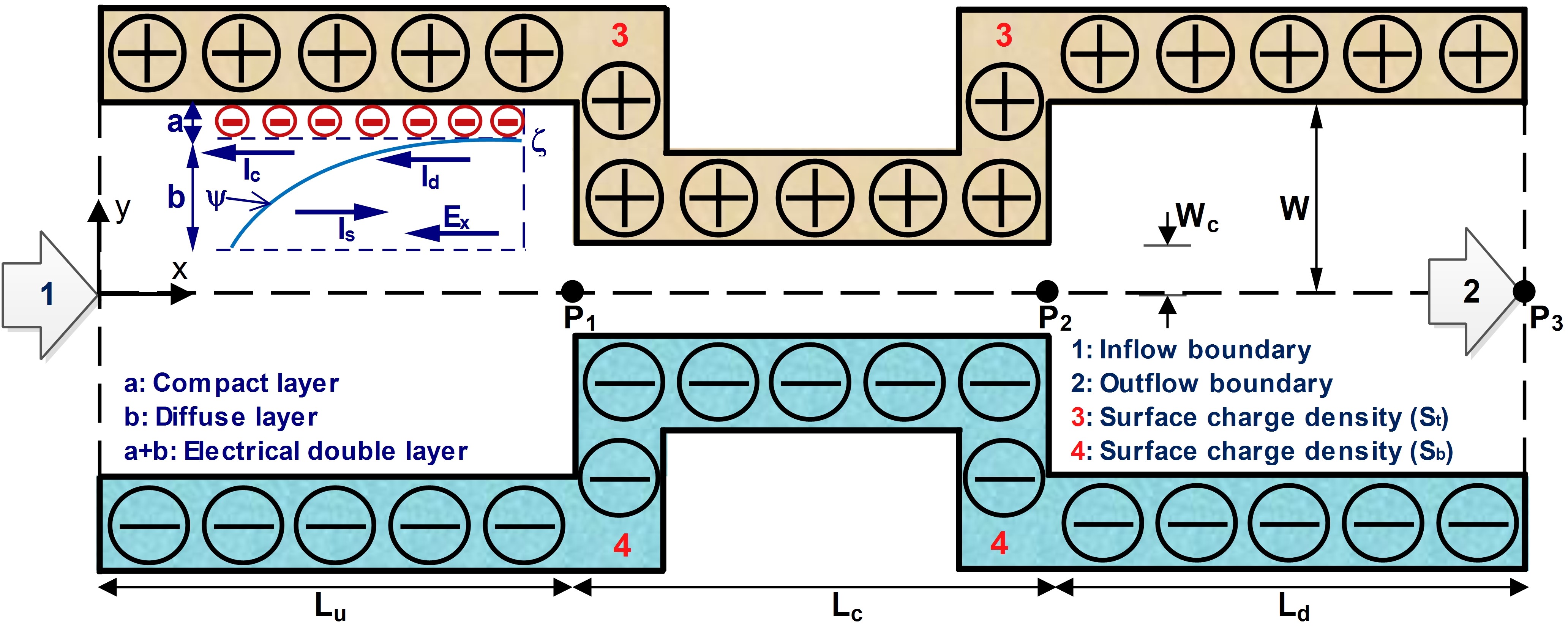}
	\caption{Schematic of electroviscous flow  (EVF) through oppositely charged non-uniform contraction-expansion microfluidic slit. }
	\label{fig:1}
\end{figure}

\noindent
Electroviscous flow (EVF) manifests when charged solid surfaces come into contact with an electrolyte solution under imposed pressure-driven flow (PDF). The surface charges modify the local ion distribution near the walls, leading to the formation of an electrical double layer (EDL)  \citep{atten1982electroviscous,hunter2013zeta,li2001electro}, as shown in \fig\ref{fig:1}. The EDL consists of a compact Stern layer and a diffuse layer, separated by the slipping (shear) plane, which carries the zeta ($\zeta$) potential that decays within the diffuse layer away from the charged walls. Due to imposed PDF,  the convective transport of counter-ions in the diffuse layer generates a streaming current ($I_\text{s}$) and a corresponding streaming potential ($\phi$). This potential, in turn, drives counter-ions against the direction of the PDF, producing a conduction current ($I_\text{c}$) which induces backflow. As a consequence, the net flow rate in the primary flow direction is reduced, giving rise to the phenomenon known as the electroviscous effect \citep{atten1982electroviscous,hunter2013zeta,dhakar2022electroviscous,dhakar2023cfd,dhakar2023contraction,dhakar2023uniform}. 

\noindent 
Recent studies \citep{dhakar2022electroviscous,dhakar2023cfd,dhakar2023contraction,dhakar2023uniform} have comprehensively reviewed electroviscous effects in symmetrically charged  (i.e., uniform surface charge density, $\sigma_\text{r} = 1$) microfluidic geometries of uniform cross-section. These include cylindrical channels \citep{rice1965electrokinetic,levine1975theory,bowen1995electroviscous,brutin2005modeling,bharti2009electroviscous,jing2016electroviscous}, slit channels \citep{burgreen1964electrokinetic,mala1997flow,mala1997heat,chun2003electrokinetic,ren2004electroviscous,chen2004developing,joly2006liquid,wang2010flow,jamaati2010pressure,zhao2011competition,tan2014combined,jing2015electroviscous,matin2016electrokinetic,jing2017non,matin2017electroviscous,kim2018analysis,mo2019electroviscous,li2021combined,li2022electroviscous}, elliptical channels \citep{hsu2002electrokinetic}, and rectangular channels \citep{yang1998modeling,li2001electro,ren2001electro}. Furthermore, electroviscous behavior in non-uniform geometries has been widely investigated, including contraction-expansion cylindrical microchannels \citep{bharti2008steady,davidson2010electroviscous}, slit microchannels \citep{davidson2007electroviscous,Berry2011,dhakar2022slip,dhakar2022electroviscous,dhakar2023contraction,dhakar2024fmfp}, and rectangular microchannels \citep{davidson2008electroviscous}.
These studies \citep{davidson2007electroviscous,davidson2008electroviscous,bharti2008steady,bharti2009electroviscous,davidson2010electroviscous,Berry2011,dhakar2022electroviscous,dhakar2022slip} demonstrate that electrokinetic flow fields, such as total potential ($U$), pressure ($P$), induced electric field strength ($E_\text{x}$), and excess charge ($n^\ast$), in symmetrically charged microchannels are strongly influenced by surface charge density ($4 \le S \le 16$), inverse Debye length ($2 \le K \le 20$), and slip length ($0 \le B_0 \le 0.20$) at a constant volumetric flow rate ($Q$). Moreover, pseudo-analytical models developed to predict pressure drop (and hence the electroviscous correction factor) generally overpredict numerical results by about 5-10\% \citep{davidson2007electroviscous,davidson2008electroviscous,bharti2008steady,bharti2009electroviscous,davidson2010electroviscous,Berry2011,dhakar2022electroviscous,dhakar2022slip}. 

\begin{table}[!b]
	\caption{Classification of surface charge asymmetry in microfluidic device walls.}\label{tab:srf}
	\scalebox{0.85}
	{\renewcommand{\arraystretch}{1.5}\begin{tabular}{|l|c|c|c|p{0.73\linewidth}|}
			\hline
			Case & $S_\text{r}$ (or $\sigma_r$) & $S_\text{t}$ &  $S_\text{b}$ & Remarks\\\hline
			A1 & \multirow{2}{*}{$S_\text{r} > 0$} & $S_\text{t}>0$ &  $S_\text{b}\ge 0$ & \multirow{2}{\linewidth}{\textbf{Case A} {(Similar-charge asymmetry)}: Both opposing (top/bottom) walls are similarly (either positive or negative) charged \citep{dhakar2023cfd}}\\ \cline{1-1}\cline{3-4}
			A2&  & $S_\text{t}<0$ &  $S_\text{b}\le 0$ & \\\hline
			B1 & \multirow{2}{*}{$S_\text{r} < 0$} & $S_\text{t}>0$ &  $S_\text{b}\le 0$ & \multirow{2}{\linewidth}{\textbf{Case B} {(Opposite-charge asymmetry)$^{\#}$}: The opposing walls are  contrastively (one wall positively, while another negatively) charged}\\\cline{1-1}\cline{3-4}
			B2&  & $S_\text{t}<0$ &  $S_\text{b}\ge 0$ & \\\hline
			B11 & $S_\text{r}=0$ & $S_\text{t}>0$ &  $S_\text{b}=0$ & The top wall is positively charged, and bottom wall is electrically neutral\\\hline
			B12 & $-1 < S_\text{r} < 0$ & $|S_\text{t}|> |S_\text{b}|$ & $S_\text{b} < 0$ & The charge at the top wall dominates over that of the bottom wall\\\hline
			B13 & $S_\text{r}=-1$ & $|S_\text{t}|= |S_\text{b}|$ &  $S_\text{b} < 0$& Both walls are equally, but oppositely, charged. The core region of the microfluidic device effectively behaves as electrically neutral\\\hline
			B14 & $-2 \le S_\text{r} < -1$ &$|S_\text{t}|< |S_\text{b}|$ &  $S_\text{b} < 0$ & The charge at the bottom wall dominates over that of the top wall\\\hline
			\multicolumn{5}{|p{1.2\linewidth}|}{$S_\text{r}\ (= S_\text{b} / S_\text{t})$ denotes the ratio of the dimensionless surface charge densities of the bottom ($S_\text{b}$) and top ($S_\text{t}$) walls;   \newline $^{\#}$The engineering parameters (such as electrical potential and pressure drop) will remain the same for cases B1 and B2; however, the local profiles will display the mirror image features about the horizontal centerline ($x, 0$).}\\\hline
	\end{tabular}}
\end{table}
\noindent
When the walls of a microfluidic device are fabricated using different materials with surface charge densities $\sigma_\text{b}$ and $\sigma_\text{t}$ for the bottom and top walls, respectively, surface charge asymmetry ($\sigma_\text{r}$) is defined by $\sigma_\text{r} = (\sigma_\text{b}/\sigma_\text{t}) \neq 1$. This asymmetry can be broadly classified into two categories: \textbf{Case A}, similar-charge asymmetry ($\sigma_\text{r} > 0$), where both walls carry charges of the same sign (either positive or negative) \citep{dhakar2023cfd}; and \textbf{Case B}, opposite-charge asymmetry ($\sigma_\text{r} < 0$), where the walls carry charges of opposite sign (one positive and one negative) \citep{dhakar2024achmt}. A detailed classification of surface charge asymmetry on microfluidic device walls in terms of the dimensionless surface charge density ($S$) is summarized in \tab\ref{tab:srf}.
Fewer studies have examined electroviscous effects in asymmetrically charged ($\sigma_\text{r} \neq 1$) uniform microchannels \citep{xuan2008streaming,soong2003theoretical,wang2010flow,sailaja2019electroviscous,Li_2025}. These investigations consistently show that surface charge asymmetry plays a significant role in shaping the hydrodynamics of uniform microfluidic geometries. For instance, \citet{xuan2008streaming} analytically examined streaming potential and electroviscous effects by considering surface charge variations both parallel ($q \parallel \nabla P$, i.e., charge heterogeneity) and perpendicular ($q \perp \nabla P$, i.e., charge asymmetry, case A) to the applied pressure gradient in a uniform microchannel. Their results revealed that charge heterogeneity notably affects the streaming potential and electroviscous behavior at small values of $K$ ($<50$), while this dependence becomes negligible at larger values of $K$ ($>50$).
\citet{soong2003theoretical} analyzed the electrokinetic flow and thermal behavior of Newtonian fluids in parallel-plate microchannels under asymmetric boundary conditions, incorporating wall slip, unequal zeta potentials (case A), and non-uniform heat fluxes at the walls. They reported that unequal zeta potentials significantly altered the distributions of both streaming and electric potentials. Similarly, \citet{wang2010flow} investigated electroviscous effects in Newtonian fluid flow through asymmetrically charged (case A) microchannels, where the hydrophobic polydimethylsiloxane (PDMS) wall exhibited slip while the opposing glass wall did not. Their results showed that the extent of wall slip strongly governed the electroviscous effects. Furthermore, \citet{sailaja2019electroviscous} demonstrated that asymmetric wall zeta potentials ($-1 \le \zeta_\text{r} \le 2$; cases A and B) significantly influenced the streaming potential and flow characteristics of power-law fluid flow in a uniform micro-slit channel. 

\noindent
In addition to the aforementioned studies on asymmetrically charged uniform microchannels \citep{xuan2008streaming,soong2003theoretical,wang2010flow,sailaja2019electroviscous}, recent investigations \citep{dhakar2023cfd,dhakar2023contraction,dhakar2023uniform,dhakar2023nonuniform,dhakar2024achmt} have focused on electroviscous flow of liquid electrolytes through asymmetrically ($\sigma_\text{r} \neq 1$) and heterogeneously charged non-uniform micro-slits. These studies evaluated the distributions of electrical potential, excess charge, induced electric field strength, and pressure drop, and concluded that surface charge asymmetry ($0 \le \sigma_\text{r} \le 2$; case A), charge heterogeneity ($0 \le S_\text{rh} \le 2$), and contraction size ($0.25 \le d_\text{c} \le 1$) significantly affect electrokinetic flow fields in the micro-slit under a constant volumetric flow rate ($Q$).

\noindent 
In summary, the literature provides substantial insight into electroviscous (EV) flow in symmetrically, homogeneously, asymmetrically, and heterogeneously charged microchannels, both uniform and non-uniform. While several studies have explored the effects of like-charge asymmetry (case A) in uniform \citep{xuan2008streaming,soong2003theoretical,wang2010flow} and non-uniform \citep{dhakar2023cfd} microchannel geometries, investigations of opposite charge asymmetry (case B) remain scarce, with only a single study addressing its influence in uniform microchannels \citep{sailaja2019electroviscous}.

\noindent
To the best of our knowledge, electroviscous flow in oppositely charged (case B) non-uniform microchannel geometries remains largely underexplored. While prior studies have predominantly focused on symmetrically charged configurations or like-charge asymmetry (case A), opposite charge asymmetry (case B), where walls carry unequal charges of opposite polarity, has received very limited attention, despite its potential to drastically modify electrohydrodynamic interactions and provide new opportunities for regulating electrokinetic transport. To bridge this gap, our recent work \citep{dhakar2023cfd} on similar charge asymmetry (case A; $0 \le \sigma_\text{r} \le 2$, $4 \le \sigma_\text{t} \le 16$, $2 \le K \le 20$) in non-uniform contraction-expansion (4:1:4) microchannels provides a foundation for extending the analysis to case B, thereby enhancing the understanding of electroviscous effects in complex microfluidic geometries.

\noindent
This work presents a numerical investigation of electroviscous flow (EVF) of liquid electrolytes through an oppositely charged (case B) non-uniform (contraction-expansion) microfluidic slit. The electrokinetic flow is modeled using the coupled Poisson, Navier-Stokes, and Nernst-Planck (P-NS-NP) equations, and solved using the finite element method (FEM) to obtain detailed distributions of electrical potential, pressure, induced electric field, excess charge, and the electroviscous correction factor. The numerical results are systematically analyzed over a wide range of dimensionless parameters, including the surface charge density ratio ($-2 \le \sigma_r \le 0$; case B), top-wall surface charge density ($4 \le \sigma_t \le 16$), inverse Debye length ($2 \le K \le 20$), Reynolds number ($Re = 10^{-2}$), Schmidt number ($\mathit{Sc} = 10^3$), and contraction ratio ($d_c = 0.25$). For completeness and comparative analysis, results for like-charged configurations ($0 \le \sigma_r \le 2$; case A) are also presented and examined. Additionally, a pseudo-analytical model is developed to predict the pressure drop and corresponding electroviscous correction factor in oppositely charged non-uniform microchannels. This framework provides a comprehensive platform to capture the interplay between electrostatic interactions, hydrodynamic transport, and geometric confinement under charge asymmetry.
\section{Physical modelling}\label{sec:mode}
\noindent
\fig\ref{fig:1} depicts the steady, fully developed laminar flow  (with average inflow velocity $\overline{V}$, \si{\metre\per\second}) of a liquid electrolyte through a two-dimensional (2D) contraction-expansion (4:1:4) microfluidic slit. The geometry comprises three serially connected sections, each characterized by its width and length (in \si{\micro \metre}): the upstream section ($2W$, $L_\text{u}$), the contraction section ($2W_\text{c}$, $L_\text{c}$), and the downstream section ($2W$, $L_\text{d}$). The overall device length is $L = (L_\text{u} + L_\text{c} + L_\text{d})$, and the contraction ratio is defined as $d_\text{c} = (W_\text{c}/W)$.

\noindent
The opposite charge asymmetry (case B) is imposed on the walls ($w$) of the microfluidic device, where the top ($t$) and bottom ($b$) walls are assumed to be fabricated from different materials to maintain contrasting surface charge densities ($\sigma_\text{w}$, \si{\coulomb\per\square\metre}), as illustrated in \fig\ref{fig:1}. For instance,  materials such as planar glass \citep{Behrens2001}, PDMS (polydimethylsiloxane) \citep{Germane2023}, and EGaIn (gallium-based alloys) \citep{Song_2024} typically exhibit negative surface charge density ($\sigma_\text{w} < 0$), whereas several other materials \citep{Liang_2023} are known to possess positive surface charge density ($\sigma_\text{w} > 0$). Consequently, the relative surface charge density on the slit walls becomes negative ($\sigma_\text{r} < 0$), where the surface charge density ratio is defined as $\sigma_\text{r} = (\sigma_\text{b}/\sigma_\text{t})$ or $S_\text{r} = (S_\text{b}/S_\text{t})$. In this study, the top wall is prescribed a positive surface charge density ($\sigma_\text{t} > 0$) and the bottom wall a negative surface charge density ($\sigma_\text{b} < 0$), corresponding to case B1 (see \tab\ref{tab:srf}). The reverse configuration (case B2), in which the wall charges are interchanged ($\sigma_\text{t} < 0$, $\sigma_\text{b} > 0$), produces identical results due to geometric and physical symmetry and can be interpreted by mirror reflecting the computational domain about the horizontal axis.

\noindent
A symmetric ($z:z$) electrolyte solution is considered and modeled as an incompressible (density $\rho$, \si{\kilogram\per\meter\cubed}) and Newtonian (viscosity $\mu$, \si{\pascal\second}) fluid. The ionic species are assumed to possess equal and opposite valences ($z_{+} = -z_{-} = z$) and identical diffusivities ($\mathcal{D}_{+} = \mathcal{D}_{-} = \mathcal{D}$, \si{\metre\squared\per\second}), with a fixed bulk ion concentration ($n_0$, \si{\per\meter\cubed}) \citep{harvie2012microfluidic,davidson2016numerical,dhakar2022electroviscous,dhakar2023uniform,dhakar2023cfd,dhakar2023contraction}. The liquid phase is further assumed to have a spatially uniform relative permittivity (dielectric constant) $\varepsilon_{\text{r}}$, which is significantly higher than that of the microchannel wall material ($\varepsilon_{\text{r,w}} \ll \varepsilon_{\text{r}}$).
\section{Mathematical modelling}\label{sec:mmode}
\noindent
The present problem is governed by a coupled set of equations \citep{davidson2007electroviscous,davidson2008electroviscous,bharti2008steady,bharti2009electroviscous,davidson2010electroviscous,Berry2011,dhakar2022electroviscous,dhakar2022slip,dhakar2023cfd,dhakar2023uniform,dhakar2023contraction}, consisting of the Poisson equation (PE, \eqn\ref{eq:1}) describing the distribution of the electrical potential, the Nernst-Planck equation (NPE, \eqn\ref{eq:5}) governing ionic species transport, and the Navier-Stokes equations (NSE, \eqns\ref{eq:9} and \ref{eq:10}) representing mass and momentum conservation in incompressible electrolyte flow under the influence of electrical body forces.
%

\noindent
Since both the dimensional and dimensionless forms of P-NP-NS model have been comprehensively presented in earlier studies \citep{davidson2007electroviscous,dhakar2022electroviscous,dhakar2023cfd,dhakar2023uniform,dhakar2023contraction}, the present work focuses on the dimensionless mathematical formulation to ensure clarity and completeness while avoiding redundancy.
The scaling variables employed to non-dimensionalize the governing equations and boundary conditions, along with the resulting dimensionless groups, are summarized in \tab\ref{tab:scaling}. Here, $e$ denotes the elementary charge of a proton, $k_\text{B}$ the Boltzmann constant, $T$ the absolute temperature, and $\varepsilon_{0}$ the permittivity of free space. Unless otherwise specified, all variables, quantities, and parameters are henceforth expressed in their dimensionless form.
The dimensionless form of the coupled governing equations (P-NP-NS model, \eqns\ref{eq:1}–\ref{eq:10}) describing the electroviscous flow fields is subsequently presented \citep{davidson2007electroviscous,bharti2008steady,dhakar2022electroviscous,dhakar2023cfd,dhakar2023contraction,dhakar2023uniform} as follows.
\begin{table}[!b]
	\caption{Scaling variables and dimensionless groups.}\label{tab:scaling}
	\scalebox{0.85}
	{\renewcommand{\arraystretch}{1.5}\begin{tabular}{|c|c|c|c|c|c|}
			\hline
			\multicolumn{6}{|l|}{(a) Scaling variables:}\\			\hline
			Length & Time   & Velocity & Pressure & Electrical potential & Ionic concentration \\ 
			$\displaystyle W$& $\displaystyle W/\overline{V}$ &
			$\displaystyle \overline{V}$  & $\displaystyle \rho\overline{V}^2$ &
			$\displaystyle U_{\text{c}} =\frac{k_{\text{B}}T}{ze}$ & $\displaystyle n_{0}$ \\ \hline
			\multicolumn{6}{|l|}{(b) Dimensionless groups:}\\			\hline
			Reynolds number & Schmidt number & Peclet  number & Inverse Debye length  & Liquid parameter & Surface charge density \\ 
			$\displaystyle Re=\frac{\rho\overline{V}W}{\mu}$& 
			$\displaystyle \mathit{Sc}=\frac{\mu}{\rho \mathcal{D}}$  & 
			$\displaystyle Pe =Re\times\mathit{Sc}$ & 
			$\displaystyle K=\sqrt{\frac{2W^2zen_{0}}{\varepsilon_{0}\varepsilon_{\text{r}} U_\text{c}}}$ & 
			$\displaystyle \beta=\frac{\rho\varepsilon_{0}\varepsilon_{\text{r}}U_\text{c}^2}{2\mu^2}$ &
			$\displaystyle \mathit{S_{\text{w}}}=\frac{\sigma_{\text{w}}W}{\varepsilon_{0}\varepsilon_\text{r} U_\text{c}}$ \\ \hline
			\multicolumn{6}{|l|}{(c) Constants:}\\			\hline
			\multicolumn{6}{|c|}{$e= \qty{1.602176634e-19}{\coulomb}$;\qquad $k_\text{B} = \qty{1.380649e-23}{\joule\per\kelvin}$;\qquad  $T=298$ K; \qquad $\varepsilon_{0} = \qty{8.8541878188e-12}{\coulomb\squared\per\newton\per\metre\squared}$} 
\\ \hline
	\end{tabular}}	
\end{table}
\begin{gather}
\nabla^2U=-(K^2/2)\ \rho_{\text{e}} 
\label{eq:1}\\
\left[\frac{\partial n_{\text{j}}}{\partial t}+\nabla\cdot(\myvec{V}n_{\text{j}})\right]=(1/Pe)\left[\nabla^2n_{\text{j}}\pm\nabla\cdot(n_{\text{j}}\nabla U)\right]
	\label{eq:5}\\
\left[\frac{\partial \mathbf{V}}{\partial t}+\nabla\cdot(\myvec{V}\myvec{V})\right] = 
	-\nabla P+(1/Re)\nabla \cdot\left[\nabla\myvec{V}+(\nabla\myvec{V})^T\right] + {\myvec{F}_{\text{e}}}
	\label{eq:9}\\
\nabla\cdot\myvec{V}=0 \label{eq:10} \\
 \rho_{\text{e}} \equiv n^\ast=(n_{+}-n_{-}), \qquad \myvec{F}_{\text{e}} = \beta (K/Re)^2n^\ast(-\nabla U)	\label{eq:4b}\\ 
U(x,y)=\psi(y)+\phi(x), 	\qquad \phi=-xE_\text{x}, \qquad 	E_\text{x}=-(\p U/\p x)
	\label{eq:4a}
\end{gather}
Here, $U$, $n_{\text{j}}$, $\mathbf{V}$, $P$, $t$, $\mathbf{F}{\text{e}}$, $\psi$, $\phi$, $\rho_{\text{e}}$ and $E_{\text{x}}$ represent the total electrical potential, number density of the $j^{\text{th}}$ ionic species, velocity vector $(V_{\text{x}}, V_{\text{y}})$, pressure, time, electrical body force, electrical double layer (EDL) potential, streaming potential, local charge density, and the axially induced electric field strength, respectively. Furthermore, the local charge density ($\rho_{\text{e}}$), as defined in \eqn\eqref{eq:4b}, is directly related to the excess charge ($n^{\ast}$), which, for a symmetric (1:1) electrolyte, corresponds to the excess ion concentration given by the difference between the cation and anion number densities ($n_{+} - n_{-}$). It should also be noted that the total electrical potential ($U$) can be decomposed into the EDL potential and the streaming potential, as expressed in \eqn\eqref{eq:4a}, but only for homogeneously charged uniform microchannels.

\noindent
The mathematical model (P-NP-NS, \eqns\ref{eq:1}-\ref{eq:10}) is solved subject to boundary conditions (BCs) specified at the inflow ($x = 0$), outflow ($x = L$), and wall boundaries of the oppositely charged non-uniform microfluidic slit \citep{davidson2007electroviscous,bharti2008steady,dhakar2022electroviscous,dhakar2023cfd,dhakar2023contraction,dhakar2023uniform}. Here, the subscripts ‘i’, ‘o’, and ‘w’ denote the inflow, outflow, and wall boundaries, respectively, while ‘b’ and ‘t’ are used to indicate the bottom and top walls.

\noindent
{[BC-1]}: 
For the potential field governed by the Poisson equation (\eqn\ref{eq:1}), a uniform axial potential gradient is applied at the inflow and outflow, and the walls are prescribed uniform surface charge densities, which may be either equal or unequal but of opposite polarity. For instance,
\begin{gather}
	\left.\frac{\partial U}{\partial x}\right|_\text{i} = C_{\text{i}}, 
	\qquad \left.\frac{\partial U}{\partial x}\right|_\text{o} = C_{\text{o}}, 
	\qquad \text{and} \qquad
	\left.(\nabla U\cdot\myvec{n})\right|_\text{w} = S_\text{w}
	\qquad \text{with} \qquad
	S_\text{r}=\frac{S_\text{b}}{S_\text{t}}
	\label{eq:3}
\end{gather}
Here, $\mathbf{n}$ represents the unit vector normal at the wall, and $S_{\text{r}}$ is the surface charge density ratio, which in this study is restricted to $S_{\text{t}} > 0$ and $S_{\text{b}} \le 0$, corresponding to case B1 as defined in \tab\ref{tab:srf}.
Furthermore, the constants ($C_{\text{i}}$ and $C_{\text{o}}$) at the inlet and outlet boundaries are determined \citep{davidson2007electroviscous,bharti2008steady,dhakar2022electroviscous,dhakar2023cfd,dhakar2023contraction,dhakar2023uniform} by enforcing the \textit{current continuity condition}, which ensures both local current conservation ($\nabla \cdot \mathbf{I} = 0$) and global current balance across the cross-section, i.e., zero net axial current ($I_{\text{net}} = 0$), as expressed below.
\begin{gather} 
	I_{\text{net}} = (I_{\text{s}} + I_{\text{d}} + I_{\text{c}}) = 0 \label{eq:2} 
\end{gather}
Here, $I_{\text{s}}$, $I_{\text{d}}$, and $I_{\text{c}}$ denote the streaming, diffusion, and conduction currents, respectively. At steady state, the diffusion current vanishes ($I_{\text{d}} = 0$).
\begin{gather}
	I_{\text{s}} = \int_{-1}^{1} n^\ast V_{\text{x}}(y) \text{d}y, \qquad I_{\text{d}} = -\frac{1}{Pe} \int_{-1}^{1} \left( \frac{\partial n_{+}}{\partial x} - \frac{\partial n_{-}}{\partial x} \right) \text{d}y, \qquad I_{\text{c}} = -\frac{1}{Pe} \int_{-1}^{1} \left( (n_{+} + n_{-}) \frac{\partial U}{\partial x} \right) \text{d}y 
	\quad
	\label{eq:2Is} 
\end{gather}
\noindent
{[BC-2]}: 
For the ion concentration field governed by the Nernst-Planck equation (\eqn\ref{eq:5}), the concentration distribution ($n_{\text{j}}$) at the inlet is prescribed from the numerical solution of steady, fully developed electroviscous (EV) flow in a uniform microfluidic slit \citep{davidson2007electroviscous,bharti2008steady,dhakar2022electroviscous,dhakar2023cfd,dhakar2023contraction,dhakar2023uniform}. At the outflow boundary, a zero axial concentration gradient is enforced for each ionic species. Along the channel walls, the normal component of the total ionic flux ($\mathbf{f}_j$) is set to zero, ensuring no penetration of ions across the walls. 
\begin{gather}
	\left.n_{\pm} \right|_{i} = \exp({\mp\psi_{\text{0}}}), \qquad  
	\left.\frac{\partial n_{\text{j}}}{\partial x}\right|_{o} = 0,
	\qquad \text{and} \qquad
	\left.(\myvec{f}_{\text{j}}\cdot \myvec{n})\right|_{w} = 0
	\label{eq:6} 
\end{gather}
Here, $\psi_{0}(y)$ represents the fully developed electric double layer (EDL) potential at inlet. The ionic flux ($\mathbf{f}_{\text{j}}$) is expressed by the Einstein relation \citep{davidson2007electroviscous,dhakar2022electroviscous} as follows.
\begin{gather}
\myvec{f}_{\text{j}} = n_{\text{j}}	\myvec{V} - ({1}/{Pe}) \left(\nabla n_{\text{j}}  + n_{\text{j}} \nabla U\right) 	\label{eq:6a} 
\end{gather}
%
{[BC-3]}: 
For the flow field governed by the Navier–Stokes equations (\eqns\ref{eq:9} - \ref{eq:10}), the inlet boundary condition is specified by the fully developed velocity profile, obtained from the numerical solution of electroviscous (EV) flow in a uniform microfluidic slit \citep{davidson2007electroviscous,bharti2008steady,dhakar2022electroviscous,dhakar2023cfd,dhakar2023contraction,dhakar2023uniform}. At the outflow boundary, which is exposed to the ambient, a zero-gradient condition is applied to the velocity field. Along the slit walls, the no-slip condition is enforced, expressed as follows.
\begin{gather}
	\left.\myvec{V}\right|_{i}=(V_{\text{0}}(y), \, 0), \qquad
	\left.\frac{\partial \myvec{V}}{\partial x}\right|_{o} = 0,
	\qquad
	\left.P\right|_{o} =0,
	\qquad
	\left.\myvec{V}\right|_{w}=(0,\, 0), \qquad
	\label{eq:11} 
\end{gather}
The coupled mathematical model, together with the corresponding boundary conditions,  (\eqns\ref{eq:1}–\ref{eq:11}) is numerically solved using the finite element method (FEM) to obtain the flow fields ($U$, $n_{\pm}$, $P$, $\mathbf{V}$) within the oppositely charged microfluidic slit, as functions of the governing flow parameters. It is noteworthy that the present mathematical model (\eqns\ref{eq:1}–\ref{eq:11}) is largely similar to that employed in our recent study \citep{dhakar2023cfd}, with the primary distinction arising from the treatment of the surface charge density ratio ($S_{\text{r}}$, \eqn\ref{eq:3}). In particular, the current work focuses on oppositely charged walls (case B, see \tab\ref{tab:srf}), whereas the previous study \citep{dhakar2023cfd} considered similarly charged walls (case A) in a non-uniform contraction–expansion microfluidic slit.
%
\section{Numerical approach}
\label{sec:sanp}
%
\noindent
In the present study, a commercial computational fluid dynamics (CFD) solver, COMSOL Multiphysics, based on the finite element method (FEM), is employed to numerically solve the mathematical model (\eqns\ref{eq:1} - \ref{eq:11}) describing electroviscous (EV) flow of a liquid electrolyte through an oppositely charged, non-uniform contraction-expansion microfluidic slit. The detailed numerical methodology has been thoroughly discussed in our recent works \citep{dhakar2022electroviscous,dhakar2023cfd,dhakar2023contraction,dhakar2023uniform}; therefore, only the essential features are briefly summarized here to avoid redundancy.
The mathematical model is implemented in COMSOL Multiphysics using appropriate physics interfaces such as the \textit{electrostatic} (es) module for the Poisson equation (\eqn\ref{eq:1}), the \textit{transport of dilute species} (tds) module for the Nernst-Planck equation (\eqn\ref{eq:5}), and the \textit{laminar flow} (spf) module for the Navier-Stokes equations (\eqns\ref{eq:9} - \ref{eq:10}). The integrals in \eqn(\ref{eq:2Is}) are evaluated using the \texttt{intop} operator, defined within the global definitions section of the model coupling framework in COMSOL.
The two-dimensional computational domain is discretized using a uniform, rectangular, structured mesh (see \tab\ref{tab:pm}). An iterative solution strategy is employed, wherein simulations are carried out using fully coupled solvers, PARDISO for the linear systems and Newton's method for nonlinear systems. The steady-state numerical solution yields the total electrical potential ($U$), ionic concentrations ($n_{\pm}$), velocity field ($\myvec{V}$), and pressure field ($P$).

\noindent 
The numerical parameters used in this study have been systematically optimized in our recent works and are summarized in \tab\ref{tab:pm}; the practical relevance of the chosen parameter ranges has been thoroughly justified in these studies \citep{dhakar2022electroviscous,dhakar2023cfd,dhakar2023contraction,dhakar2023uniform}. Furthermore, the present numerical modeling approach has been validated in our previous investigations \citep{dhakar2022electroviscous,dhakar2023cfd,dhakar2023contraction,dhakar2023uniform} against the literature \citep{davidson2007electroviscous} for electroviscous (EV) flow through a symmetrically charged ($S_{\text{r}} = 1$) contraction-expansion microchannel. To avoid redundancy, the validation is not repeated here. Based on these prior studies, the new numerical results presented in the following section are expected to be accurate within $\pm 1-2\%$ numerical uncertainty.
\begin{table}[t!]
	\centering
	\caption{Dimensionless parameters considered in the present study.}\label{tab:pm}
	\scalebox{0.83}
	{\renewcommand{\arraystretch}{1.5}
		\begin{tabular}{|l|c|c|c|c|c|c|}
			\hline
			Geometrical parameters & {$W$} & {$L_\text{u}=L_\text{d}=L_\text{c}$} &  {$L$} & \multicolumn{3}{c|}{$d_\text{c} = (W_\text{c}/W)$} \\\cline{2-7}
			& 1 & 5 & 15 & \multicolumn{3}{c|}{0.25} \\\hline
			Mesh parameters & \multicolumn{2}{c|}{Mesh points per unit length ($\Delta$)} & Number of elements ($N_\text{e}$) & \multicolumn{3}{c|}{Degree of freedom (DoF)} \\\cline{2-7}
			&  \multicolumn{2}{c|}{100}  & 333600 & \multicolumn{3}{c|}{3018814}  \\\hline
			Flow parameters & $K$&${S_\text{t}}$&${S_\text{r}}$& $Re$  &$\mathit{Sc}$ & $\beta$\\\cline{2-7}
			& $2\le K\le 20$ & $4\le \sut\le 16$ & $-2\le \sur\le 0$ &  $10^{-2}$ & $10^3$ & $2.34\times10^{-4}$\\\cline{2-7}
			&$2K~|~K\in[1..10]$&$2{S_\text{t}}~|~{S_\text{t}}\in[2^1..2^3]$&$-0.25{S_\text{r}}~|~{S_\text{r}}\in[0..8]$  & & &  \\\hline
		\end{tabular}
	}	
\end{table}
%
\section{Results and discussion}
%
\noindent
This section examines the hydrodynamic characteristics of electroviscous (EV) flow of a symmetric electrolyte through an oppositely charged (case B) contraction-expansion microfluidic slit across a broad range of governing parameters ($K$, $S_{\text{t}}$, $S_{\text{r}}$, see \tab\ref{tab:pm}). The study provides new insights into electrokinetic behavior by analyzing the total electrical potential ($U$), ionic concentrations ($n_{\pm}$), excess charge ($n^\ast$), induced electric field ($E_{\text{x}}$), velocity field ($\myvec{V}$), pressure distribution ($P$), and the electroviscous correction factor ($Y$) as functions of these parameters (\tab\ref{tab:pm}).
Furthermore, the influence of the oppositely charged surface ratio ($S_\text{r} < 0$) on hydrodynamic quantities is examined by introducing a scaling factor ($\lambda_\text{n}$) and a normalization parameter ($\theta_\text{r}$). The flow characteristics are represented as $\lambda = (U, n^{\ast}, E_{\text{x}}, P)$ and their normalized counterparts as $\theta = (\Delta U, n^\ast_\text{c}, E_{\text{x,c}}, \Delta P)$, under identical conditions, as expressed below.
\begin{gather}
	\lambda_\text{n} = \left(\frac{\lambda - \lambda_{\text{max}}}{\lambda_{\text{max}} - \lambda_{\text{min}}}  \right), 
	\label{eq:MR}
	\qquad\text{and}\qquad \theta_\text{r} = \left( \frac{\theta}{\theta_{\text{ref}}}  \right)
\end{gather}
The subscripts \textit{min}, \textit{max}, and \textit{ref} denote the minimum, maximum, and reference ($S_\text{r} \ge 0$; \cite{dhakar2023cfd}) values of the quantities $\lambda$ and $\theta$, evaluated under identical flow-governing conditions ($K$, $S_{\text{t}}$, $S_{\text{r}}$).
%
\subsection{Total electrical potential (\mm{U})}
\label{sec:potential}
%
\begin{figure}[!b]
	\centering\includegraphics[width=1\linewidth]{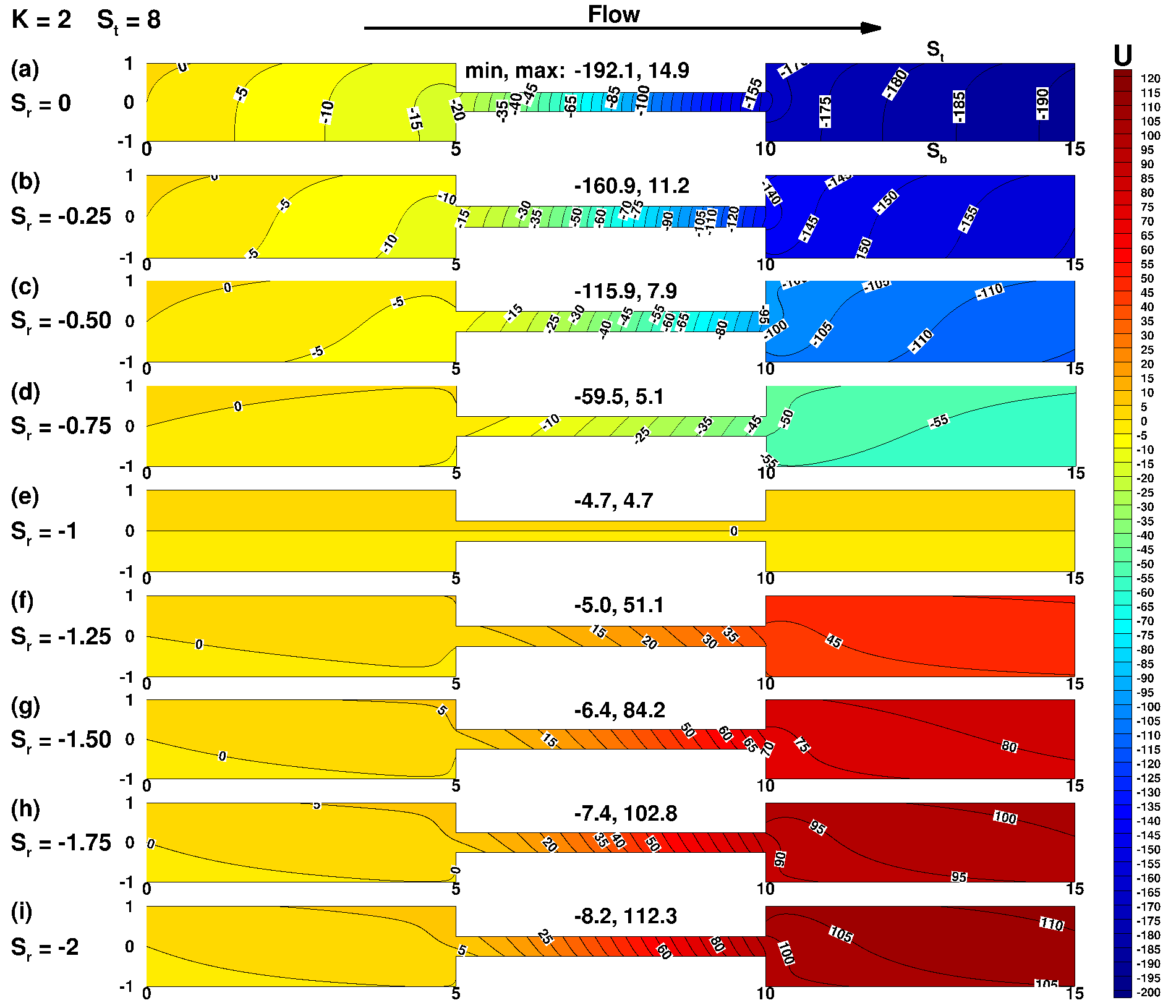}
	\caption{Distribution of total electrical potential (\mm{U}) in the oppositely charged microfluidic contraction-expansion slit for \mm{-2\le \sur\le 0}, \mm{K=2} and \mm{\sut=8}.}
	\label{fig:2}
\end{figure} 
\fig\ref{fig:2} illustrates the distribution of the total electrical potential ($U$, \eqn\ref{eq:1}) in the oppositely charged  contraction-expansion microfluidic slit for a fixed set of conditions ($-2\le S_{\text{r}}\le 0$, $K=2$, $S_{\text{t}}=8$); contours for other conditions (\tab\ref{tab:pm}) are qualitatively similar and are therefore not presented here. At $S_{\text{r}}=-1$ (case B13), the device tends toward electrical neutrality due to equally but oppositely charged walls, as seen in \fig\ref{fig:2}e. In this case, the minimum and maximum electrical potentials are recorded as $U_{\text{min}}=-1.6$ and $U_{\text{max}}=7.7$,  
respectively, for $K=2$ and $S_{\text{t}}=8$.
While the contours of $U$ are inclined in the pressure-driven flow (PDF) direction for cases B11 and B12 ($-1 < S_{\text{r}} \le 0$, see \figs\ref{fig:2}a–\ref{fig:2}d), a reverse inclination is observed for case B14 ($-2 \le S_{\text{r}} < -1$, see \figs\ref{fig:2}f–\ref{fig:2}i). This behavior arises because the top wall maintains a fixed positive charge ($S_{\text{t}}>0$), while the surface charge density ratio ($S_{\text{r}}$) varies from $0$ to $-2$.
Furthermore, potential ($U$) decreases for case B12 ($-1 < S_{\text{r}} < 0$) and increases for case B14 ($-2 \le S_{\text{r}} < -1$), due to the advection of excess charge ($n^\ast<0$ and $n^\ast>0$, respectively) along the slit length ($x,0$). The potential gradient is higher in the contraction region because the reduced cross-sectional area leads to an increase in both flow velocity ($\mathbf{V}$) and excess charge ($n^{\ast}$) clustering, thereby intensifying the streaming potential ($\phi$). Consequently, the maximum potential gradient occurs within the contraction. Furthermore, the total potential ($U$) increases with decreasing $S_{\text{r}}$. For example, $U$ reaches a minimum of $-191.9$ 
at $S_{\text{r}}=0$ (\fig\ref{fig:2}a) and a maximum of $116.4$  
at $S_{\text{r}}=-2$ (\fig\ref{fig:2}i) for $S_{\text{t}}=8$ and $K=2$. 
%
\begin{figure}[htbp]
	\centering
	\subfigure[$U=f(x,K)$ for $-2 \le S_{\text{r}}\le 0$ and $4 \le S_{\text{t}}\le 16$] {\includegraphics[width=0.9\linewidth]{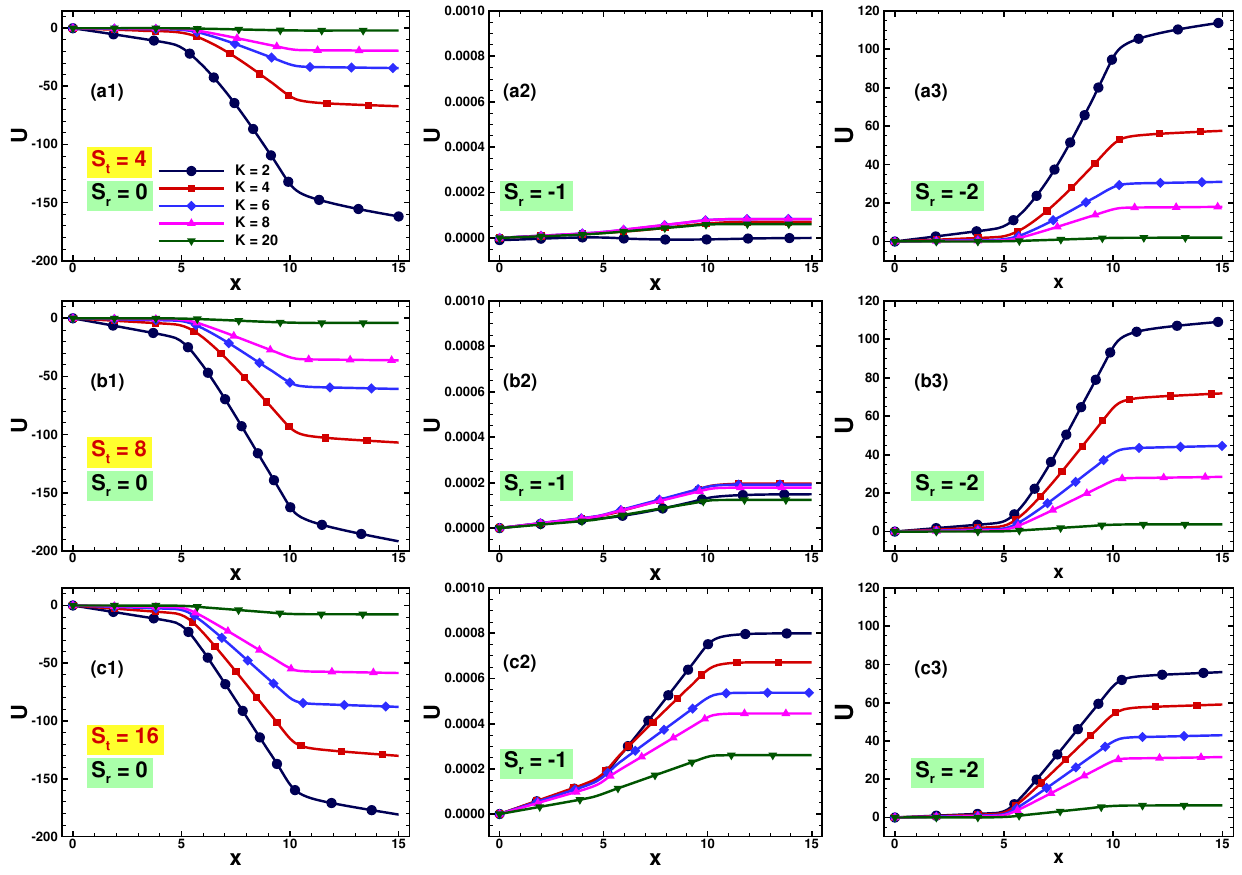}}
	\subfigure[$U=f(x,S_{\text{r}})$  for $2 \le K \le 20$ and $4 \le S_{\text{t}}\le 16$] {\includegraphics[width=0.9\linewidth]{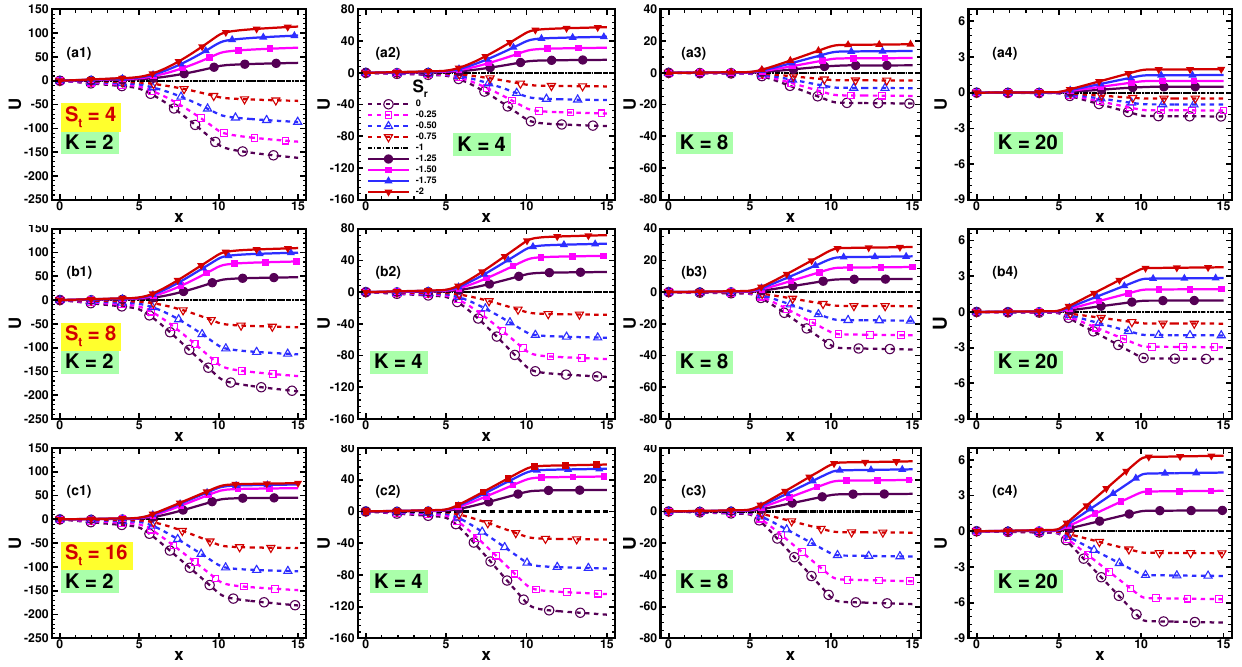}}
	\caption{Centreline profiles of the total electrical potential ($U$) in the oppositely charged micro-slit.}
	\label{fig:3}
\end{figure} 

\noindent
Subsequently, the centerline profiles of the total potential ($U$) in the oppositely charged micro-slit are presented in \fig\ref{fig:3} over the considered range of parameters (\tab\ref{tab:pm}). At $S_{\text{r}} = -1$ (case B13), the potential approaches zero owing to overall electrical neutrality, irrespective of the other parameters ($K$, $S_{\text{t}}$). The centerline profiles of $U$ exhibit qualitatively similar trends for cases B13 and B14 ($-2 \leq S_{\text{r}} \leq -1$), whereas opposite trends are observed for cases B11 and B12 ($-1 < S_{\text{r}} \leq 0$). These observations are consistent with earlier limited studies \citep{davidson2007electroviscous,dhakar2022electroviscous,dhakar2023cfd} across a range of $K$ and $S_{\text{t}}$ conditions.
Broadly, the potential ($U$) along the centreline of the oppositely charged slit decreases for case B12 ($-1 < S_{\text{r}} < 0$) and increases for case B14 ($-2 \leq S_{\text{r}} < -1$). The maximum potential gradient appears in the contraction region, relative to the upstream and downstream sections, due to the reduced cross-sectional area that accelerates the flow and concentrates excess charge. The potential drop ($\Delta U$) decreases for case B14 (and increases for case B12) with decreasing $K$ (i.e., thickening of EDL), irrespective of $S_{\text{t}}$ (\fig\ref{fig:3} and \tab\ref{tab:1}). 
%
\begin{sidewaystable}
	\centering
	\caption{The total potential drop ({$\Delta U$}), critical value of excess charge ({$n^\ast_\text{c}$}), critical value of induced electric field strength ({$E_{\text{x,c}}$}), and pressure drop ({$10^{-5}|\Delta P|$}) on the centreline of the oppositely charged micro slit. The maximum value of quantities ({$\Delta U$}, {$n^\ast_\text{c}$}, {$E_{\text{x,c}}$}, {$10^{-5}|\Delta P|$}) are highlighted in bold for each combination of {$\sut$} and {$K$}.} \label{tab:1}
	\scalebox{0.6}
	{\renewcommand{\arraystretch}{1.5}
		\begin{tabular}{|r|r|r|r|r|r|r|r|r|r|r|r|r|r|r|r|r|r|r|r|}
			\hline
			 {$S_{\text{t}}$} &	{$K$}	&	{$S_{\text{r}}$=} 0	&	$-0.25$	& $-0.50$ &	$-0.75$	& $-1$ & $-1.25$ & $-1.50$ & $-1.75$ & $-2$ & {$S_{\text{r}}$=} 0	&	$-0.25$	& $-0.50$ &	$-0.75$	& $-1$ & $-1.25$ & $-1.50$ & $-1.75$ & $-2$ \\\cline{3-20} 
			&	&	\multicolumn{9}{c|}{$\Delta U$} &	\multicolumn{9}{c|}{$n^\ast_\text{c}$}	\\\hline 
			0	&	{$\infty$}	&	\multicolumn{9}{c|}{0}	&	\multicolumn{9}{c|}{0}     	  \\\hline
			4	&	2	&	$-161.9300$	&	$-127.9900$	&	$-86.5040$	&	$-42.2380$	&	{$-8.9001\times10^{-11}$}	&	37.6170	&	69.5340	&	95.0680	&	\textbf{113.9700}	&	$-3.0562$	&	$-1.9615$	&	$-1.0570	$&	$-0.4419$	&	{$-1.9297\times10^{-6}$}	&	0.4206	&	0.9416	&	1.6151	&	\textbf{2.3557}	\\
			&	4	&	$-67.1180$	&	$-51.2230$	&	$-34.3620$	&	$-17.0980$	&	{$7.0868\times10^{-5}$}	&	16.4040	&	31.6680	&	45.4580	&	\textbf{57.5750}	&	$-0.7520$	&	$-0.5525$	&	$-0.3603$	&	$-0.1766$	&	{$-1.0713\times10^{-6}$}	&	0.1723	&	0.3418	&	0.5080	&	\textbf{0.6682}	\\
			&	6	&	$-34.3130$	&	$-25.9000$	&	$-17.2790$	&	$-8.5965$	&	{$8.3455\times10^{-5}$}	&	8.3704	&	16.3920	&	23.9630	&	\textbf{31.0150}	&	$-0.3006$	&	$-0.2249$	&	$-0.1492$	&	$-0.0740$	&	{$-6.2878\times10^{-7}$}	&	0.0724	&	0.1428	&	0.2107	&	\textbf{0.2759}	\\
			&	8	&	$-19.3580$	&	$-14.5380$	&	$-9.6751$	&	$-4.8141$	&	{$8.2752\times10^{-5}$}	&	4.7256	&	9.3245	&	13.7630	&	\textbf{18.0150}	&	$-0.1353$	&	$-0.1013$	&	$-0.0673$	&	$-0.0335$	&	{$-3.6344\times10^{-7}$}	&	0.0329	&	0.0652	&	0.0966	&	\textbf{0.1272}	\\
			&	20	&	$-1.9993$	&	$-1.4987$	&	$-0.9982$	&	$-0.4985$	&	{$6.1060\times10^{-5}$}	&	0.4968	&	0.9913	&	1.4830	&	\textbf{1.9715}	&	$-0.0027$	&	$-0.0020$	&	$-0.0013$	&	$-0.0007$	&	{$-1.0961\times10^{-8}$}	&	0.0007	&	0.0013	&	0.0020	&	\textbf{0.0026}	\\\hline
			8	&	2	&	$-191.8400$	&	$-159.7500$	&	$-113.9700$	&	$-56.9810$	&	{$1.4877\times10^{-4}$}	&	47.9980	&	81.1500	&	99.7090	&	\textbf{109.2600}	&	$-6.1007$	&	$-4.2743$	&	$-2.3557$	&	$-0.8640$	&	{$-3.3591\times10^{-6}$}	&	0.7972	&	1.7527	&	2.5729	&	\textbf{3.2824}	\\
			&	4	&	$-106.8800$	&	$-84.3260$	&	$-57.5750$	&	$-28.5620$	&	{$1.9630\times10^{-4}$}	&	25.3960	&	45.8590	&	61.0930	&	\textbf{71.8870}	&	$-1.4315$	&	$-1.0447$	&	$-0.6682$	&	$-0.3199$	&	{$-1.9058\times10^{-6}$}	&	0.2967	&	0.5623	&	0.7873	&	\textbf{0.9730}	\\
			&	6	&	$-60.7930$	&	$-46.4760$	&	$-31.0150$	&	$-15.2290$	&	{$1.9165\times10^{-4}$}	&	13.9200	&	26.0580	&	36.2520	&	\textbf{44.5810}	&	$-0.5650$	&	$-0.4205$	&	$-0.2759$	&	$-0.1348$	&	{$-1.1743\times10^{-6}$}	&	0.1261	&	0.2413	&	0.3442	&	\textbf{0.4348}	\\
			&	8	&	$-36.1470$	&	$-27.2180$	&	$-18.0150$	&	$-8.8444$	&	{$1.7810\times10^{-4}$}	&	8.2776	&	15.8230	&	22.5520	&	\textbf{28.4460}	&	$-0.2589$	&	$-0.1931$	&	$-0.1272$	&	$-0.0624$	&	{$-6.9222\times10^{-7}$}	&	0.0592	&	0.1146	&	0.0038	&	\textbf{0.2127}	\\
			&	20	&	$-3.9648$	&	$-2.9673$	&	$-1.9713$	&	$-0.9807$	&	{$1.2377\times10^{-4}$}	&	0.9676	&	1.9180	&	2.8483	&	\textbf{3.7560}	&	$-0.0053$	&	$-0.0039$	&	$-0.0026$	&	$-0.0013$	&	{$-2.1528\times10^{-8}$}	&	0.0013	&	0.0026	&	0.0038	&	\textbf{0.0050}	\\\hline
			16	&	2	&	$-180.9200$	&	$-148.7300$	&	$-109.2600$	&	$-60.3270$	&	{$7.9961\times10^{-4}$}	&	45.6050	&	65.5950	&	73.3390	&	\textbf{76.0880}	&	$-9.6773$	&	$-6.3706$	&	$-3.2824$	&	$-1.2989$	&	{$-3.5223\times10^{-6}$}	&	0.9840	&	1.6194	&	2.0671	&	\textbf{2.4217}	\\
			&	4	&	$-130.1300$	&	$-104.2300$	&	$-71.8860$	&	$-35.4190$	&	{$6.7165\times10^{-4}$}	&	27.1350	&	44.0940	&	53.7830	&	\textbf{59.1340}	&	$-2.3184$	&	$-1.5865$	&	$-0.9730$	&	$-0.4657$	&	{$-2.2853\times10^{-6}$}	&	0.3814	&	0.6492	&	0.8370	&	\textbf{0.9769}	\\
			&	6	&	$-87.7650$	&	$-67.7610$	&	$-44.5800$	&	$-21.0180$	&	{$5.3680\times10^{-4}$}	&	16.6820	&	28.8540	&	37.3180	&	\textbf{43.0510}	&	$-0.9403$	&	$-0.6812$	&	$-0.4348$	&	$-0.2068$	&	{$-1.7307\times10^{-6}$}	&	0.1756	&	0.3158	&	0.4253	&	\textbf{0.5112}	\\
			&	8	&	$-58.4520$	&	$-43.9750$	&	$-28.4450$	&	$-13.3790$	&	{$4.4497\times10^{-4}$}	&	11.0900	&	19.8510	&	26.5620	&	\textbf{31.6070}	&	$-0.4519$	&	$-0.3315$	&	$-0.2127$	&	$-0.1010$	&	{$-1.1337\times10^{-6}$}	&	0.0881	&	0.1628	&	0.2251	&	\textbf{0.2769}	\\
			&	20	&	$-7.6769$	&	$-5.7108$	&	$-3.7557$	&	$-1.8431$	&	{$2.6037\times10^{-4}$}	&	1.7533	&	3.4014	&	4.9358	&	\textbf{6.3530}	&	$-0.0103$	&	$-0.0076$	&	$-0.0050$	&	$-0.0025$	&	{$-4.0116\times10^{-8}$}	&	0.0024	&	0.0046	&	0.0067	&	\textbf{0.0086}	\\\hline
			&	& \multicolumn{9}{c|}{$E_{\text{x,c}}$} &	\multicolumn{9}{c|}{$10^{-5}|\Delta P|$}	\\\hline 
			0	&	{$\infty$}	&	\multicolumn{9}{c|}{0}	&	\multicolumn{9}{c|}{1.0616}     	  \\\hline
			4	&	2	&	\textbf{28.6420}	&	24.0290	&	16.9370	&	8.5316	&	{$1.5162\times10^{-5}$}	&	$-7.9614$	&	$-14.8660$	&	$-20.2000$	&	$-23.6110$	&	\textbf{1.0931}	&	1.0777	&	1.0678	&	1.0629	&	1.0616	&	1.0628	&	1.0664	&	1.0726	&	1.0812	\\
			&	4	&	\textbf{14.1750}	&	11.0210	&	7.5012	&	3.7685	&	{$8.4143\times10^{-6}$}	&	$-3.6283$	&	$-6.9619$	&	$-9.8847$	&	$-12.3350$	&	\textbf{1.0774}	&	1.0705	&	1.0655	&	1.0626	&	1.0616	&	1.0625	&	1.0652	&	1.0694	&	1.0748	\\
			&	6	&	\textbf{7.1829}	&	5.4513	&	3.6502	&	1.8193	&	{$6.1257\times10^{-6}$}	&	$-1.7672$	&	$-3.4472$	&	$-5.0126$	&	$-6.4453$	&	\textbf{1.0706}	&	1.0667	&	1.0639	&	1.0622	&	1.0617	&	1.0622	&	1.0638	&	1.0663	&	1.0696	\\
			&	8	&	\textbf{3.9032}	&	2.9352	&	1.9549	&	0.9729	&	{$4.6740\times10^{-6}$}	&	$-0.9537$	&	$-1.8792$	&	$-2.7684$	&	$-3.6154$	&	\textbf{1.0667}	&	1.0645	&	1.0629	&	1.0620	&	1.0617	&	1.0620	&	1.0629	&	1.0644	&	1.0663	\\
			&	20	&	\textbf{0.3821}	&	0.2864	&	0.1908	&	0.0953	&	{$1.6644\times10^{-6}$}	&	$-0.0949$	&	$-0.1894$	&	$-0.2834$	&	$-0.3767$	&	\textbf{1.0619}	&	1.0618	&	1.0617	&	1.0616	&	1.0616	&	1.0616	&	1.0617	&	1.0618	&	1.0620	\\\hline
			8	&	2	&	\textbf{31.4090}	&	28.9440	&	23.6110	&	12.9820	&	{$2.0787\times10^{-5}$}	&	$-11.3050$	&	$-18.0670$	&	$-20.6500$	&	$-21.4580$	&	\textbf{1.1489}	&	1.1110	&	1.0812	&	1.0656	&	1.0616	&	1.0652	&	1.0756	&	1.0896	&	1.1038	\\
			&	4	&	\textbf{20.9100}	&	17.3080	&	12.3350	&	6.2886	&	{$8.3464\times10^{-6}$}	&	$-5.5654$	&	$-9.7899$	&	$-12.6330$	&	$-14.4240$	&	\textbf{1.1140}	&	1.0917	&	1.0748	&	1.0648	&	1.0617	&	1.0645	&	1.0717	&	1.0816	&	1.0927	\\
			&	6	&	\textbf{12.3020}	&	9.5507	&	6.4453	&	3.1811	&	{$7.9007\times10^{-6}$}	&	$-2.8833$	&	$-5.3389$	&	$-7.3330$	&	$-8.8999$	&	\textbf{1.0931}	&	1.0796	&	1.0696	&	1.0637	&	1.0619	&	1.0636	&	1.0682	&	1.0747	&	1.0824	\\
			&	8	&	\textbf{7.2044}	&	5.4483	&	3.6154	&	1.7760	&	{$7.0415\times10^{-6}$}	&	$-1.6549$	&	$-3.1502$	&	$-4.4672$	&	$-5.6040$	&	\textbf{1.0802}	&	1.0721	&	1.0663	&	1.0629	&	1.0619	&	1.0630	&	1.0658	&	1.0701	&	1.0753	\\
			&	20	&	\textbf{0.7576}	&	0.5670	&	0.3767	&	0.1874	&	{$3.1817\times10^{-6}$}	&	$-0.1849$	&	$-0.3664$	&	$-0.5441$	&	$-0.7175$	&	\textbf{1.0629}	&	1.0624	&	1.0620	&	1.0617	&	1.0617	&	1.0618	&	1.0621	&	1.0625	&	1.0630	\\\hline
			16	&	2	&	\textbf{29.2120}	&	26.1980	&	21.4580	&	13.5040	&	{$1.4973\times10^{-5}$}	&	$-10.0490$	&	$-13.1870$	&	$-14.0330$	&	$-14.1550$	&	\textbf{1.2222}	&	1.1554	&	1.1038	&	1.0719	&	1.0617	&	1.0692	&	1.0833	&	1.0970	&	1.1090	\\
			&	4	&	\textbf{23.3610}	&	19.7430	&	14.4240	&	7.4291	&	{$7.9317\times10^{-6}$}	&	$-5.5806$	&	$-8.7436$	&	$-10.3560$	&	$-11.1470$	&	\textbf{1.1839}	&	1.1325	&	1.0927	&	1.0691	&	1.0622	&	1.0673	&	1.0781	&	1.0901	&	1.1014	\\
			&	6	&	\textbf{16.7450}	&	13.2550	&	8.8999	&	4.2350	&	{$6.9166\times10^{-6}$}	&	$-3.3115$	&	$-5.6445$	&	$-7.1966$	&	$-8.1986$	&	\textbf{1.1470}	&	1.1100	&	1.0824	&	1.0670	&	1.0627	&	1.0662	&	1.0740	&	1.0835	&	1.0932	\\
			&	8	&	\textbf{11.3330}	&	8.6103	&	5.6040	&	2.6396	&	{$7.2685\times10^{-6}$}	&	$-2.1693$	&	$-3.8558$	&	$-5.1216$	&	$-6.0508$	&	\textbf{1.1181}	&	1.0932	&	1.0753	&	1.0655	&	1.0628	&	1.0653	&	1.0710	&	1.0784	&	1.0863	\\
			&	20	&	\textbf{1.4662}	&	1.0908	&	0.7175	&	0.3522	&	{$5.4371\times10^{-6}$}	&	$-0.3349$	&	$-0.6498$	&	$-0.9428$	&	$-1.2132$	&	\textbf{1.0668}	&	1.0646	&	1.0630	&	1.0622	&	1.0620	&	1.0624	&	1.0633	&	1.0646	&	1.0662	\\\hline		
		\end{tabular}
	}
\end{sidewaystable}
For example, the variation in $\Delta U$ with increasing $K$ (from $2$ to $20$) is found to be {98.77\percent}, {97.93\percent}, and {95.76\percent} at $S_{\text{r}} = 0$; {98.82\percent}, {98.28\percent}, and {96.94\percent} at $S_{\text{r}} = -0.75$; {98.68\percent}, {97.98\percent}, and {96.16\percent} at $S_{\text{r}} = -1.25$; and {98.27\percent}, {96.56\percent}, and {91.65\percent} at $S_{\text{r}} = -2$, corresponding to $S_{\text{t}} = 4$, $8$, and $16$, respectively, refer \tab\ref{tab:1}.
\begin{figure}[!b]
	\centering\includegraphics[width=1\linewidth]{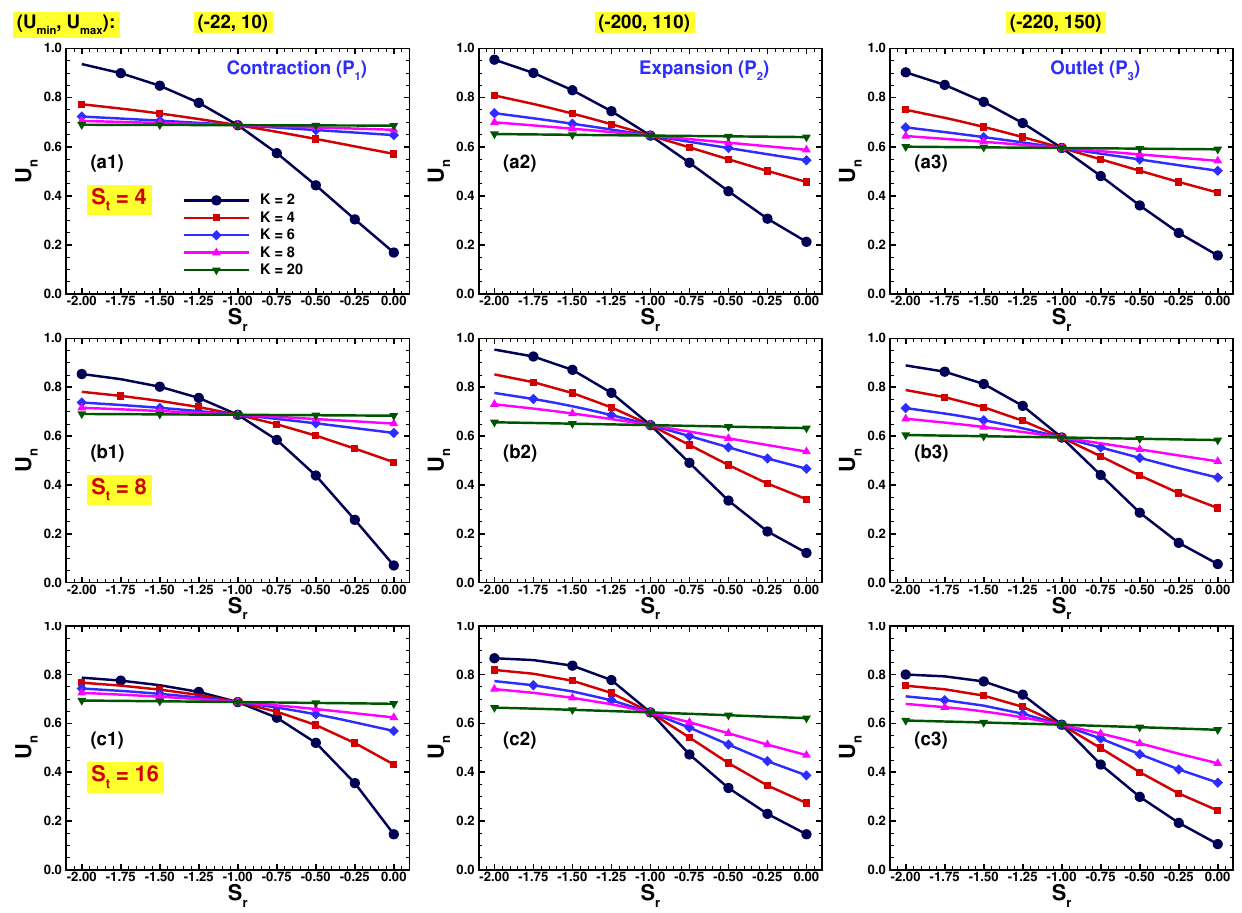}
	\caption{Influence of dimensionless parameters (${K,\sut,\sur}$) on scaled total electrical potential (${U_\text{n}}$) variation on the contraction (${P_1}$, first column), expansion (${P_2}$, second column), and outlet (${P_3}$, third column) centreline points of the oppositely charged micro slit.}
	\label{fig:3c}
\end{figure} 
The influence of $K$ on $\Delta U$ is most pronounced at higher $S_{\text{r}}$ and lower $S_{\text{t}}$.
The variation of $\Delta U$ with $S_{\text{t}}$ (increasing from $4$ to $16$) is most pronounced at higher values of $S_{\text{r}}$ and $K$ (\tab\ref{tab:1}). At $K=2$, the changes in $\Delta U$ are {11.73{\percent}}, {42.83{\percent}}, {21.24{\percent}}, and {33.24{\percent}} for $S_{\text{r}}=0$, $-0.75$, $-1.25$, and $-2$, respectively. At $K=20$, the corresponding variations become much larger, amounting to {283.98{\percent}}, {269.76{\percent}}, {252.90{\percent}}, and {222.40{\percent}}, respectively.
%
Further, the influence of $S_{\text{r}}$ on $\Delta U$ is most pronounced at the lowest $S_{\text{t}}$ and highest $K$ values. 
For example, when $S_{\text{r}}$ decreases from $0$ to $-0.75$, $\Delta U$ increases by {73.92{\percent}}, {70.30{\percent}}, and {66.66{\percent}} at $K=2$, and by {75.07{\percent}}, {75.26{\percent}}, and {75.99{\percent}} at $K=20$ for $S_{\text{t}}=4$, $8$, and $16$, respectively. A further decrease in $S_{\text{r}}$ from $-1.25$ to $-2$ enhances $\Delta U$ by {202.97{\percent}}, {127.63{\percent}}, and {66.84{\percent}} at $K=2$, and by {296.82{\percent}}, {288.19{\percent}}, and {262.35{\percent}} at $K=20$. Overall, reducing $S_{\text{r}}$ from $0$ to $-2$ increases $\Delta U$ by {170.38{\percent}}, {156.95{\percent}}, and {142.06{\percent}} at $K=2$, and by {198.61{\percent}}, {194.73{\percent}}, and {182.75{\percent}} at $K=20$ for $S_{\text{t}}=4$, $8$, and $16$, respectively (\tab\ref{tab:1}).
%
In general, $\Delta U$ decreases for $S_{\text{r}} > -1$ and increases for $S_{\text{r}} < -1$ with rising $S_{\text{t}}$, though opposite trends emerge at higher $S_{\text{t}}$ and lower $K$ (\fig\ref{fig:3}, \tab\ref{tab:1}). This behavior arises from the stronger electrostatic force near the walls, which enhances the transport of excess charge ($n^\ast < 0$ or $n^\ast > 0$), thereby altering the streaming current and potential drop. At sufficiently high $S_{\text{t}}$, however, the intensified electrostatic force suppresses excess-ion transport within the slit, leading to a reduction in $\Delta U$. Moreover, the potential drop ($\Delta U$) increases monotonically with decreasing $S_{\text{r}}$, regardless of $K$ and $S_{\text{t}}$ (\fig\ref{fig:3}, \tab\ref{tab:1}). This trend arises from the stronger electrostatic attraction at lower $S_{\text{r}}$, which enhances the convective transport of excess positive charge ($n^\ast > 0$) within the slit, thereby increasing the streaming potential and, consequently, $\Delta U$ as $S_{\text{r}}$ decreases from $0$ to $-2$.

\noindent
Furthermore, the total potential along the centreline of an oppositely charged micro-slit varies substantially, ranging from $U_\text{min} = -220$ to $U_\text{max} = 150$, depending on the parametric conditions ($S_\text{r}$, $S_\text{t}$, and $K$), as shown in \fig\ref{fig:3} and \tab\ref{tab:1}. To better understand the influence of these dimensionless parameters, the potential is scaled using \eqn\ref{eq:MR}, and the variation of the scaled potential ($U_\text{n}$) is analyzed at the centreline locations $P_1$, $P_2$, and $P_3$ of the micro-slit (\fig\ref{fig:3c}). A distinct cross-over in $U_\text{n}$ occurs at $S_\text{r} = -1$. Specifically, $U_\text{n}$ increases for $S_\text{r} < -1$ and decreases for $S_\text{r} > -1$ as $K$ decreases, regardless of $S_\text{t}$. The maximum variation with respect to $K$ is obtained at $P_1$ for $S_\text{t} = 8$ and $S_\text{r} = 0$, where $U_\text{n}$ increases by 872.63\% (from $0.0703$ to $0.6839$). Comparable increments of 420.41\% (from $0.1216$ to $0.6329$) and 667.17\% (from $0.0761$ to $0.5839$) are observed at $P_2$ and $P_3$, respectively, under the same conditions (\fig\ref{fig:3c}). Similarly, $U_\text{n}$ increases for $S_\text{r} < -1$ and decreases for $S_\text{r} > -1$ with increasing $S_\text{t}$. The influence of $S_\text{t}$ is most pronounced at $P_3$ for $S_\text{r} = 0$ and $K = 2$, where $U_\text{n}$ decreases by 32.7\% (from $0.1569$ to $0.1056$). At the same conditions, $U_\text{n}$  reduces by 13.57\% (from $0.1688$ to $0.1459$) and 31.36\% (from $0.2124$ to $0.1458$) are obtained at $P_1$ and $P_2$, respectively. This cross-over behavior of $U_\text{n}$ at $S_\text{r} = -1$ arises from the competing effects of electrostatic attraction and repulsion between oppositely charged walls, which regulate the accumulation and transport of excess ions, thereby controlling the sign and magnitude of the streaming potential.
The potential ($U_\text{n}$) increases consistently with decreasing surface charge ratio ($S_\text{r}$), irrespective of $K$ and $S_\text{t}$. The maximum sensitivity of $U_\text{n}$ to $S_\text{r}$ is observed at $P_1$ for $S_\text{t}=8$ and $K=2$. For example, the increments in $U_\text{n}$ are 1115.56\% (from $0.0703$ to $0.8547$), 685.03\% (from $0.1216$ to $0.9547$), and 1069.25\% (from $0.0761$ to $0.8899$) at $P_1$, $P_2$, and $P_3$, respectively, when $S_\text{r}$ decreases from $0$ to $-2$ at $S_\text{t}=8$ and $K=2$ (\fig\ref{fig:3c}).
Overall, these results indicate that the influence of the governing parameters differs across the centreline: the effect of $S_\text{t}$ is most pronounced at $P_3$, while the combined influence of $K$ and $S_\text{r}$ is dominant at $P_1$. This occurs because the reduction in cross-sectional area at $P_1$ enhances both the local velocity and clustering of excess charge, thereby amplifying the change in total potential ($U$, from \eqn\ref{eq:1}) and its normalized form ($U_\text{n}$) at the contraction relative to other centreline points (\fig\ref{fig:3c}). 

\noindent
Physically, this section highlights stronger electrokinetic coupling near the contraction zone, where geometric confinement intensifies charge transport and potential variation. The electrical potential ($U$) is related to the excess charge ($n^\ast$) through \eqn(\ref{eq:1}); accordingly, the dependence of $n^\ast$ on the governing parameters ($K$, $S_\text{t}$, $S_\text{r}$) is examined in the following section.
%
\subsection{Excess charge (\mm{n^\ast})}
\label{sec:charge}
\begin{figure}[t!]
	\centering\includegraphics[width=1\linewidth]{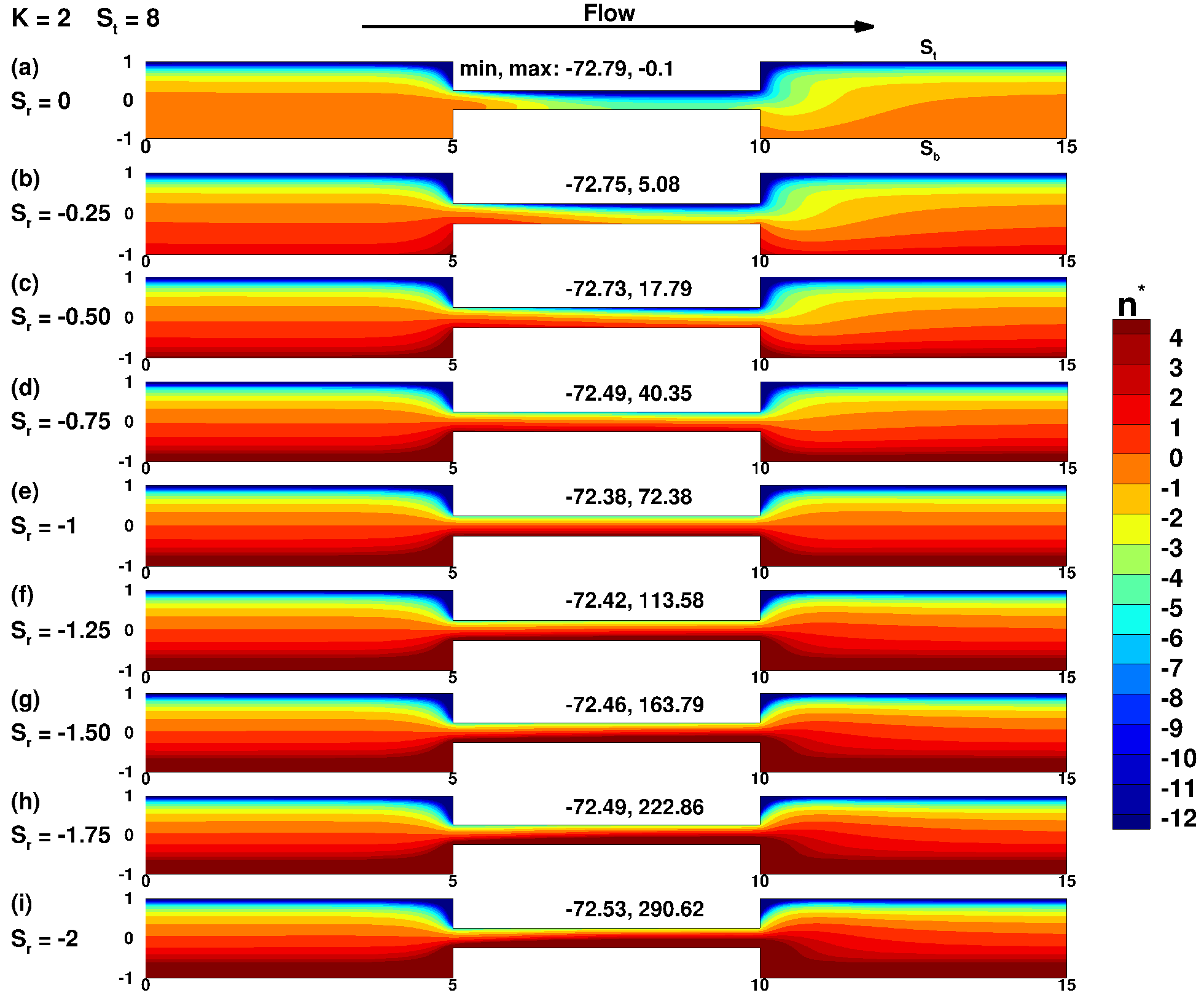}
	\caption{Distribution of excess charge (\mm{n^{\ast}}) in the oppositely charged micro slit for \mm{-2\le \sur\le 0}, \mm{K=2} and \mm{\sut=8}.}
	\label{fig:5}
\end{figure}

%
\noindent 
\fig\ref{fig:5} illustrates the distribution of excess charge ($n^\ast$, \eqn\ref{eq:1}) in the oppositely charged micro-slit for $-2 \le S_\text{r} \le 0$, $S_\text{t}=8$, and $K=2$. Qualitatively similar $n^\ast$ contour profiles are obtained under other conditions (\tab\ref{tab:pm}) and are therefore not shown. The clustering of excess charge ($n^\ast < 0$ and $n^\ast > 0$) is observed near the walls, consistent with the positively and negatively charged boundaries, and is more pronounced in the contraction due to the sudden reduction in flow area compared to other sections of micro-slit. At $S_\text{r} = -1$, equal and opposite charges are present on the walls, resulting in symmetric distributions with $n^\ast_\text{max} = -n^\ast_\text{min} = 72.38$ (\fig\ref{fig:5}e). The magnitude of $n^\ast$ increases with decreasing surface charge density ratio ($S_\text{r}$): it reaches its minimum in cases B11 and 12 ($0 \ge S_\text{r} \ge -1$) and its maximum in cases B13 and B14 ($-1 \ge S_\text{r} \ge -2$), due to the fixed $S_\text{t}$ on the top wall while $S_\text{r}$ varies on the opposite wall. Physically, a stronger charge asymmetry between the walls increases the net attractive electrostatic force, thereby intensifying excess charge accumulation in the slit. For example, the minimum and maximum values are $n^\ast_\text{min} = -72.79$ (\fig\ref{fig:5}a) and $n^\ast_\text{max} = 290.62$ (\fig\ref{fig:5}i) for $S_\text{r}=0$ and $-2$, respectively, at $S_\text{t}=8$ and $K=2$, over the range of conditions studied.
\begin{figure}[htbp]
	\centering
	\subfigure[\mm{n^{\ast}=f(K)^\#}]{\includegraphics[width=0.9\linewidth]{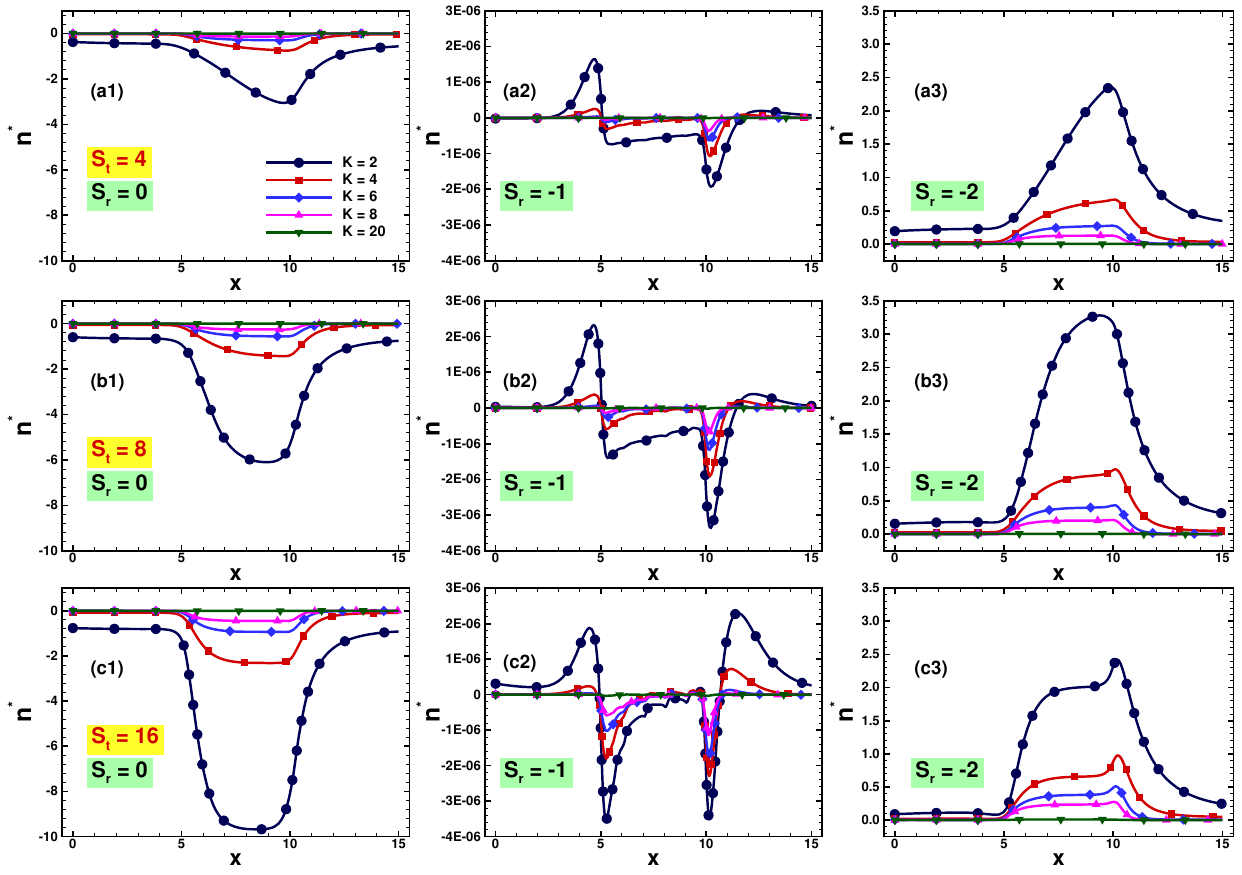}}
    \subfigure[\mm{n^{\ast}=f(\sur)^\#}]{\includegraphics[width=0.9\linewidth]{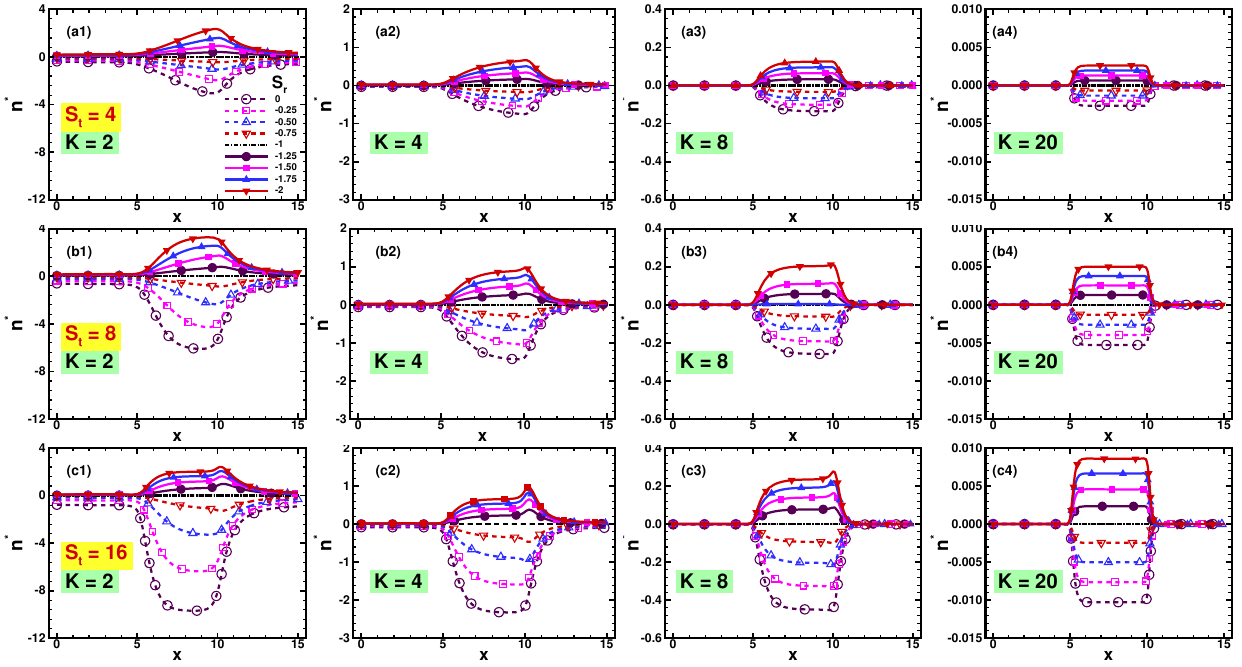}}
	\caption{Centreline profiles of excess charge (\mm{n^{\ast}}) in the oppositely charged micro slit for \mm{-2\le \sur\le 0}, \mm{4\le \sut\le 16}, and  \mm{2\le K\le 20}.}
	\label{fig:6}
\end{figure} 
\\\noindent 
Subsequently, \fig\ref{fig:6} presents the centreline profiles of excess charge ($n^\ast$) in the oppositely charged micro-slit for a range of parameters ($S_\text{r}$, $S_\text{t}$, $K$; \tab\ref{tab:pm}). At $S_\text{r}=-1$ (equal and oppositely charged walls), the centreline excess charge remains nearly zero throughout the device, with only small perturbations observed at the contraction and expansion points due to sharp geometric changes. For $S_\text{r} > -1$, the centreline $n^\ast$ profiles are qualitatively similar to earlier reports \citep{davidson2007electroviscous,dhakar2022electroviscous,dhakar2023cfd}, while for $S_\text{r} < -1$, the profiles exhibit an opposite trend, for the considered ranges of $K$ and $S_\text{t}$. In particular, smaller $n^\ast$ values are observed in the contraction for $S_\text{r} < -1$, whereas higher values are found for $S_\text{r} > -1$, compared to upstream/downstream sections. To capture these extrema, the critical excess charge ($n^\ast_\text{c}$, \tab\ref{tab:1}) is defined as the maximum (or minimum) value of $n^\ast$ on the centerline. The magnitude $|n^\ast_\text{c}|$ increases with decreasing $K$ (corresponding to a thicker EDL), with the strongest variation occurring at the largest $S_\text{r}$ and lowest $S_\text{t}$, refer \fig\ref{fig:6}.  For instance, when $K$ increases from 2 to 20, refer \tab\ref{tab:1}, $n^\ast_\text{c}$ at $S_\text{r}=0$ decreases from ($-3.0562$ to $-0.0027$), ($-6.1007$ to $-0.0053$), and ($-9.6773$ to $ -0.0103$) for $S_\text{t}=4, 8, 16$, respectively, corresponding to reductions of  about $99.89-99.91\%$. Similarly, at $S_\text{r}=-0.75$, $n^\ast_\text{c}$ decreases from ($-0.4419 $ to $ -0.0007$), ($-0.8640 $ to $ -0.0013$), and ($-1.2989 $ to $ -0.0025$) for the same $S_\text{t}$ values, with reductions of  about $99.81-99.85\%$. For $S_\text{r}=-1.25$, the centreline charge reverses polarity, with $n^\ast_\text{c}$ reducing from ($0.4206 $ to $ 0.0007$), ($0.7972 $ to $ 0.0013$), and ($0.9840 $ to $ 0.0024$), corresponding to  about $99.76-99.84\%$ reductions. Finally, at $S_\text{r}=-2$, the variation is from ($2.3557 $ to $ 0.0026$), ($3.2824$ to $0.0050$), and ($2.4217$ to $0.0086$) for $S_\text{t}=4, 8, 16$, respectively, yielding reductions of about $99.64-99.89\%$.
\\\noindent 
The maximum influence of $S_\text{t}$ on $n^\ast_\text{c}$ is observed at the highest values of $S_\text{r}$ and $K$. For instance, at $K=2$, $n^\ast_\text{c}$ changes from ($-3.0562$ to $-9.6773$), ($-0.4419$ to $-1.2989$), ($0.4206$ to $0.9840$), and ($2.3557$ to $2.4217$) for $S_\text{r}=0, -0.75, -1.25, -2$, respectively, when $S_\text{t}$ increases from 4 to 16. Similarly, at $K=20$, the variation is from ($-0.0027$ to $-0.0103$), ($-0.0007$ to $-0.0025$), ($0.0007$ to $0.0024$), and ($0.0026$ to $0.0086$) for the same $S_\text{r}$ values. These correspond to percentage changes of $\sim 216\%, 194\%, 134\%,$ and $2.8\%$ at $K=2$, and $\sim 286\%, 272\%, 256\%,$ and $229\%$ at $K=20$ for $S_\text{r}=0, -0.75, -1.25, -2$, respectively (\tab\ref{tab:1}).
On the other hand, the maximum influence of $S_\text{r}$ on $n^\ast_\text{c}$ is observed at the lowest $K$ and $S_\text{t}$. For example, when $S_\text{r}$ decreases from 0 to $-0.75$, $n^\ast_\text{c}$ changes from ($-3.0562$ to $-0.4419$), ($-6.1007$ to $-0.8640$), and ($-9.6773$ to $-1.2989$) at $K=2$, corresponding to enhancements of $\sim 85.5-86.6\%$. At $K=20$, the variation is ($-0.0027$ to $-0.0007$), ($-0.0053$ to $-0.0013$), and ($-0.0103$ to $-0.0025$), yielding $\sim 75.1-76.0\%$ enhancement for $S_\text{t}=4,8,16$, respectively. When $S_\text{r}$ decreases from $-1.25$ to $-2$, $n^\ast_\text{c}$ increases more sharply: from ($0.4206$ to $2.3557$), ($0.7972$ to $3.2824$), and ($0.9840$ to $2.4217$) at $K=2$ (an increase of $\sim 146-460\%$), and from ($0.0007$ to $0.0026$), ($0.0013$ to $0.0050$), and ($0.0024$ to $0.0086$) at $K=20$ (an increase of $\sim 267-297\%$). Overall, reducing $S_\text{r}$ from 0 to $-2$ enhances $n^\ast_\text{c}$ from ($-3.0562$ to $2.3557$), ($-6.1007$ to $3.2824$), and ($-9.6773$ to $2.4217$) at $K=2$ (an increase of $\sim 117-154\%$), and from ($-0.0027$ to $0.0026$), ($-0.0053$ to $0.0050$), and ($-0.0103$ to $0.0086$) at $K=20$ (an increase of $\sim 184-199\%$) for $S_\text{t}=4,8,16$, respectively (\tab\ref{tab:1}).
\\\noindent 
In general, $n^\ast_\text{c}$ decreases for $S_\text{r} > -1$ and increases for $S_\text{r} < -1$ with increasing $S_\text{t}$, irrespective of $K$. This behavior arises because negative ions dominate when $S_\text{r} > -1$, while positive ions dominate when $S_\text{r} < -1$. Furthermore, $n^\ast_\text{c}$ increases with $S_\text{r}$, regardless of $S_\text{t}$ and $K$ (\fig\ref{fig:6}, \tab\ref{tab:1}), as the enhanced attractive force strengthens the accumulation of excess positive charge ($n^\ast > 0$) in the micro-slit with increasing charge asymmetry ($S_\text{r}$).
\begin{figure}[t!]
	\centering\includegraphics[width=1\linewidth]{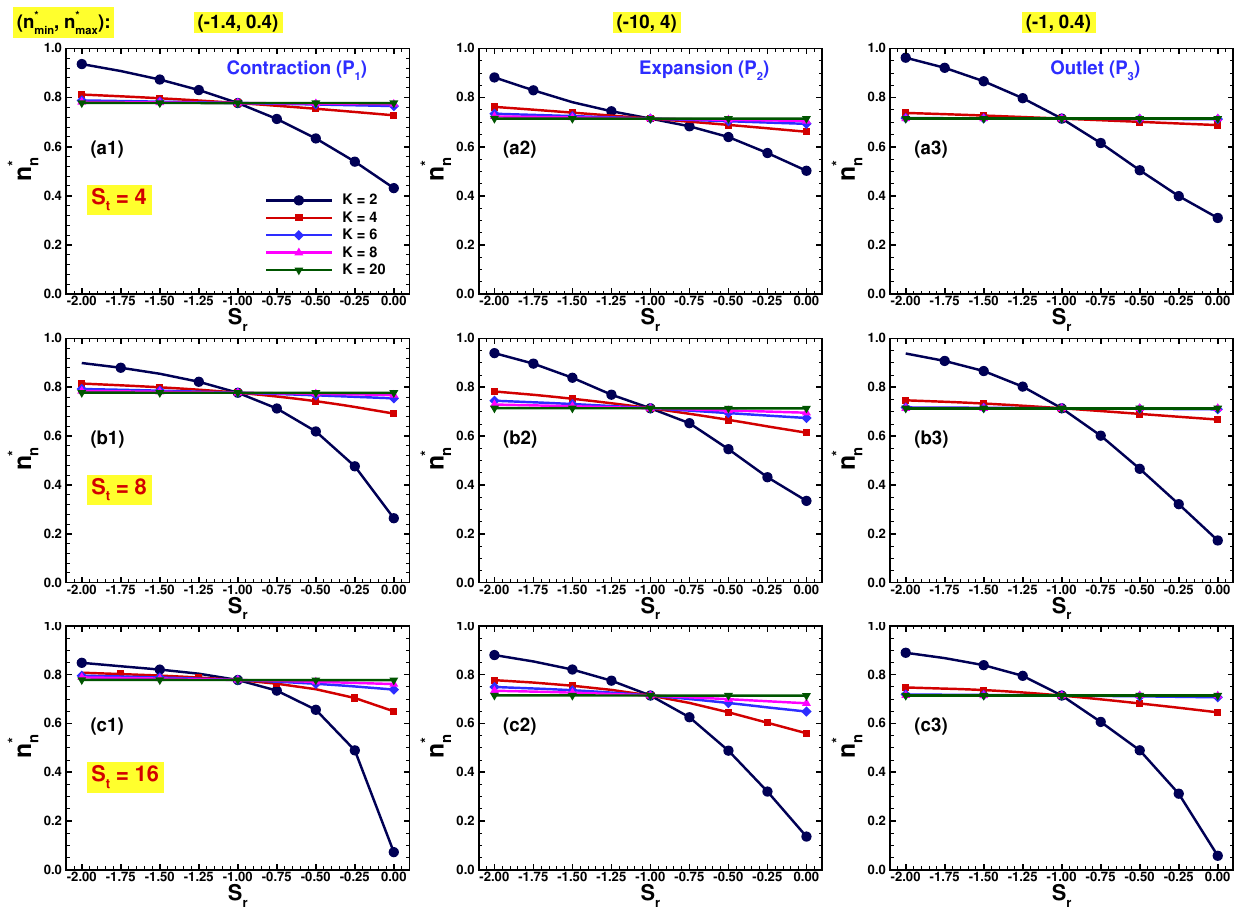}
	\caption{Influence of dimensionless parameters (\mm{K,\sut,\sur}) on scaled excess charge (\mm{n^{\ast}_\text{n}}) variation on the contraction (\mm{P_1}, first column), expansion (\mm{P_2}, second column), and outlet (\mm{P_3}, third column) centreline points of the oppositely charged micro slit.}
	\label{fig:6b}
\end{figure} 
\\\noindent 
To further examine the detailed influence of the dimensionless parameters ($K, S_\text{t}, S_\text{r}$) on the excess charge, it is scaled in the same manner as the potential, $U_\text{n}$ (see section \ref{sec:potential}). The corresponding variations of the scaled excess charge ($n^\ast_\text{n}$, \eqn\ref{eq:MR}) at the centreline points ($P_1, P_2, P_3$) of the micro-slit are shown in \fig\ref{fig:6b}. The dependence of $n^\ast_\text{n}$ on the governing parameters ($K, S_\text{t}, S_\text{r}$) is qualitatively similar to that of excess charge ($n^\ast$). For instance, $n^\ast_\text{n}$ decreases for $S_\text{r} > -1$ and increases for $S_\text{r} < -1$ with decreasing $K$. The strongest influence of $K$ is observed at $S_\text{t} = 16$ and $S_\text{r} = 0$ at $P_3$, where the maximum increments in $n^\ast_\text{n}$ are recorded as from $0.0726$ to $0.7772$ ($\sim 970.43\%$), from $0.1364$ to $0.7136$ ($\sim 423.29\%$), and from $0.0577$ to $0.7143$ ($\sim 1137.62\%$) at $P_1, P_2$, and $P_3$, respectively, when $K$ increases from 2 to 20 (\fig\ref{fig:6b}). Similarly, $n^\ast_\text{n}$ decreases for $S_\text{r} > -1$ and increases for $S_\text{r} < -1$ with increasing $S_\text{t}$, irrespective of $K$. The maximum influence of $S_\text{t}$ is obtained at $S_\text{r} = 0$ and $K = 2$ at $P_1$, where $n^\ast_\text{n}$ reduces most strongly with increasing $S_\text{t}$ (from 4 to 16), by $83.16\%$ (from $0.4311$ to $0.0726$), $72.82\%$ (from $0.5018$ to $0.1364$), and $81.36\%$ (from $0.3096$ to $0.0577$) at $P_1, P_2$, and $P_3$, respectively (\fig\ref{fig:6b}).
\\\noindent 
The excess charge ($n^{\ast}_\text{n}$) consistently increases with an enhancement in $S\text{r}$, independent of $S_\text{t}$ and $K$. The strongest influence of $S_\text{r}$ is observed at $S_\text{t}=16$ and $K=2$ at $P_3$. Under these conditions, $n^{\ast}\text{n}$ increases substantially, from $0.0726$ to $0.8482$ ($\sim 1068\%$), from $0.1364$ to $0.8801$ ($\sim 545\%$), and from $0.0577$ to $0.8895$ ($\sim 1441\%$) at $P_1$, $P_2$, and $P_3$, respectively, as $S\text{r}$ decreases from $0$ to $-2$ (\fig\ref{fig:6b}). Overall, the sensitivity of $n^{\ast}_\text{n}$ to $S\text{t}$ is most pronounced at $P_1$, while the effects of $K$ and $S_\text{r}$ dominate at $P_3$. This distinct behaviour arises from the advection of excess ions along the centreline of the oppositely charged micro slit, which amplifies the charge accumulation at downstream locations, leading to stronger accumulation at $P_3$ compared to other positions (\fig\ref{fig:6b}).
%
\subsection{Induced electric field strength ($E_{\text{x}}$)}
\label{sec:electric}
\begin{figure}[htbp]
	\centering
	\subfigure[\mm{E_{\text{x}}=f(K)^\#}]{\includegraphics[width=0.9\linewidth]{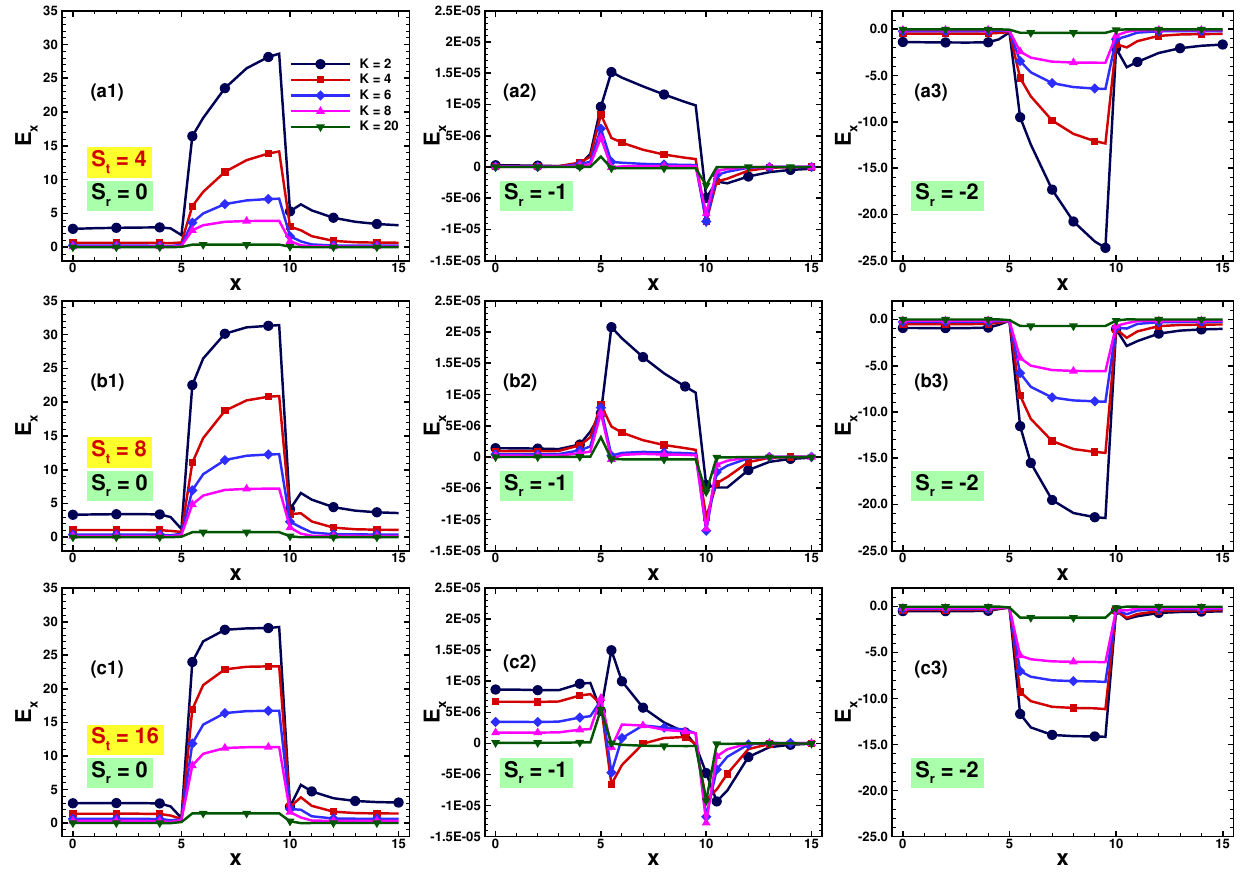}}
	\subfigure[\mm{E_{\text{x}}=f(\sur)^\#}]{\includegraphics[width=0.9\linewidth]{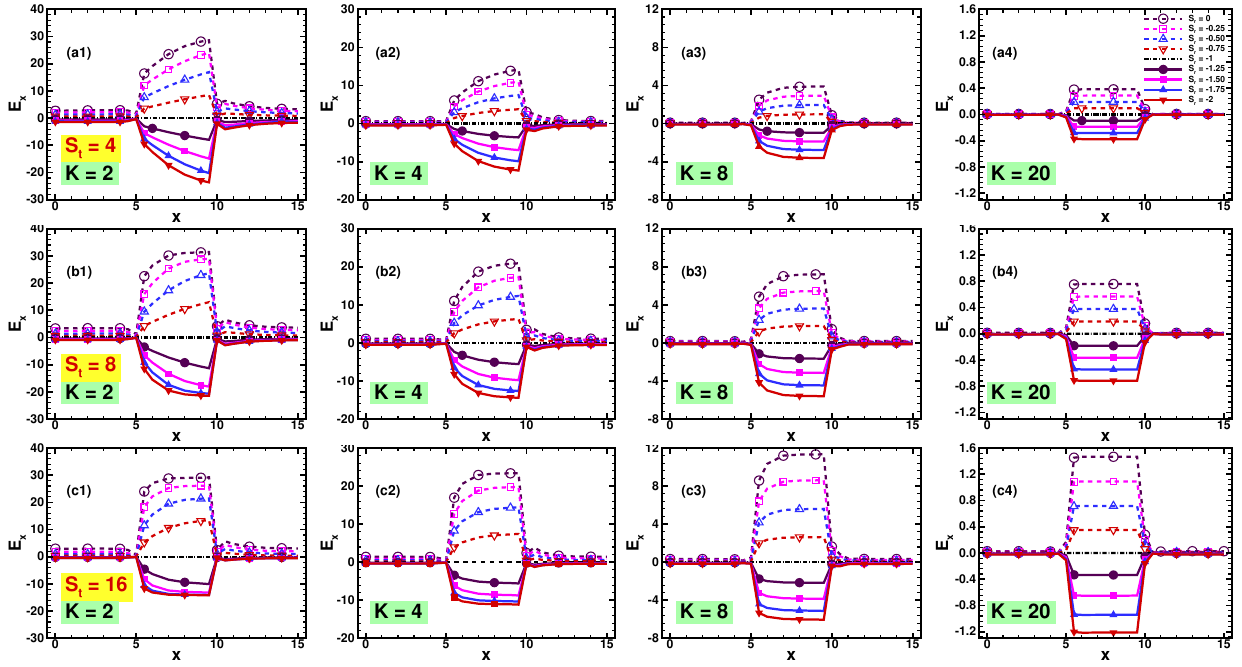}}
	\caption{Centreline profiles of induced electric field strength (\mm{E_{\text{x}}}) in the oppositely charged micro slit for \mm{-2\le \sur\le 0}, \mm{4\le \sut\le 16}, and  \mm{2\le K\le 20}.}
	\label{fig:7}
\end{figure}

%
The transport of excess charge ($n^\ast$) under the imposed pressure-driven flow generates an induced electric field ($E_\text{x}$, \eqn\ref{eq:4a}). Figure \ref{fig:7} presents the centreline profiles of the electric field ($E_\text{x}$) in the oppositely charged micro slit over a broad range of governing parameters (\tab\ref{tab:pm}). At $S_\text{r}=-1$, the electric field nearly vanishes because equal and opposite wall charges neutralize each other. For $S_\text{r}>-1$, the centreline profiles of $E_\text{x}$ follow trends qualitatively consistent with the literature \citep{dhakar2022electroviscous,dhakar2023cfd}, while for $S_\text{r}<-1$ they exhibit opposite behavior. 

\noindent 
The induced electric field ($E_\text{x}$) reaches its maximum (for $S_\text{r}>-1$) or minimum (for $S_\text{r}<-1$) in the contraction region of the micro slit, relative to other regions. The extremum (i.e., minima or maxima) of this field is denoted as the critical induced electric field strength ($E_{\text{x,c}}$, \tab\ref{tab:1}). For $S_\text{r}>-1$, $E_{\text{x,c}}$ decreases with decreasing $K$ (i.e., EDL thickening), while the reverse trend holds for $S_\text{r}<-1$, irrespective of $S_\text{t}$ (\fig\ref{fig:7}, \tab\ref{tab:1}). The strongest influence of $K$ is found at the highest $S_\text{r}$ and lowest $S_\text{t}$. For instance, when $K$ varies from 2 to 20, $E_{\text{x,c}}$ decreases (\tab\ref{tab:1}) from $28.6420$ to $0.3821$ (98.67\%), $31.4090$ to $0.7576$ (97.59\%), and $29.2120$ to $1.4662$ (94.98\%) at $S_\text{r}=0$; from $8.5316$ to $0.0953$ (98.88\%), $12.9820$ to $0.1874$ (98.56\%), and $13.5040$ to $0.3522$ (97.39\%) at $S_\text{r}=-0.75$; from $-7.9614$ to $-0.0949$ (98.81\%), $-11.3050$ to $-0.1849$ (98.36\%), and $-10.0490$ to $-0.3349$ (96.67\%) at $S_\text{r}=-1.25$; and from $-23.6110$ to $-0.3767$ (98.40\%), $-21.4580$ to $-0.7175$ (96.66\%), and $-14.1550$ to $-1.2132$ (91.43\%) at $S_\text{r}=-2$ for $S_\text{t}=4,8,16$, respectively.

\noindent 
The variation of $E_\text{x,c}$ with $S_\text{t}$ is most pronounced at the highest $S_\text{r}$ and $K$. 
For instance, at $K=2$, $E_\text{x,c}$ changes (in \%) by 1.99, 58.28, 26.22, and 40.05 for $S_\text{r}=0,-0.75,-1.25,-2$, respectively, when $S_\text{t}$ increases from 4 to 16; whereas at $K=20$, the corresponding changes (in \%) are from 283.73, 269.66, 252.81, and 222.05. In contrast, the influence of $S_\text{r}$ on $E_\text{x,c}$ is strongest at the lowest $S_\text{t}$ and highest $K$. For example, $E_\text{x,c}$ decreases (in \%)  by 70.21, 58.67, and 53.77 at $K=2$, and by 75.07, 75.26, and 75.98 at $K=20$ for $S_\text{t}=4,8,16$, respectively, when $S_\text{r}$ decreases from 0 to $-0.75$. A further reduction of $S_\text{r}$ from $-1.25$ to $-2$ causes $E_\text{x,c}$ to decrease  (in \%)  by 196.57, 89.81, and 40.96 at $K=2$, and by 293.81, 288.16, and 262.21 at $K=20$. Overall, the reduction of $S_\text{r}$ from 0 to $-2$ yields a net decrease  (in \%) in $E_\text{x,c}$ by 182.43, 168.32, and 148.46 at $K=2$, and by 198.59, 194.71, and 182.74 at $K=20$ for $S_\text{t}=4,8,16$, respectively (\tab\ref{tab:1}).
In general, $E_\text{x,c}$ increases for $S_\text{r}>-1$ and decreases for $S_\text{r}<-1$ with increasing $S_\text{t}$. However, at lower $K$ and higher $S_\text{t}$, the trend reverses (\fig\ref{fig:7}, \tab\ref{tab:1}). This occurs because the enhancement in convective excess charge ($|n^\ast| > 0$) alters the streaming current, thereby modifying $E_\text{x,c}$ with increasing $S_\text{t}$. At low $K$ and high $S_\text{t}$, the overlap of electric double layers (EDLs) suppresses the flow, leading to this reversal. Furthermore, $E_\text{x,c}$ consistently decreases with decreasing $S_\text{r}$, irrespective of $K$ and $S_\text{t}$ (\fig\ref{fig:7}, \tab\ref{tab:1}). This is because a reduction in $S_\text{r}$ increases the available excess charge ($|n^\ast| > 0$) for transport, which diminishes the streaming current and, consequently, the electric field within the device.
\begin{figure}[t!]
	\centering\includegraphics[width=1\linewidth]{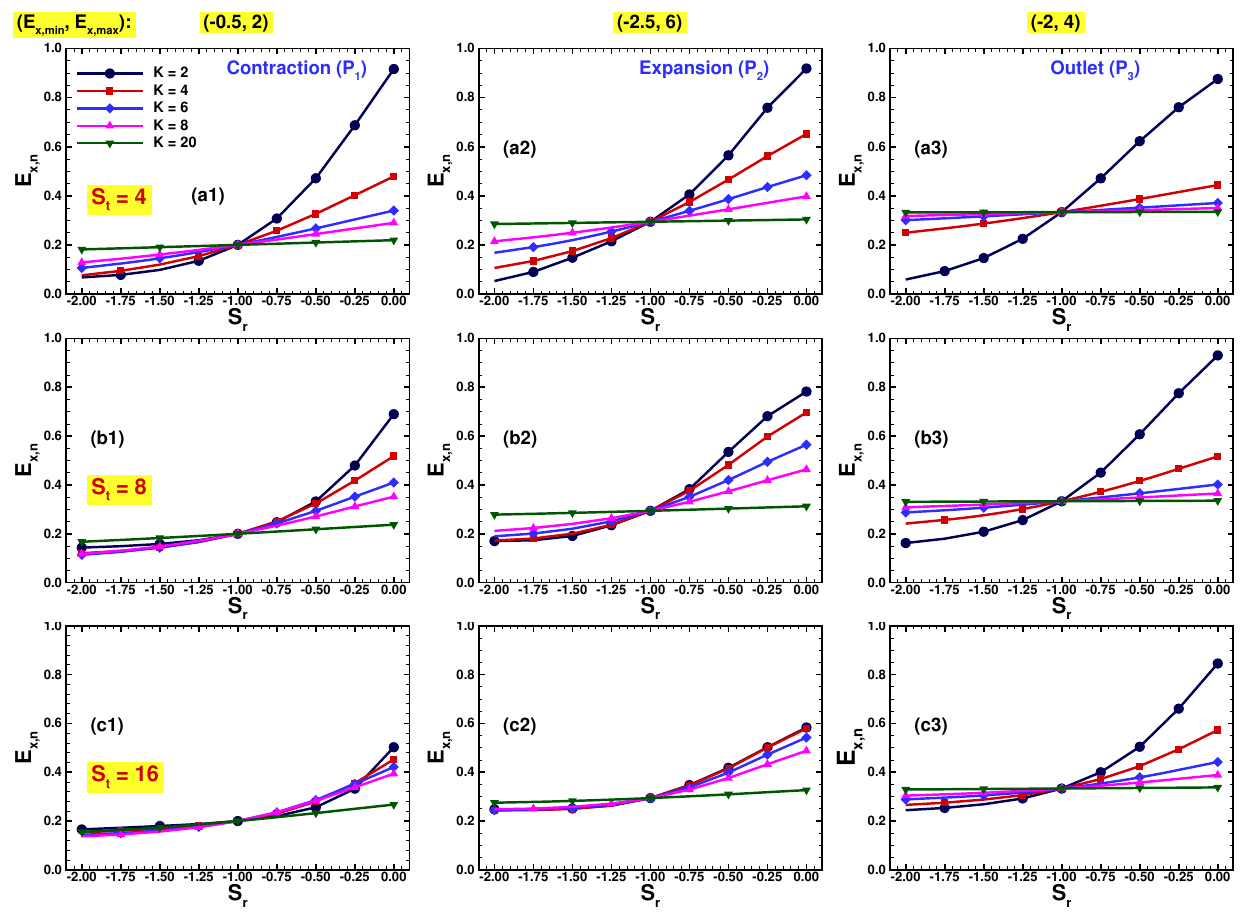}
	\caption{Influence of dimensionless parameters (\mm{K,\sut,\sur}) on scaled induced electric field strength (\mm{E_{\text{x,n}}}) variation on the contraction (\mm{P_1}, first column), expansion (\mm{P_2}, second column), and outlet (\mm{P_3}, third column) centreline points of the oppositely charged micro slit.}
	\label{fig:8}
\end{figure} 

\noindent 
Further, the induced electric field is scaled (similar to $U_\text{n}$, section \ref{sec:potential}) to examine the detailed influence of the dimensionless parameters ($K$, $S_\text{t}$, $S_\text{r}$). \fig\ref{fig:8} presents the scaled induced electric field strength ($E_{\text{x,n}}$, \eqn\ref{eq:MR}) at the centreline locations ($P_1$, $P_2$, $P_3$) of the micro slit as a function of $S_\text{r}$, $S_\text{t}$, and $K$. Qualitatively, $E_{\text{x,n}}$ exhibits the same dependence on these parameters as $E_\text{x}$. For instance, $E_{\text{x,n}}$ decreases for $S_\text{r}>-1$ and increases for $S_\text{r}<-1$ with increasing $K$, regardless of $S_\text{t}$, with the maximum variation observed at $P_3$ for $S_\text{t}=4$ and $S_\text{r}=-2$.  
Specifically, $E_{\text{x,n}}$ increases from $0.0676$ to $0.1815$ (168.32\%), $0.0528$ to $0.2850$ (440.19\%), and $0.0590$ to $0.3322$ (462.99\%) at $P_1$, $P_2$, and $P_3$, respectively, when $K$ is raised from 2 to 20 at $S_\text{t}=4$ and $S_\text{r}=-2$ (\fig\ref{fig:8}). Similarly, $E_{\text{x,n}}$ increases for $S_\text{r}>-1$ and decreases for $S_\text{r}<-1$ with higher $S_\text{t}$; however, opposite trends emerge at high $S_\text{t}$ and low $K$. The effect of $S_\text{t}$ on $E_{\text{x,n}}$ is most pronounced at $S_\text{r}=-2$ and $K=2$, particularly at $P_2$. For example, when $S_\text{t}$ increases from 4 to 16 under these conditions ($S_\text{r}=-2$, $K=2$), $E_{\text{x,n}}$ rises from $0.0676$ to $0.1657$ (144.92\%) at $P_1$, $0.0528$ to $0.2483$ (370.57\%) at $P_2$, and $0.0590$ to $0.2445$ (314.40\%) at $P_3$ (\fig\ref{fig:8}).
%
Moreover, $E_{\text{x,n}}$ decreases with decreasing $S_\text{r}$, irrespective of $S_\text{t}$ and $K$, with the strongest variation observed at $P_2$ for $K=2$ and $S_\text{t}=4$. For example, when $S_\text{r}$ decreases from 0 to $-2$, $E_{\text{x,n}}$ reduces from 0.9163 to 0.0676 (92.62\%) at $P_1$, from 0.9183 to 0.0528 (94.25\%) at $P_2$, and from 0.8755 to 0.0590 (93.26\%) at $P_3$ (\fig\ref{fig:8}). Overall, the influence of $K$ on $E_{\text{x,n}}$ is most significant at $P_3$, whereas the combined effect of $S_\text{t}$ and $S_\text{r}$ is dominant at $P_2$. This is attributed to the reduced cross-sectional area at $P_2$, which enhances both excess charge clustering ($n^\ast$, see section \ref{sec:charge}) and fluid velocity, thereby amplifying the change in $E_{\text{x,n}}$ compared to other centreline locations (\fig\ref{fig:8}).

\noindent 
The preceding sections analyzed the detailed influence of the dimensionless parameters ($S_\text{r}, S_\text{t}, K$) on the electrical and ionic fields, namely the total electrical potential ($U$), excess ionic charge ($n^\ast$), and induced electric field strength ($E_\text{x}$). In the following sections, attention is shifted to the flow fields ($\myvec{V}$, $P$) to examine their dependence on the same dimensionless parameters ($S_\text{r}, S_\text{t}, K$). 
%
\subsection{Velocity ($\myvec{V}$)}
\label{sec:velocity}
%
\begin{figure}[t!]
	\centering\includegraphics[width=1\linewidth]{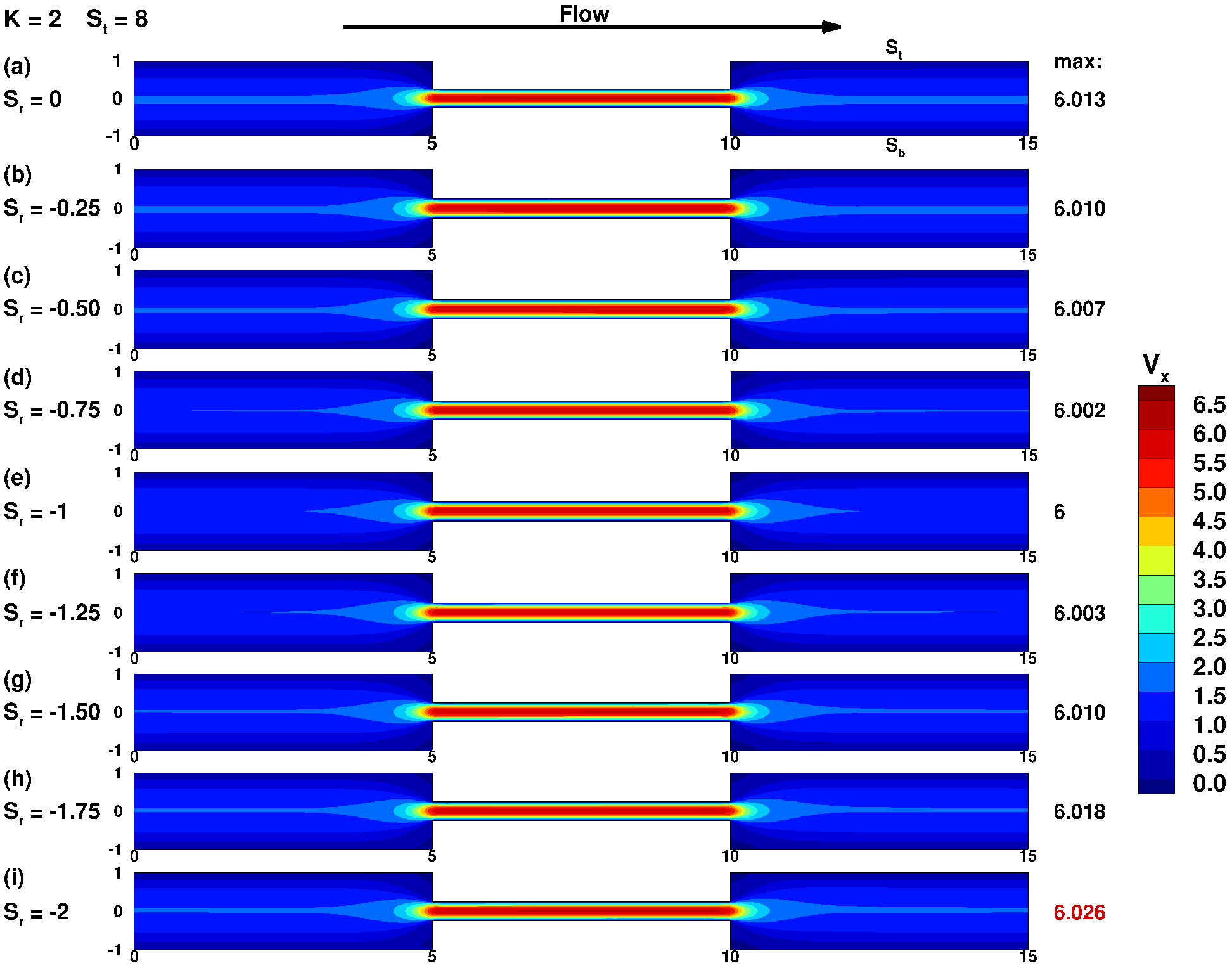}
	\caption{Distribution of velocity (\mm{V_\text{x}}) in the oppositely charged micro slit for \mm{-2\le \sur\le 0}, \mm{K=2} and \mm{\sut=8}.}
	\label{fig:9}
\end{figure} 
\noindent
\fig\ref{fig:9} presents the velocity distribution ($V_\text{x}$) in the micro-slit for $-2 \leq S_\text{r} \leq 0$, $S_\text{t}=8$, and $K=2$. Qualitatively similar contour velocity profiles are obtained for other parameter ranges (\tab\ref{tab:pm}) and are therefore not shown here. The maximum velocity ($V_\text{max}$) consistently occurs in the contraction region, where the sudden reduction in cross-sectional area accelerates the flow while maintaining a constant volumetric flow rate ($Q$). With decreasing $S_\text{r}$, $V_\text{max}$ exhibits a smaller increment for $-2 \leq S_\text{r} \leq -1$ and a decrement for $-1 \leq S_\text{r} \leq 0$. The highest velocity, $V_\text{x}=6.026$, is observed at $S_\text{r}=-2$, $S_\text{t}=8$, and $K=2$ (\fig\ref{fig:9}(i)).
\begin{figure}[t!]
	\centering\includegraphics[width=1\linewidth]{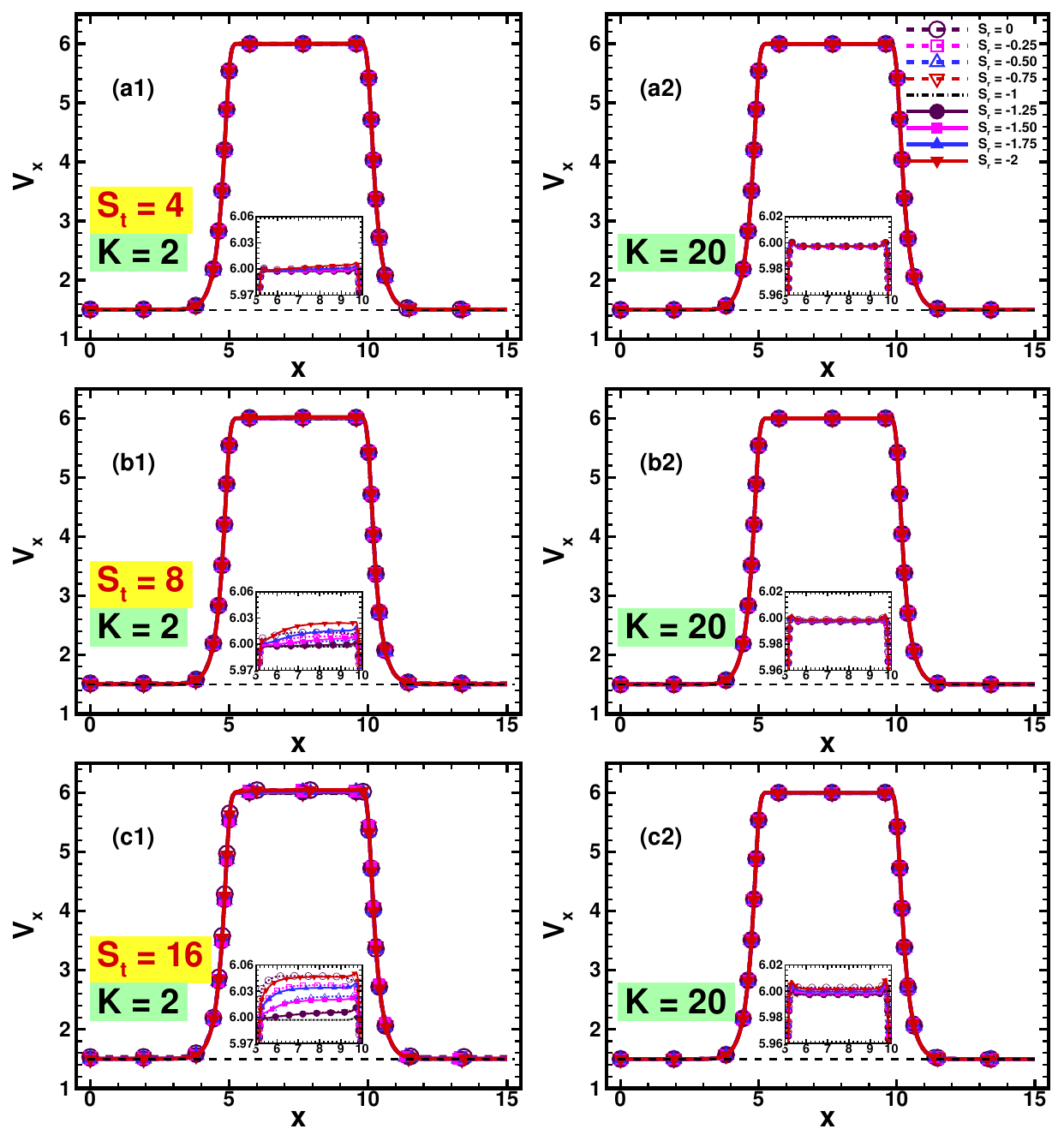}
	\caption{Centreline profiles of velocity (\mm{V_\text{x}}) in the oppositely charged micro slit for \mm{-2\le \sur\le 0}, \mm{4\le \sut\le 16}, and  \mm{2\le K\le 20}.}
	\label{fig:10}
\end{figure} 
\begin{table}[t!]
	\centering
	\caption{The maximum velocity (\mm{V_\text{max}}, \mm{V^\ast_\text{max}}) on the centreline of the oppositely charged micro slit. The highest value of maximum velocity (\mm{V_\text{max}}, \mm{V^\ast_\text{max}}) is underlined for each \mm{\sut} and \mm{K}.}\label{tab:Vmax}
	\scalebox{1}
	{
		\begin{tabular}{rrrrrrrrrrr}
			\hline 
			\mm{\sut} & \mm{K}	&	\multicolumn{9}{c}{\mm{V_\text{max}}}	\\\cline{3-11} 
			& 	&	\mm{\sur=} 0	&	-0.25	& -0.50 &	-0.75	& -1 & -1.25 & -1.50 & -1.75 & -2 \\\hline  
			0   & 	{$\infty$}    & 	6	    & 6	        & 6          & 	6	    & 6         & 	6	&   6	&   6	 &   6  \\\hline  
			4	&	2	&	6.0036	&	6.0025	&	6.0011	&	6.0002	&	6	&	6.0003	&	6.0016	&	6.0041	&	{\underline{6.0073}}	\\
			&	20	&	6.0004	&	6.0003	&	6.0002	&	6.0001	&	6.0002	&	6.0002	&	6.0003	&	6.0005	&	{\underline{6.0007}}	\\\hline
			8	&	2	&	6.0134	&	6.0103	&	6.0073	&	6.0022	&	6.0001	&	6.0027	&	6.0101	&	6.0180	&	{\underline{6.0255}}	\\
			&	20	&	6.0016	&	6.0010	&	6.0007	&	6.0005	&	6.0005	&	6.0008	&	6.0012	&	6.0019	&	{\underline{6.0027}}	\\\hline
			16	&	2	&	6.0475	&	6.0369	&	6.0255	&	6.0111	&	6.0012	&	6.0111	&	6.0260	&	6.0394	&	{\underline{6.0515}}	\\
			&	20	&	6.0064	&	6.0041	&	6.0027	&	6.0021	&	6.0023	&	6.0033	&	6.0048	&	6.0070	&	{\underline{6.0095}}	\\\hline
			& 	& \multicolumn{9}{c}{\mm{V^\ast_\text{max}} (\mm{=V_{max}/V_{max,{S_{\text{k}}}=0}})}	\\\cline{3-11}
			& 	&	\mm{\sur=} 0	&	-0.25	& -0.50 &	-0.75	& -1 & -1.25 & -1.50 & -1.75 & -2 \\\cline{3-11}
			0   & 	{$\infty$}    & 	1	    & 1	        & 1          & 	1	    & 1         & 	1	&   1	&   1	 &   1  \\\hline 
			4	&	2	&	1.0006	&	1.0004	&	1.0002	&	1	&	1	&	1.0001	&	1.0003	&	1.0007	&	{\underline{1.0012}}	\\
			&	20	&	1.0001	&	1.0001	&	1	&	1	&	1	&	1	&	1.0001	&	1.0001	&	{\underline{1.0001}}	\\\hline
			8	&	2	&	1.0022	&	1.0017	&	1.0012	&	1.0004	&	1	&	1.0005	&	1.0017	&	1.0030	&	{\underline{1.0043}}	\\
			&	20	&	1.0003	&	1.0002	&	1.0001	&	1.0001	&	1.0001	&	1.0001	&	1.0002	&	1.0003	&	{\underline{1.0005}}	\\\hline
			16	&	2	&	1.0079	&	1.0062	&	1.0043	&	1.0019	&	1.0002	&	1.0019	&	1.0043	&	1.0066	&	{\underline{1.0086}}	\\
			&	20	&	1.0011	&	1.0007	&	1.0005	&	1.0004	&	1.0004	&	1.0006	&	1.0008	&	1.0012	&	{\underline{1.0016}}	\\\hline		
		\end{tabular}
	}
\end{table}
\begin{figure}[t!]
	\centering\includegraphics[width=1\linewidth]{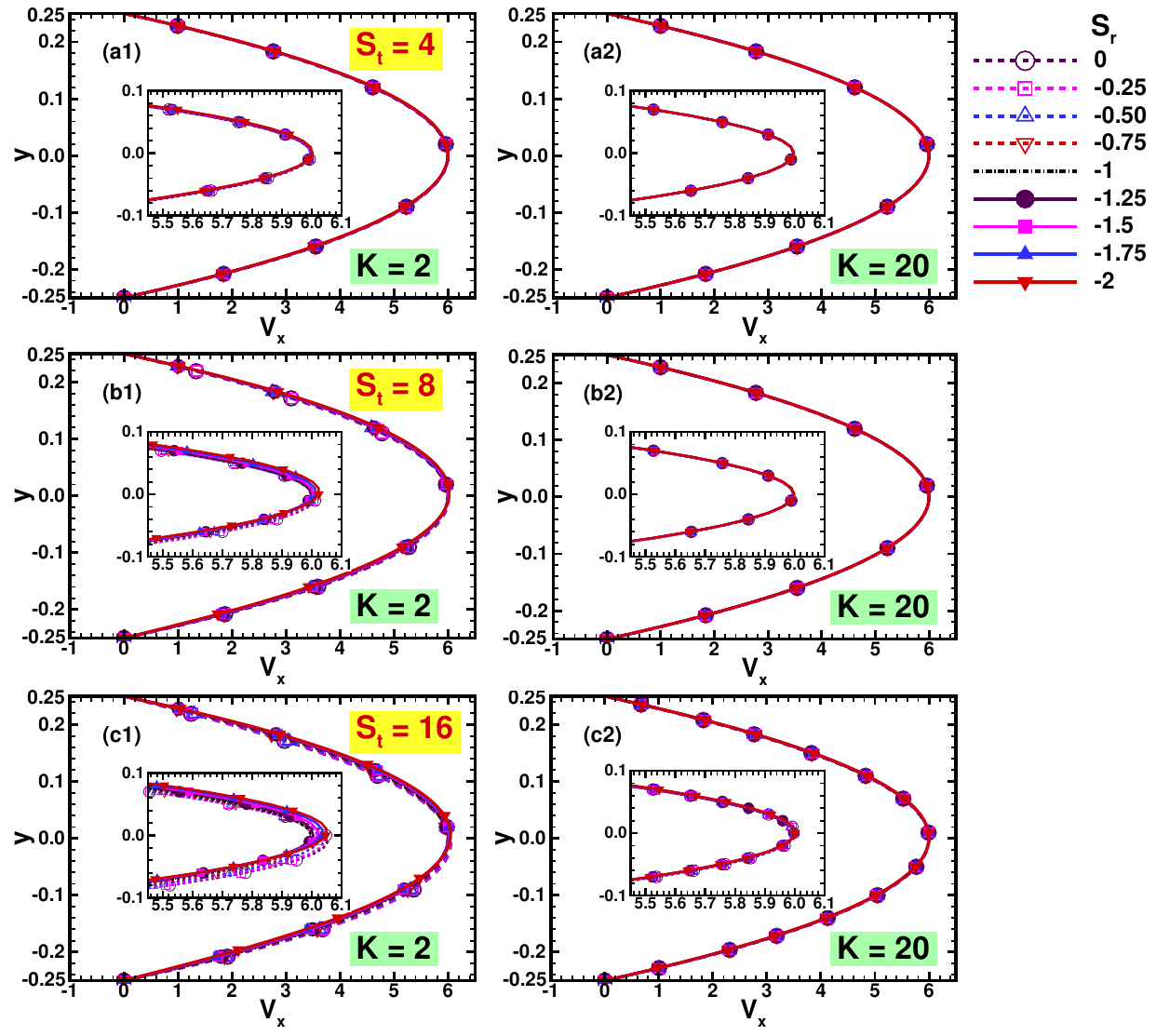}
	\caption{Velocity ($V_\text{x}$) variation over the vertical length ($y$) at the contraction section ($x=L/2$) of oppositely charged micro slit for $-2\le \sur\le 0$, $4\le \sut\le 16$, and  $2\le K\le 20$.}
	\label{fig:11}
\end{figure} 
\\\noindent
\fig\ref{fig:10} shows the centreline velocity profiles ($V_\text{x}$) in the micro slit as a function of dimensionless parameters ($K, S_\text{t}, S_\text{r}$). The velocity consistently peaks ($V_\text{max}$) in the contraction region relative to the upstream and downstream sections. With increasing $K$ (thinning of the EDL), $V_\text{max}$ decreases, and the influence of $K$ is strongest at the highest $S_\text{t}$ and lowest $S_\text{r}$. 
%
For example, $V_\text{max}$ decreases by $\sim0.05\%$,  $\sim0.19\%$, $\sim0.68\%$ at $S_\text{r}=0$, and by $\sim0.11\%$, $\sim0.38\%$, $\sim0.69\%$ at $S_\text{r}=-2$ for ($S_\text{t}=4,8,16$), respectively, as $K$ increases from $2$ to $20$ (\tab\ref{tab:Vmax}). In contrast, $V_\text{max}$ increases with $S_\text{t}$, with the strongest effect at low $S_\text{r}$ and $K$. For instance, it increases by $\sim0.73\%$ and $\sim0.74\%$ at $K=2$, and by $\sim0.10\%$ and $\sim0.15\%$ at $K=20$ for ($S_\text{r}=0,-2$), respectively, when $S_\text{t}$ rises from $4$ to $16$. Additionally, $V_\text{max}$ decreases for $0 \geq S_\text{r} \geq -1$ but increases for $-1 \geq S_\text{r} \geq -2$, with the strongest $S_\text{r}$ effect at higher $S_\text{t}$ and lower $K$  (\tab\ref{tab:Vmax}). For example, $V_\text{max}$ increases by $\sim0.06\%$, $\sim0.20\%$, $\sim0.07\%$ at $K=2$, and by $\sim0.005\%$, $\sim0.02\%$, $\sim0.05\%$ at $K=20$ for ($S_\text{t}=4,8,16$), respectively, as $S_\text{r}$ decreases from $0$ to $-2$.

\noindent
The centreline velocity profiles ($V_\text{x}$) in the micro-slit as functions of the dimensionless parameters ($K, S_\text{t}, S_\text{r}$) are presented in \fig\ref{fig:10}. The velocity attains its maximum ($V_\text{max}$) in the contraction region, higher than in the upstream and downstream sections. With increasing $K$ (thinning of the EDL), $V_\text{max}$ decreases, with the strongest influence observed at the highest $S_\text{t}$ and lowest $S_\text{r}$ (\fig\ref{fig:10}, \tab\ref{tab:Vmax}). 
For example, $V_\text{max}$ decreases by $\sim0.05\%$, $\sim0.19\%$, and $\sim0.68\%$ at $S_\text{r}=0$, and by $\sim0.11\%$,  $\sim0.38\%$, and $\sim0.69\%$ at $S_\text{r}=-2$ for $S_\text{t}=(4,8,16)$, respectively, as $K$ increases ($2$ to $20$). Conversely, $V_\text{max}$ increases with $S_\text{t}$, with the strongest variation at the lowest $S_\text{r}$ and $K$. For instance, $V_\text{max}$ increases by $\sim0.73\%$ and $\sim0.74\%$ at $K=2$, and by $\sim0.10\%$ and $\sim0.15\%$ at $K=20$ for $S_\text{r}=(0,-2)$, respectively, as $S_\text{t}$ increases ($4$ to $16$). Furthermore, with decreasing $S_\text{r}$, $V_\text{max}$ first decreases for $0 \geq S_\text{r} \geq -1$ and then increases for $-1 \geq S_\text{r} \geq -2$, with the strongest effect at higher $S_\text{t}$ and lower $K$. For instance, $V_\text{max}$ increases by $\sim0.06\%$, $\sim0.20\%$, and $\sim0.07\%$ at $K=2$, and by $\sim0.005\%$, $\sim0.02\%$, and $\sim0.05\%$ at $K=20$ for $S_\text{t}=(4,8,16)$, respectively, as $S_\text{r}$ decreases ($0$ to $-2$). 

\noindent
To further quantify the electroviscous effects on velocity field, the maximum velocity is normalized by the non-EVF case. The normalized maximum velocity is defined as $V^\ast_\text{max} = (V_\text{max}/V_{\text{max},S_\text{k}=0})$, and its dependence on the governing parameters ($K, S_\text{t}, S_\text{r}$) is summarized in \tab\ref{tab:Vmax}. The variations in $V^\ast_\text{max}$ with $K,,S_\text{t}$, and $S_\text{r}$ are qualitatively similar to those of $V_\text{max}$. 
The maximum decrement in $V^\ast_\text{max}$ is noted as $0.69\%$ at $S_\text{t}=16$, $S_\text{r}=-2$; $0.74\%$ at $K=2$, $S_\text{r}=-2$; and $0.67\%$ at $K=2$, $S_\text{t}=16$, when $K$ is varied ($2$ to $20$), $S_\text{t}$ is increased ($4$ to $16$), and $S_\text{r}$ is decreased ($-1.25$ to $-2$), respectively (\tab\ref{tab:Vmax}). 

\noindent 
Furthermore, \fig\ref{fig:11} shows the velocity distribution ($V_\text{x}$) along the vertical length ($-0.25\le y\le 0.25$) at the contraction region ($x=L/2$) of the micro-slit for a wide range of $S_\text{r}$, $S_\text{t}$, and $K$ (\tab\ref{tab:pm}). The velocity ($V_\text{x}$) increases from the walls ($y=\pm 0.25$) toward the centreline ($y=0$), where the maximum velocity ($V_\text{max}$) is attained. Near the walls, $V_\text{x}$ exhibits increment for $S_\text{r} > -1$ and decrement for $S_\text{r} < -1$ with increasing $S_\text{r}$; consequently, reverse trends are observed on the centreline due to the constraint of constant volumetric flow rate ($Q$). In addition, $V_\text{max}$ is found to slightly increase with decreasing $K$ and increasing $S_\text{t}$ (\fig\ref{fig:11}). Overall, the variations in $V_\text{max}$ remain small, within $\sim0.005-0.74\%$, indicating weak sensitivity of the velocity field to electrokinetic parameters compared to the electrical and ionic fields. To complement the velocity field analysis, the subsequent section presents the dependence of the pressure field on the EVF parameters ($S_\text{r}$, $S_\text{t}$, and $K$), highlighting the interplay between velocity, pressure, and potential field distributions in the micro-slit.
%
\subsection{Pressure ($P$)}
\label{sec:pressure}
\begin{figure}[t!]
	\centering\includegraphics[width=1\linewidth]{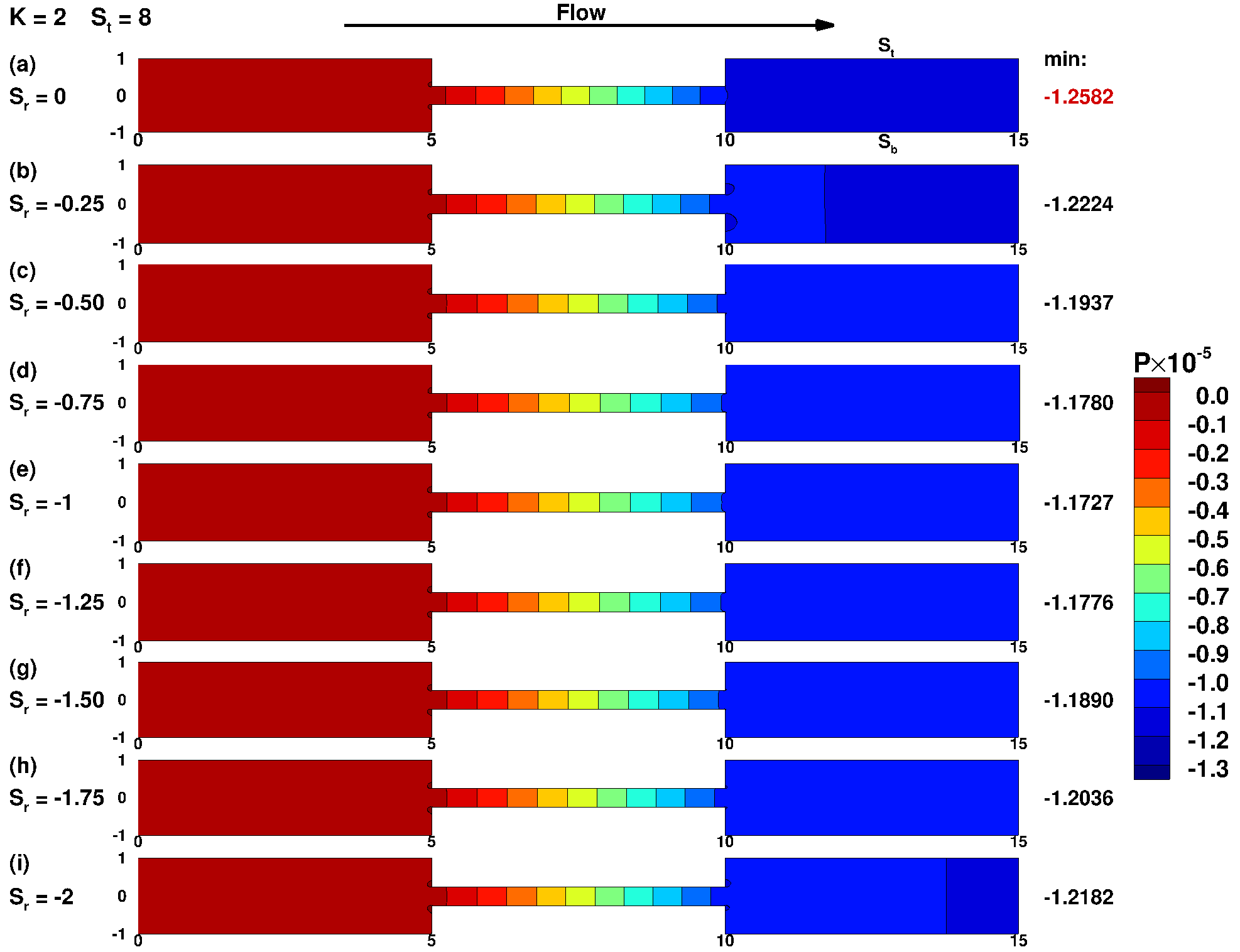}
	\caption{Distribution of pressure ($P^{\ast}=P\times10^{-5}$) in the oppositely charged micro slit for $-2 \leq S_\text{r} \leq 0$, $S_\text{t}=8$, and $K=2$.}
	\label{fig:12}
\end{figure}

%
The distribution of pressure ($P^{\ast}=P\times10^{-5}$) in the oppositely charged contraction–expansion micro slit for $-2 \leq S_\text{r} \leq 0$, $S_\text{t}=8$, and $K=2$ is shown in \fig\ref{fig:12}. Similar qualitative trends in pressure contours are obtained for other ranges of parameter conditions (\tab\ref{tab:pm}), and are therefore not presented here. 
As expected, the pressure decreases along the length of the micro slit, with a higher pressure gradient observed in the contraction region compared to the upstream and downstream sections. 
This occurs because as the cross-sectional area decreases, velocity and charge clustering increase, thereby enhancing both hydrodynamic and electrostatic resistances within the contraction. Moreover, the magnitude of the pressure drop ($|\Delta P|$) decreases for $S_\text{r}>-1$ and increases for $S_\text{r}<-1$ with decreasing $S_\text{r}$. For instance, the minimum pressure drop ($\Delta P_\text{min}^{\ast}$) is recorded as $-1.2582$ at $S_\text{t}=8$, $S_\text{r}=0$, and $K=2$ (\fig\ref{fig:12}a).
\begin{figure}[htbp]
	\centering
	\subfigure[\mm{P=f(K)^\#}]{\includegraphics[width=0.9\linewidth]{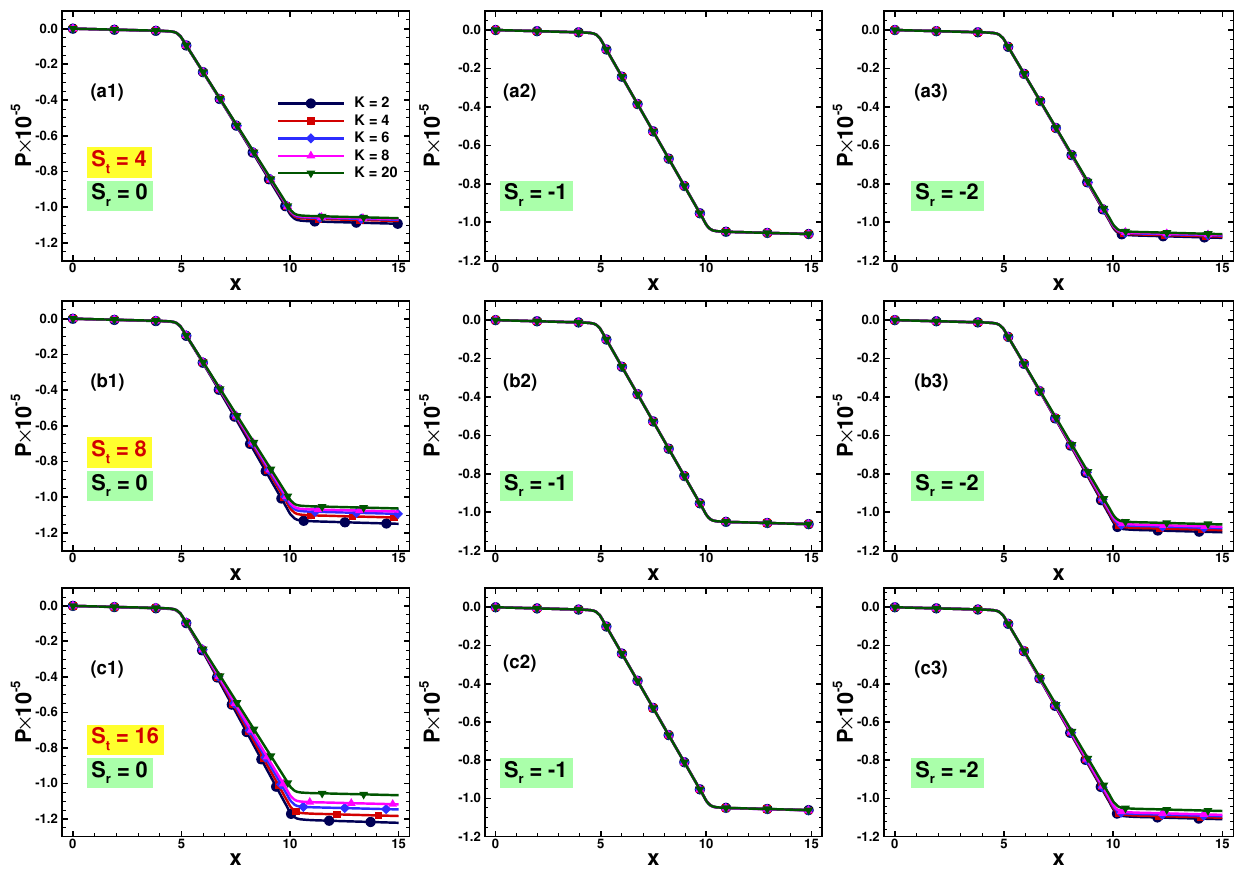}}
	\subfigure[\mm{P=f(\sur)^\#}]{\includegraphics[width=0.9\linewidth]{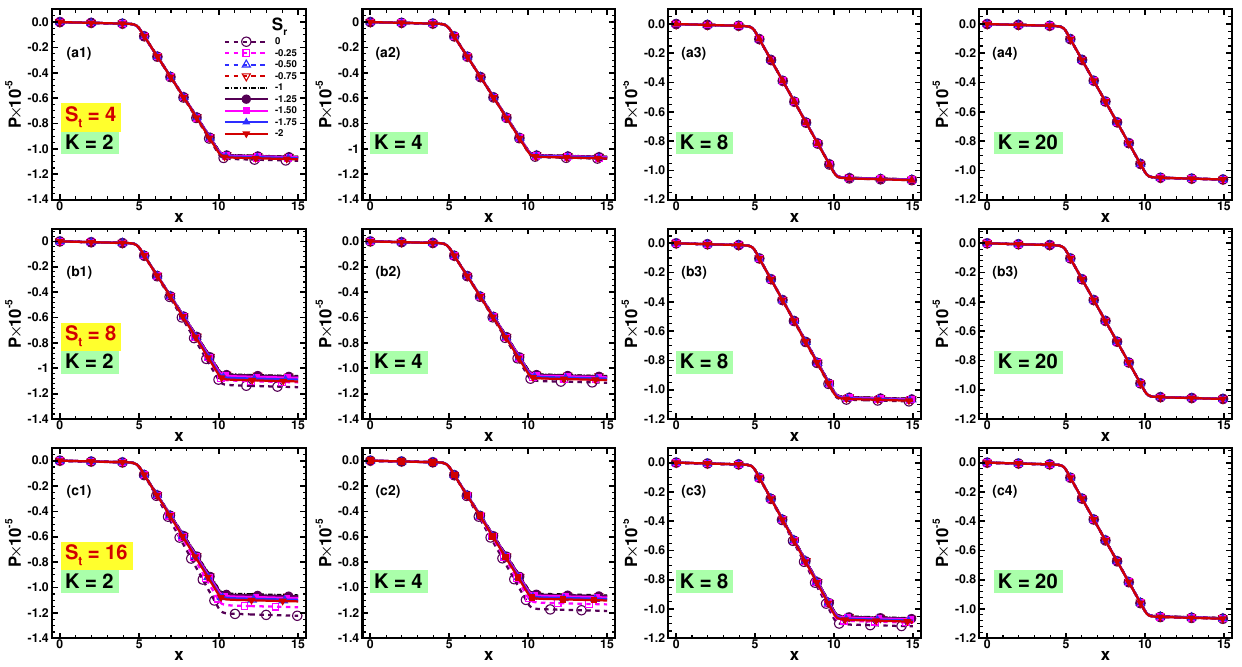}}
	\caption{Centreline profiles of pressure (\mm{P}) in the oppositely charged micro slit for \mm{-2\le \sur\le 0}, \mm{4\le \sut\le 16}, and  \mm{2\le K\le 20}.}
	\label{fig:13}
\end{figure} 

\noindent 
Subsequently, \fig\ref{fig:13} depicts the centreline pressure ($P$) profiles in the oppositely charged micro-slit for broader ranges of the flow conditions ($S_\text{r}$, $S_\text{t}$, and $K$) (\tab\ref{tab:pm}). The centreline pressure profiles are qualitatively similar for $S_\text{r}>-1$ and opposite for $S_\text{r}<-1$, consistent with previous studies \citep{davidson2007electroviscous,dhakar2022electroviscous,dhakar2023cfd}. Along the slit length ($x$), pressure ($P$) decreases as expected, since variations in potential (section \ref{sec:potential}) impose additional resistance on the flow. The pressure gradient ($\partial P/\partial x$) is relatively higher in the contraction region compared to the upstream and downstream sections. An increase in the pressure drop ($\Delta P$) is observed with increasing $K$, i.e., with EDL thinning (\fig\ref{fig:13} and \tab\ref{tab:1}).
The influence of $K$ on $|\Delta P|$ is most pronounced at the highest $S_\text{t}$ and $S_\text{r}$. 
For instance, the reduction in $|\Delta P|$ when $K$ increases from $2$ to $20$ is quantified as $2.85\%$, $7.49\%$, and $12.71\%$ at $S_\text{r}=0$; $0.12\%$,  $0.37\%$, $0.37\%$ and $0.90\%$ at $S_\text{r}=-0.75$; $0.11\%$, $0.32\%$, and $0.64\%$ at $S_\text{r}=-1.25$; and $1.78\%$, $3.70\%$, and $3.86\%$ at $S_\text{r}=-2$ for $S_\text{t}=4,8,16$, respectively (\tab\ref{tab:1}).

\noindent
Subsequently, \fig\ref{fig:13} presents the centreline pressure ($P$) profiles in the oppositely charged micro-slit over a broad range of conditions ($S_\text{r}$, $S_\text{t}$, and $K$; \tab\ref{tab:pm}). The profiles are qualitatively similar for $S_\text{r}>-1$ and opposite in trend for $S_\text{r}<-1$, consistent with the literature \citep{davidson2007electroviscous,dhakar2022electroviscous,dhakar2023cfd}. As expected, the pressure decreases, along the length ($x,0$). The decrease in pressure becomes more pronounced, compared to non-electroviscous flow, since potential variations  (refer Section \ref{sec:potential}) impose additional resistance to the electroviscous flow. The pressure gradient ($\text{d}P/\text{d}x$) is comparatively higher in the contraction than in the upstream and downstream regions. An enhancement in pressure drop ($\Delta P$) is observed with increasing $K$ (thinning of the EDL), with the strongest influence at the highest $S_\text{t}$ and $S_\text{r}$ values (\fig\ref{fig:13} and \tab\ref{tab:1}). 
For example, the decrease in $|\Delta P|$ is recorded as $2.85\%$, $7.49\%$, and $12.71\%$ at $S_\text{r}=0$; $0.12\%$, $0.37\%$, and $0.90\%$ at $S_\text{r}=-0.75$; $0.11\%$, $0.32\%$, and $0.64\%$ at $S_\text{r}=-1.25$; and $1.78\%$, $3.70\%$, and $3.86\%$ at $S_\text{r}=-2$ for ($S_\text{t}=4,8,16$), respectively, as $K$ increases (from 2 to 20).

\noindent
The variation of $|\Delta P|$ with $S_\text{t}$ is most pronounced at the highest $S_\text{r}$ and lowest $K$. For example, $|\Delta P|$ increases (for $S_\text{r}=0,-0.75,-1.25,-2$) by 11.81\%, 0.85\%, 0.60\%, 2.57\% at $K=2$, and by 0.46\%, 0.06\%, 0.08\%, 0.40\% at $K=20$, respectively, when $S_\text{t}$ varies from $4$ to $16$ (\tab\ref{tab:1}).  Moreover, the influence of $S_\text{r}$ on $|\Delta P|$ is strongest at the lowest $K$ and highest $S_\text{t}$. For instance, $|\Delta P|$ decreases (as $S_\text{r}$ reduces from $0$ to $-0.75$) by $-2.76\%$, $-7.25\%$, $-12.30\%$ at $K=2$, and by $-0.03\%$, $-0.11\%$, $-0.43\%$ at $K=20$. Corresponding changes in $|\Delta P|$ when $S_\text{r}$ decreases from $-1.25$ to $-2$ are 1.73\%, 3.62\%, 3.72\% at $K=2$, and 0.04\%, 0.11\%, 0.36\% at $K=20$. Overall, reducing $S_\text{r}$ from $0$ to $-2$ increases $|\Delta P|$ by 1.09\%, 3.93\%, 9.26\% at $K=2$, and by 0.01\%, 0.01\%, $-0.06\%$ at $K=20$ for $S_\text{t}=4,8,16$, respectively (\tab\ref{tab:1}).
%
Further, the decrement in pressure drop, $\Delta P$, is observed with increasing $S_\text{t}$ (\fig\ref{fig:13} and \tab\ref{tab:1}). This behavior is attributed to the enhancement in electrical resistance with increasing $S_\text{t}$, which increases $|\Delta P|$ according to \eqn(\ref{eq:9}). The magnitude of $|\Delta P|$ decreases (for $0 \ge S_\text{r} \ge -1$) and increases (for $-1 \ge S_\text{r} \ge -2$) as $S_\text{r}$ decreases (\fig\ref{fig:13} and \tab\ref{tab:1}). This is because the additional electrical resistance, i.e., the force contribution from \eqn\ref{eq:9}, increases or decreases with the variation of excess charge, $n^\ast<0$ (for $S_\text{r}>-1$) and $n^\ast>0$ (for $S_\text{r}<-1$), respectively.
\begin{figure}[t!]
	\centering\includegraphics[width=1\linewidth]{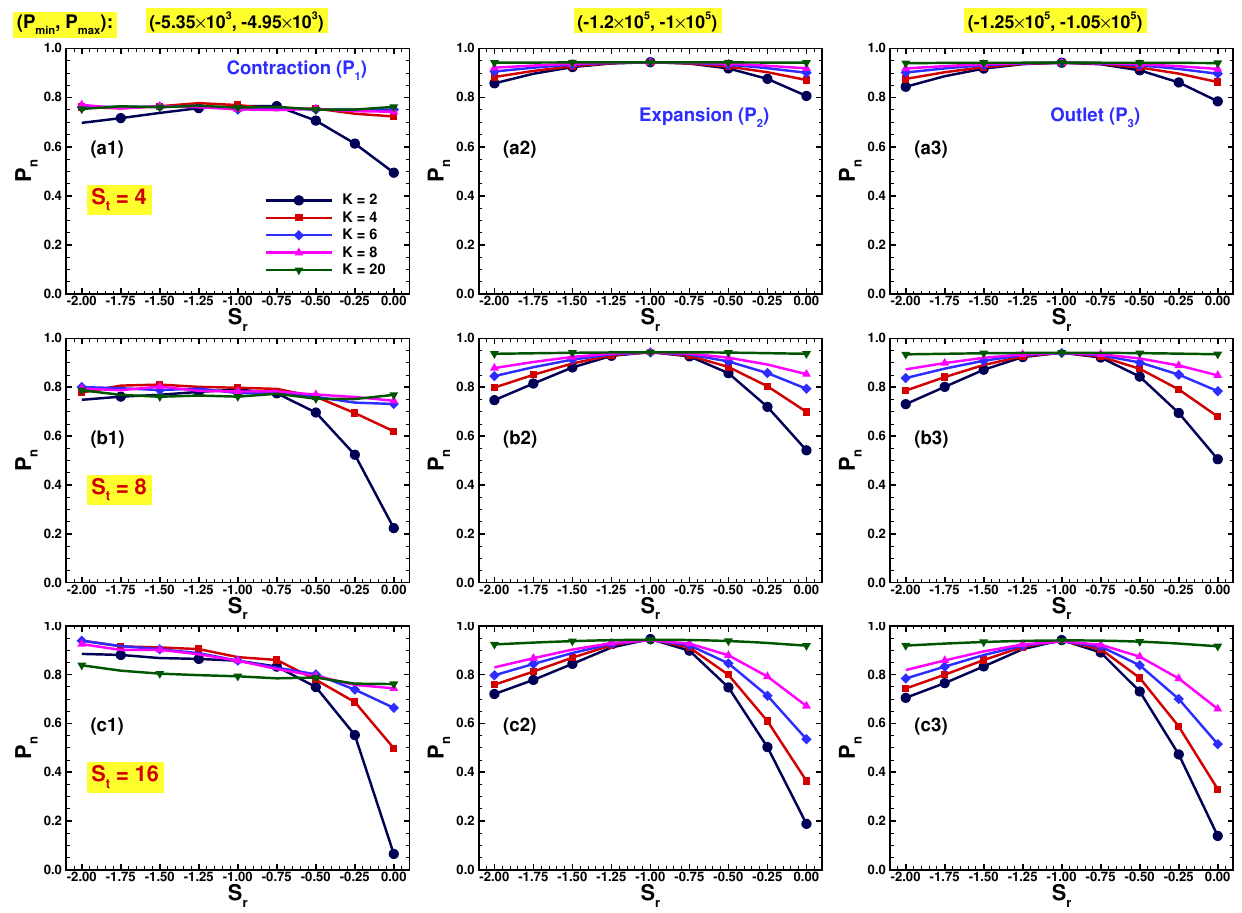}
	\caption{Influence of dimensionless parameters (\mm{K, \sut, \sur}) on scaled pressure (\mm{P_\text{n}}) variation on the contraction (\mm{P_1}, first column), expansion (\mm{P_2}, second column), and outlet (\mm{P_3}, third column) centreline locations of the oppositely charged micro slit.}
	\label{fig:14}
\end{figure} 

\noindent
Subsequently, pressure is scaled (same as $U_\text{n}$, refer section \ref{sec:potential}) to understand the detailed influence of dimensionless parameters ($S_\text{r},S_\text{t},K$). \fig\ref{fig:14} shows the scaled pressure ($P_\text{n}$, \eqn\ref{eq:MR}) variation on the centreline locations ($P_1,P_2,P_3$) of the considered micro-slit. The scaled pressure ($P_\text{n}$) variation with dimensionless parameters ($S_\text{r},S_\text{t},K$) is qualitatively similar to the pressure ($P$). For instance, $P_\text{n}$ increases with increasing $K$ (i.e., EDL thinning), irrespective of $S_\text{r}$ and $S_\text{t}$; the influence of $K$ on $P_\text{n}$ is observed maximum for $S_\text{r}=0$ and $S_\text{t}=16$ at $P_1$  (\fig\ref{fig:14}). The maximum enhancement in $P_\text{n}$ is obtained as 1066.67\% (from $0.0652$ to $0.7613$), 387.27\% (from $0.1885$ to $0.9185$), 558.99\% (from $0.1390$ to $0.9160$) for ($P_1,P_2,P_3$), respectively, when $K$ increased (from $2$ to $20$) at $S_\text{r}=0$ and $S_\text{t}=16$. 
The decrement and increment in pressure ($P_\text{n}$) is noted for $S_\text{r}>-1$ and $S_\text{r}<-1$, respectively, with increasing $S_\text{t}$, irrespective of $K$. The maximum change in $P_\text{n}$ with $S_\text{t}$ is observed for $S_\text{r}=0$ and $K=2$ at $P_1$  (\fig\ref{fig:14}). For instance, $P_\text{n}$ maximally reduces by $86.79\%$ ($0.4940$ to $0.0652$), $76.63\%$ ($0.8056$ to $0.1885$), $82.28\%$ ($0.7845$ to $0.1390$) for $P_1,P_2,P_3$, respectively, when $S_\text{t}$ varied from $4$ to $16$ at $S_\text{r}=0$ and $K=2$.
Furthermore, the pressure $P_\text{n}$ increases (for $S_\text{r}>-1$) and decreases (for $S_\text{r}<-1$) with decreasing $S_\text{r}$, irrespective of $K$. The influence of $S_\text{r}$ on $P_\text{n}$ is most significant for $K=2$ and $S_\text{t}=16$ at $P_1$  (\fig\ref{fig:14}). For example, $P_\text{n}$ increases maximally by 1255.94\% (from $0.0652$ to $0.8848$), 282.33\% (from $0.1885$ to $0.7205$), and 407.19\% (from $0.1390$ to $0.7050$) at ($P_1,P_2,P_3$), respectively, when $S_\text{r}$ decreases (from $0$ to $-2$) at $S_\text{t}=16$ and $K=2$ (refer \fig\ref{fig:14}). Hence, the impact of the dimensionless parameters ($S_\text{r},S_\text{t},K$) on pressure ($P$) is found to be strongest at $P_1$, compared to $P_2$ and $P_3$  (\fig\ref{fig:14}). This can be attributed to the maximum variation in potential ($U_\text{n}$) at $P_1$ (refer section \ref{sec:potential}), which imposes relatively higher flow resistance, thereby producing the largest change in pressure ($P$) at the contraction point ($P_1$) than at the other locations ($P_2,P_3$). 
%
\subsection{Electroviscous correction factor ($Y$)}
\label{sec:ECF}
%
In electroviscous flows (EVFs), the streaming potential ($\phi$) develops from the convective transport of excess charge near the microchannel walls due to imposed pressure-driven flow. This potential induces an additional hydrodynamic resistance, represented by the electrical force ($\myvec{F_\text{e}}$, \eqn\ref{eq:9}), which intensify the pressure drop ($\Delta P$) along the micro-slit compared to the corresponding pressure drop ($\Delta P_{0}$) in the absence of EVFs (for $S_\text{k}=0$ or $K=\infty$) at the same volumetric flow rate ($Q$). 
This influence is generally referred to as the \textit{electroviscous effect} and is commonly quantified \citep{davidson2007electroviscous,davidson2008electroviscous,bharti2008steady,dhakar2022electroviscous,dhakar2023cfd} in terms of the apparent or effective viscosity ($\mu_\text{eff}$), which is defined as the viscosity that produces the pressure drop ($\Delta P$) in a non-EVF case ($S_\text{k}=0$ or $K=\infty$).
In steady laminar microfluidic flow at low Reynolds number (e.g., $Re = 10^{-2}$), the nonlinear advection term in the momentum equation (\eqn\ref{eq:9}) becomes negligible. Consequently, the relative enhancement in pressure drop ($\Delta P / \Delta P_0$) directly corresponds to the relative enhancement in viscosity ($\mu_\text{eff} / \mu$) under identical conditions. Hence, the \textit{electroviscous correction factor} ($Y$) is expressed as follows:
\begin{gather}
	Y=\frac{\mu_{\text{eff}}}{\mu}=\frac{\Delta P}{\Delta P_{\text{0}}}
	\label{eq:27}
\end{gather}
where $\mu$ is the viscosity of the liquid yielding the pressure drop $\Delta P_0$ in the non-EVF case ($S_k = 0$ or $K = \infty$). A value of $Y = 1$ corresponds to this non-EVF baseline, whereas $Y > 1$ indicates the presence of electroviscous effects arising from excess charge transport and the resulting induced streaming potential.
Experimentally, \textit{electroviscous correction factor} ($Y$) can be determined by measuring the volumetric flow rate ($Q$) and corresponding pressure drop ($\Delta P$) in the microchannel. By comparing $\Delta P$ in the presence of electroviscous effects to the baseline pressure drop ($\Delta P_0$) in a channel where $S_k = 0$ or $K \to \infty$, the effective viscosity ($\mu_\text{eff}$) can be inferred, providing a direct quantification of the electroviscous effect.
\begin{figure}[t!]
	\centering\includegraphics[width=1\linewidth]{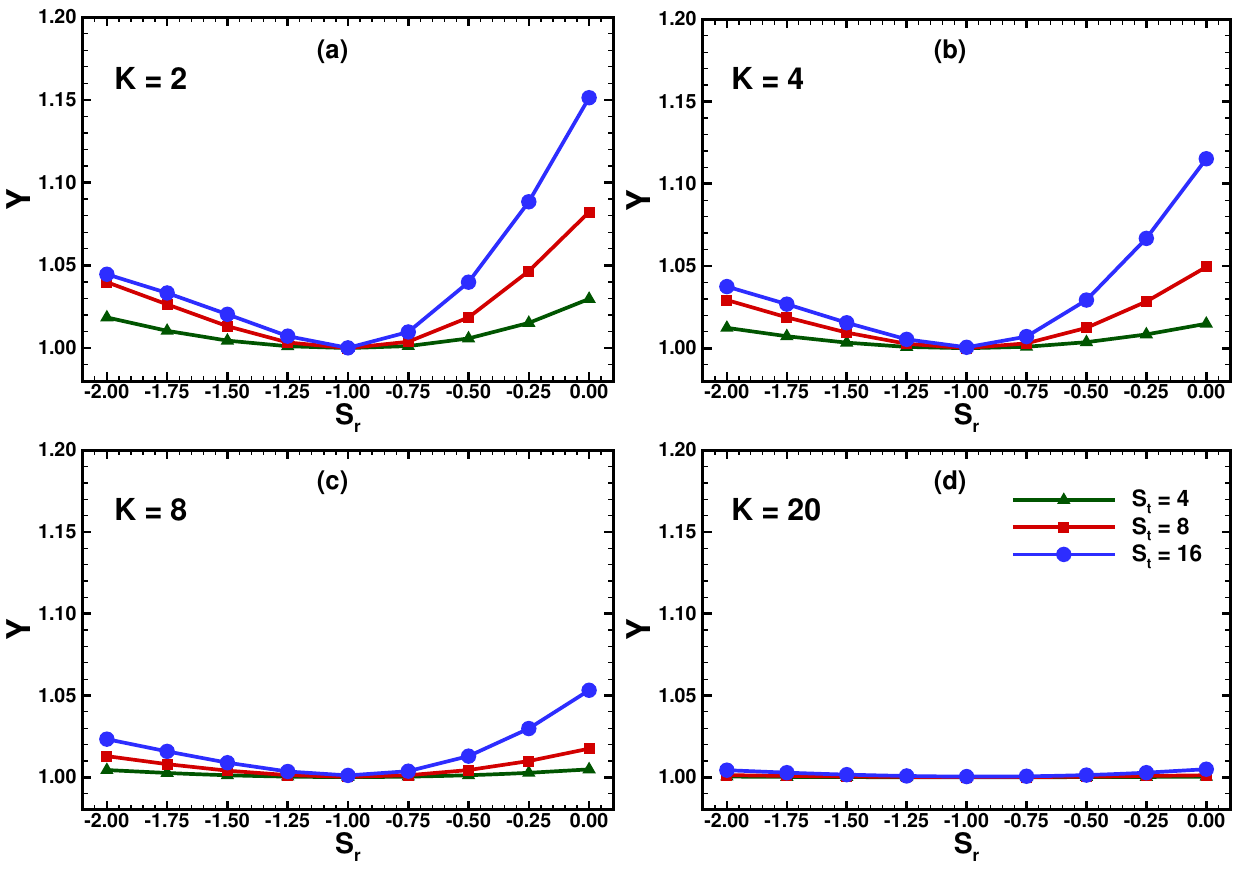}
	\caption{Electroviscous correction factor (\mm{Y}).}
	\label{fig:17}
\end{figure} 

\noindent
\fig\ref{fig:17} depicts the electroviscous correction factor ($Y$, \eqn\ref{eq:27}) as a function of the dimensionless parameters ($S_\text{r}, S_\text{t}, K$). Electroviscous effects are absent when $Y = 1$ and become stronger as $Y > 1$. The correction factor shows an inverse dependence on the inverse Debye length ($Y \propto K^{-1}$) and a direct dependence on surface charge density ($Y \propto S_\text{t}$); i.e., $Y$ increases with decreasing $K$ (EDL thickening) and increasing $S_\text{t}$, irrespective of $S_\text{r}$. This occurs because the increased charge-attractive force near the slit walls enhances the pressure drop $|\Delta P|$ with increasing $S_\text{t}$ (refer section \ref{sec:pressure}), leading to a higher $Y$ from \eqn\eqref{eq:27}. For instance, the maximum enhancement in $Y$ is 14.57\% (at $S_\text{t}=16$, $S_\text{r}=0$) and 11.81\% (at $K=2$, $S_\text{r}=0$) when $K$ reduces (from 20 to 2) and $S_\text{t}$ increases (from 4 to 16), respectively (\fig\ref{fig:17}). Furthermore, $Y$ exhibits an inverse dependence on the charge density ratio ($Y \propto S_\text{r}^{-1}$ for $-2 \le S_\text{r} \le -1$) and a direct dependence ($Y \propto S_\text{r}$ for $-1 \le S_\text{r} \le 0$), i.e., as $S_\text{r}$ increases, $Y$ decreases for $S_\text{r} < -1$ and increases for $S_\text{r} > -1$, respectively, regardless of $S_\text{t}$ and $K$. This behavior arises from the variation in pressure drop ($\Delta P$) with $S_\text{r}$ (increases for $S_\text{r} < -1$ and decreases for $S_\text{r} > -1$) as discussed in section \ref{sec:pressure}. For example, the correction factor increases maximally by 14.02\% (at $K=2$,  $S_\text{t}=16$) when $S_\text{r}$ is increased from $-0.75$ to $0$ (\fig\ref{fig:17}).
Overall, the maximum enhancement in $Y$ is 15.13\% (at $K=2$, $S_\text{t}=16$, $S_\text{r}=0$) relative to the non-EVF case ($S_k=0$ or $K = \infty$) (\fig\ref{fig:17}). This indicates that opposite charge asymmetry enhances electroviscous effects in contraction-expansion micro-slits, which can be exploited to control and manipulate microfluidic flows in practical applications.

\noindent 
The functional dependence of the correction factor ($Y$) on the dimensionless parameters ($S_\text{r}, S_\text{t}, K$) is expressed as follows.
\begin{gather}
Y = 
	\beta_1 + (\beta_2 + \beta_4 \sut)\sut + (\beta_3 + \beta_5 \sur)\sur + \beta_6 \sut\sur
	\label{eq:Ys}
	\\
	\text{where,} \quad  
	\beta_{\text{i}} = \sum_{{j}=1}^5 \left(\alpha_{\text{ij}}\ X^{({j}-1)}\right);
	\nonumber
	\qquad  X = K^{-1};  
	\qquad \alpha = 
	\begin{cases}	
		M \quad \text{for} \quad -2\le \sur\le 0
		\\N \quad \text{for} \quad -2\le \sur\le 2
	\end{cases} \nonumber
	\\
%
	\alpha = \left[\begin{array}{c|c} 		M & N 	\end{array} \right]  \label{M:Y} \\\nonumber
	= \begin{bmatrix}
		\begin{array}{rrrrr|rrrrr}
			0.9971	&	0.0971	&	-0.4745	&	1.2318	&	-1.0602	& 1.0153	&	-0.3419	&	-0.8425	&	6.8176	&	-8.4501\\
			0.0009	&	-0.0460	&	0.6742	&	-2.1604	&	2.1232	&0.0016	&	-0.1008	&	1.7302	&	-5.9500	&	6.0740\\
			-0.0060	&	0.0481	&	3.5492	&	-13.2430	&	13.8020 &0.0198	&	-0.6571	&	4.8509	&	-13.0500	&	11.9340	\\
			-0.0001	&	0.0021	&	-0.0177	&	0.0471	&	-0.0420	& -0.0003	&	0.0114	&	-0.1047	&	0.3137	&	-0.3025\\
			-0.0029	&	0.0010	&	2.3158	&	-8.6188	&	8.9768	&-0.0095	&	0.2813	&	-1.1867	&	2.2988	&	-1.6802	\\
			0.0002	&	-0.0165	&	0.2856	&	-0.9843	&	1.0044	&	-0.0039	&	0.1139	&	-0.4725	&	0.8640	&	-0.5879	
		\end{array}
	\end{bmatrix} 
\end{gather}
where, $1\le i\le 5$. The correlation coefficients ($\alpha_\text{ij}$, i.e., $M_\text{ij}$ and $N_\text{ij}$, \eqn\ref{M:Y}) are determined statistically from 135 data points through non-linear regression analysis carried out using DataFit 9.x (Free Trial), along with the statistical parameters (\tab\ref{tab:dev}) evaluated between the predicted values (\eqn\ref{eq:Ys}) and the numerical results (\fig\ref{fig:17} for $-2 \le S_\text{r} \le 0$, and Figure 15 \citep{dhakar2023cfd} for $0 \le S_\text{r} \le 2$).
%
%
%
\begin{table}[t!]
	\caption{Statistical parameters for pressure drop ($\Delta P$) and electroviscous correction factor ($Y$) over the considered ranges of ($K, S_\text{t}$) (refer \tab\ref{tab:pm}).}
	\label{tab:dev}
	\centering\renewcommand{\arraystretch}{1.5}
	\scalebox{1}{
		\begin{tabular}{|c|c|c|c|c|c|}
			\hline
{Quantity} & Parameters & $\delta_{\min}$ (\%) & $\delta_{\max}$ (\%) & $\delta_{\text{avg}}$ (\%) & $R^2$ \\ \hline
			$Y$ (\eqn\ref{eq:Ys}) & $-2 \le S_\text{r} \le 0$ & -2.98 & 2.04 & -0.06 & 86.56 \\\cline{2-6}
			& $-2 \le S_\text{r} \le 2$ & -5.76 & 6.27 & 0.03 & 95.79 \\ \hline
			$\Delta P$ (\eqn\ref{eq:38}) & $-2 \le S_\text{r} \le 0$ & -2.50 & 2.40 & -0.05 & 86.59 \\\cline{2-6}
			& $-2 \le S_\text{r} \le 2$ & -5.15 & 6.68 & 0.05 & 95.77 \\ \hline
	\end{tabular}}
\end{table}
%
%
\subsection{Pseudo-analytical model for pressure drop and electroviscous correction factor}
%
In this study, the pressure drop ($\Delta P$) is computed numerically (section \ref{sec:pressure}) for the oppositely charged contraction-expansion micro-slit. 
To address broader ranges of governing parameters and to improve the practical applicability of the results in microchip design for microfluidic applications, it is also beneficial to predict the pressure drop analytically. 

\noindent
Previous studies \citep{davidson2007electroviscous,bharti2008steady,dhakar2022electroviscous,dhakar2023cfd} have proposed pseudo-analytical models to estimate the pressure drop in symmetrically ($S_\text{r}=1$) and asymmetrically ($S_\text{r}\neq1$) charged contraction-expansion ($d_\text{c}=0.25$) micro-slits, under both no-slip and charge-dependent boundary conditions. 
These pseudo-analytical models, based on classical hydrodynamic theory, express the pressure drop in steady, fully developed laminar (low $Re$) flow of an incompressible, Newtonian liquid through a contraction-expansion micro slit, under non-EVF conditions ($S_\text{k}=0$ or $K=\infty$), as the sum of two contributions: (i) the pressure drop in each uniform section (upstream, contraction, and downstream), evaluated using Poiseuille flow, and (ii) the excess pressure drop arising from sudden expansion and contraction, modeled by creeping flow through a thin orifice \citep{davidson2007electroviscous,bharti2008steady,dhakar2022electroviscous,dhakar2023cfd}, as follows.
\begin{gather}
	\Delta P_{\text{0,m}}=\left(\sum_{i=u,c,d}\Delta P_{\text{0,i}}\right) +\Delta P_{\text{0,e}} 	\label{eq:33a} 
	\\
	\Delta P_{\text{0,i}} = \left(\frac{3}{Re}\right)\left(\frac{{\Delta L_{\text{i}}}}{d_\text{i}^3}\right);\qquad 
	\Delta P_{\text{0,e}} =\left(\frac{16}{Re}\right)\left(\frac{1}{\pi d_{\text{c}}^2}\right); \qquad d_\text{i}=\frac{W_\text{i}}{W}
	\label{eq:33b}
\end{gather}
where $u$, $c$, and $d$ denote the upstream, contraction, and downstream sections; $0$ indicates the non-EVF case; $e$ denotes extra; and $m$ for mathematical.

\noindent 
Building on this foundation, the present study extends the pseudo-analytical model framework (\eqn\ref{eq:33a}) to both similarly (case A: $S_\text{r}>0$) and oppositely (case B: $S_\text{r}<0$) charged contraction-expansion micro slits, thereby enabling analytical prediction of the pressure drop ($\Delta P$) across a broader range of electrohydrodynamic conditions ($-2 \leq S_\text{r} \leq 2$) for a contraction ratio ($d_c$), as formulated in \eqn\eqref{eq:38}.
\begin{gather}
	\Delta P_{\text{m}}= 
    \Gamma_\text{ab}\ \Delta P_{0,\text{m}}
	= \Gamma_\text{ab}\left[\frac{3}{Re} \left(L_{\text{u}} +  \frac{L_{\text{c}}}{d_{\text{c}}^3} + L_{\text{d}} + \frac{16}{3\pi d_{\text{c}}^2} \right)\right]
	\label{eq:38}
\end{gather}
where, the correction factor ($\Gamma_\text{ab}$) incorporates electroviscous effects into the non-EVF pressure drop ($\Delta P_{0,\text{m}}$), thereby providing the actual pressure drop ($\Delta P_\text{m}$) under electroviscous conditions. The subscript ‘ab’ denotes that the numerical data from both Case A and Case B of charge asymmetry ($-2 \leq S_r \leq 2$; refer \tab\ref{tab:srf}) are considered.
\begin{figure}[b!]
	\centering
	\subfigure[Case B]{\includegraphics[width=0.49\linewidth]{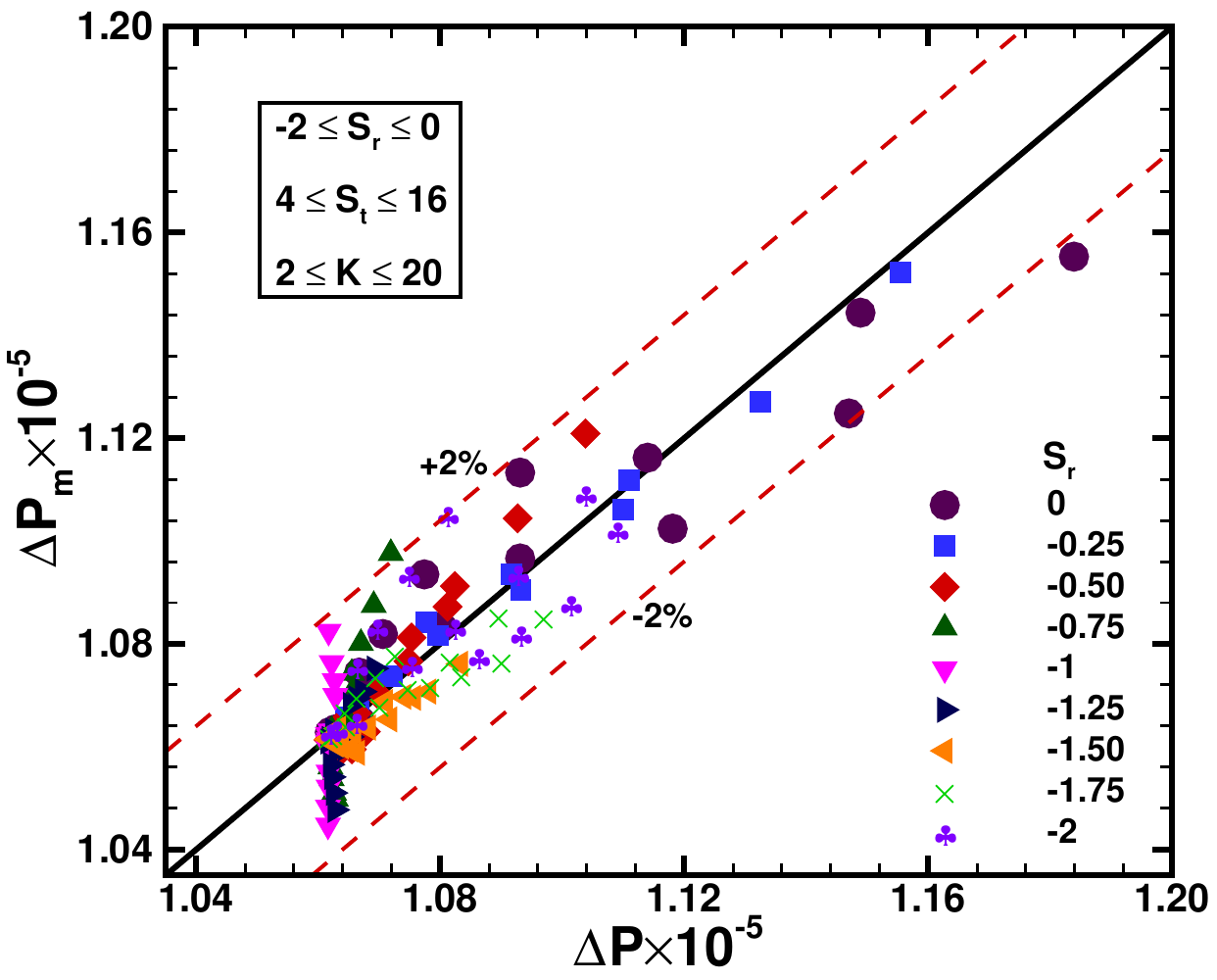}}
	\subfigure[Case B]{\includegraphics[width=0.49\linewidth]{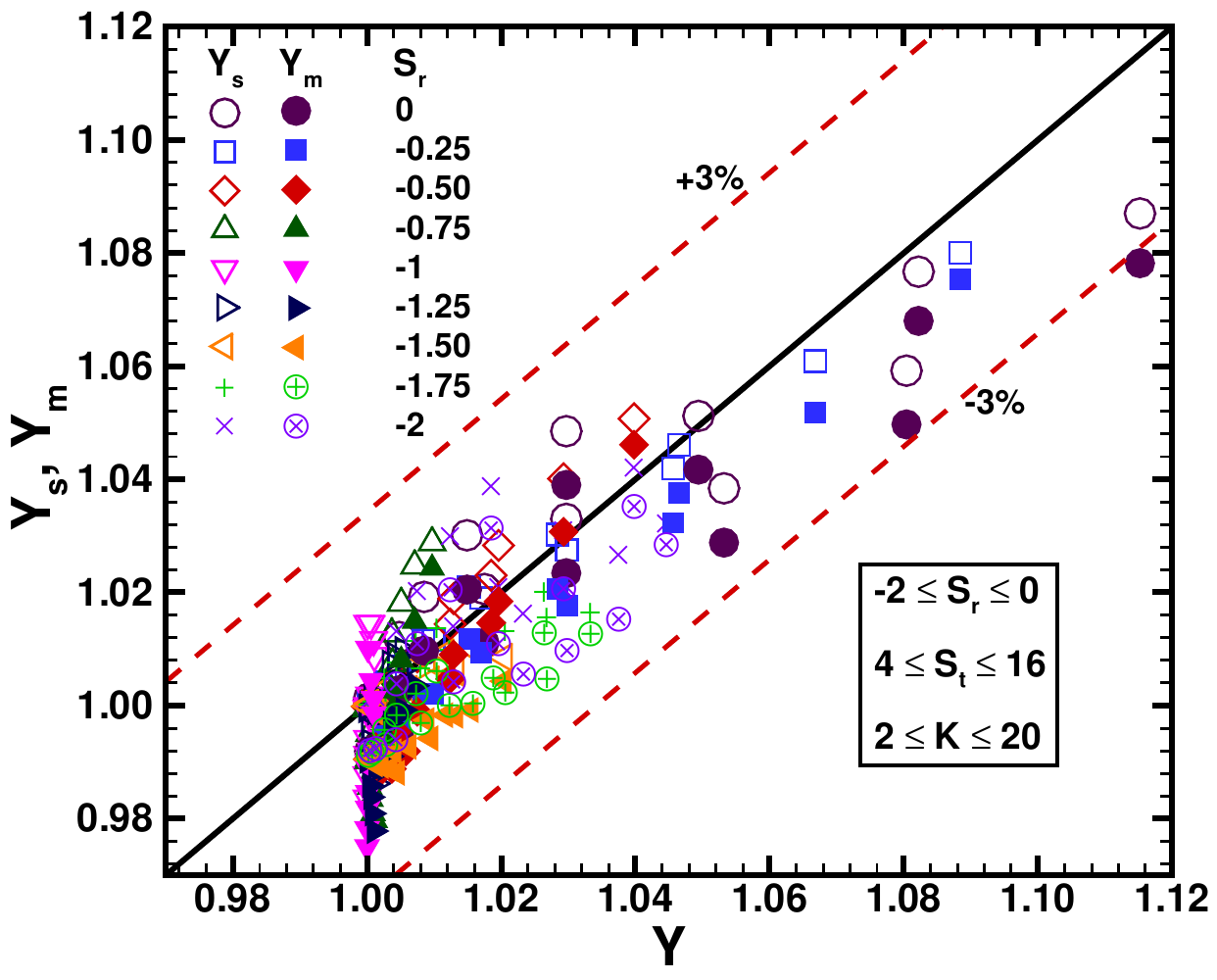}}
	\subfigure[Combined Case A \citep{dhakar2023cfd} and Case B] {\includegraphics[width=0.49\linewidth]{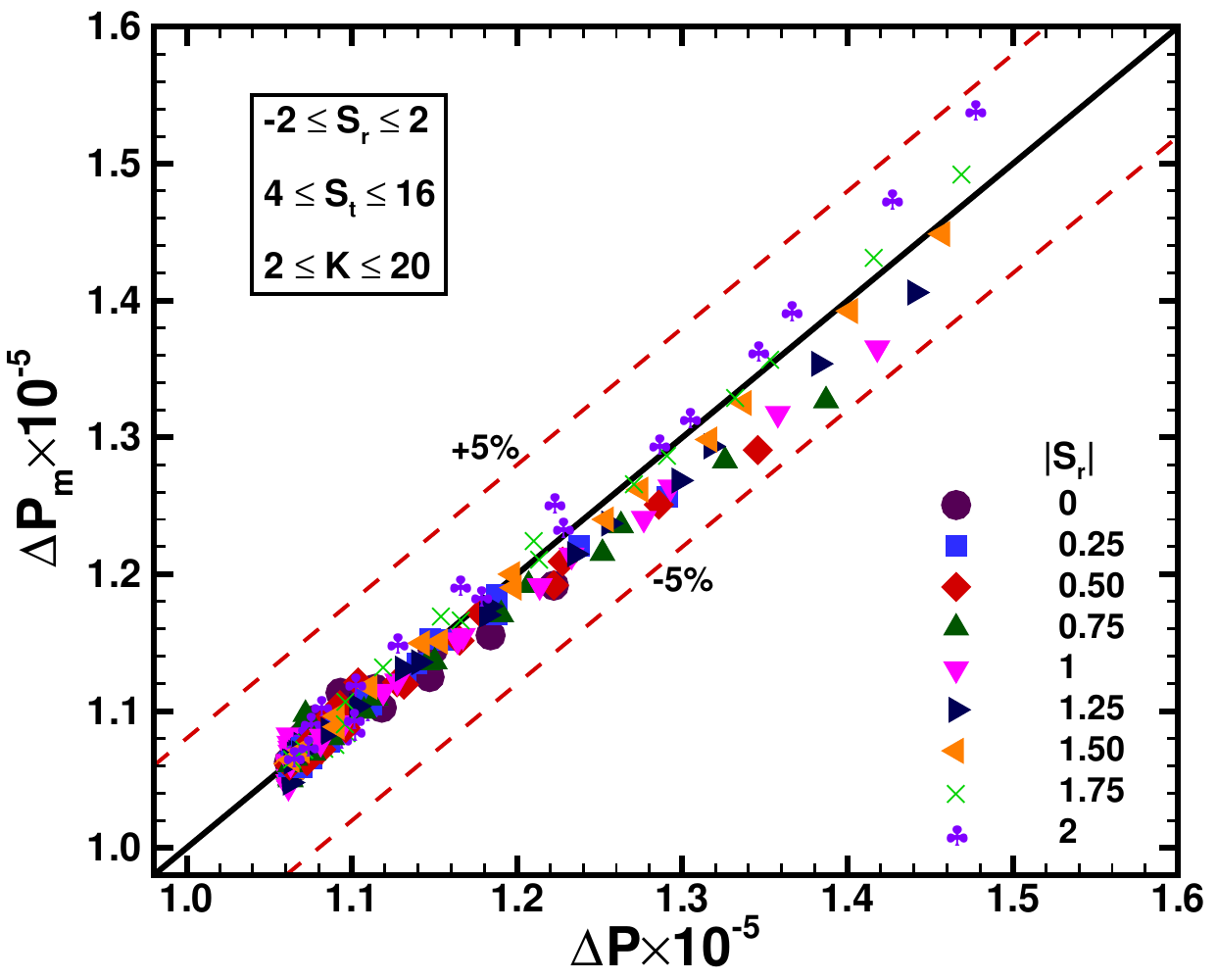}}
	\subfigure[Combined Case A \citep{dhakar2023cfd} and Case B] {\includegraphics[width=0.49\linewidth]{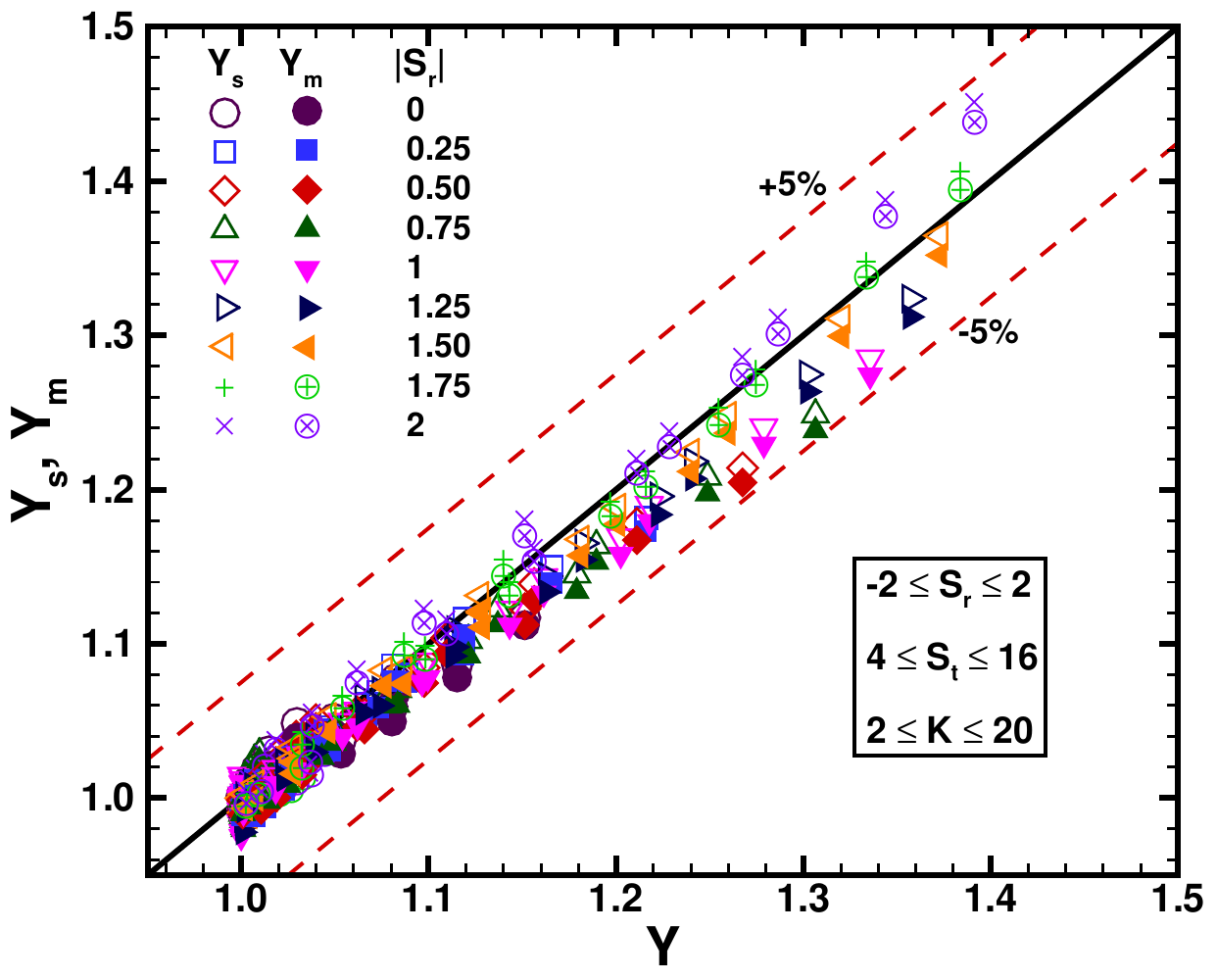}}
	\caption{Parity charts of (a, c) pressure drop ($\Delta P$ vs. $\Delta P_\text{m}$) and (b, d) correction factor ($Y$ vs. $Y_\text{s}, Y_\text{m}$), over the parameter ranges ($K, \Sigma_t, S_r$; see \tab\ref{tab:pm}). Panels (a) and (b) correspond to {Case B}, while (c) and (d) show combined results of {Case A} and {Case B}.}
	\label{fig:18}
\end{figure} 
%
Subsequently, the functional dependence of the correction factor ($\Gamma_\text{ab}$)  on the dimensionless parameters ($S_\text{r}, S_\text{t}, K$) is statistically expressed as follows.
\begin{gather}
    \Gamma_\text{ab} = 
	\beta_1 + (\beta_2 + \beta_4 \sut)\sut + (\beta_3 + \beta_5 \sur)\sur + \beta_6 \sut\sur
\label{M:model}
\\
\text{where,} \qquad \beta_{\text{i}} = \sum_{{j}=1}^5 \alpha_{\text{ij}} X^{({j}-1)};\quad X = K^{-1}; 
\qquad \alpha = 
	\begin{cases}	
			m \quad \text{for} \quad -2\le \sur\le 0
			\\n \quad \text{for} \quad -2\le \sur\le 2
		\end{cases} \nonumber
\\ 
\alpha = \left[\begin{array}{c|c} 		m & n 	\end{array} \right]  \label{M:gevac} \\\nonumber
= \begin{bmatrix}
	\begin{array}{rrrrr|rrrrr}
		0.9874	&	0.1098	&	-0.5979	&	1.654	&	-1.5076  &1.0056	&	-0.3308	&	-0.9077	&	7.004	&	-8.639	\\
		0.00095	&	-0.04688	&	0.6795	&	-2.1802	&	2.1461 & 0.00159	&	-0.1008	&	1.7228	&	-5.9248	&	6.0498	\\
		 -0.006	&	0.0516	&	3.4762	&	-12.989	&	13.541 & 0.0197	&	-0.6542	&	4.8341	&	-13.022	&	11.918	\\
		 -6\times10^{-5}	&	0.0021	&	-0.01753	&	0.04698	&	-0.042 & -0.000309	&	0.01135	&	-0.104	&	0.3115	&	-0.3001	\\
		 -0.0029	&	0.0017	&	2.2975	&	-8.5519	&	8.9095 & -0.0094	&	0.2781	&	-1.1714	&	2.2662	&	-1.6553 \\
		 0.0002	&	-0.0157	&	0.2855	&	-0.987	&	1.005	& 	-0.00393	&	0.1136	&	-0.4735	&	0.8766	&	-0.6093
	\end{array}
\end{bmatrix} 
\end{gather}
%
%
The correlation coefficients ($\alpha_\text{ij}$, i.e., $m_\text{ij}$ and $n_\text{ij}$, \eqn\ref{M:gevac}) are determined statistically from 135 data points through non-linear regression analysis carried out using DataFit 9.x (Free Trial), along with the statistical parameters (\tab\ref{tab:dev}) evaluated between the predicted values (\eqn\ref{eq:38}) and the numerical results (\tab\ref{tab:1} for $-2 \le S_\text{r} \le 0$, and Table 5 \citep{dhakar2023cfd} for $0 \le S_\text{r} \le 2$).

\noindent 
Subsequently, the pseudo-analytical model (\eqns\ref{eq:33a} and \ref{eq:38}) is extended to determine the correction factor ($Y$), as given below.
\begin{gather}
	Y_{\text{m}}=\frac{\Delta P_{\text{m}}}{\Delta P_{0,\text{m}}}
	\label{eq:40}
\end{gather}
\noindent
\fig\ref{fig:18}(a) and \ref{fig:18}(b)  present parity charts for the pressure drop ($\Delta P$ vs. $\Delta P_\text{m}$) and the electroviscous correction factor ($Y$ vs. $Y_\text{m}$ and $Y_\text{s}$), comparing numerical ($\Delta P$, \tab\ref{tab:1}; $Y$, \eqn\ref{eq:27}), statistical ($Y_\text{s}$, \eqn\ref{eq:Ys}) and pseudo-analytical ($\Delta P_\text{m}$, \eqn\ref{eq:38}; $Y_\text{m}$, \eqn\ref{eq:40}) predictions over a broad range of dimensionless parameters (\tab\ref{tab:pm}) for the oppositely charged micro slit ($-2 \leq S_r \leq 0$, {Case B}). Furthermore, \fig\ref{fig:18}(c) and \ref{fig:18}(d) also show parity charts of the same quantities ($\Delta P$ and $Y$) for both similar-charge cases ($S_r \geq 0$, {Case A}) \citep{dhakar2023cfd} and opposite-charge cases ($S_r < 0$, {Case B}) under otherwise identical conditions. Broadly, the deviation between numerical and analytical results decreases with increasing inverse Debye length ($K$), and with decreasing surface charge density ratio ($S_\text{r}$) and surface charge density ($S_\text{t}$).
\begin{figure}[b!]
	\centering\includegraphics[width=0.8\linewidth]{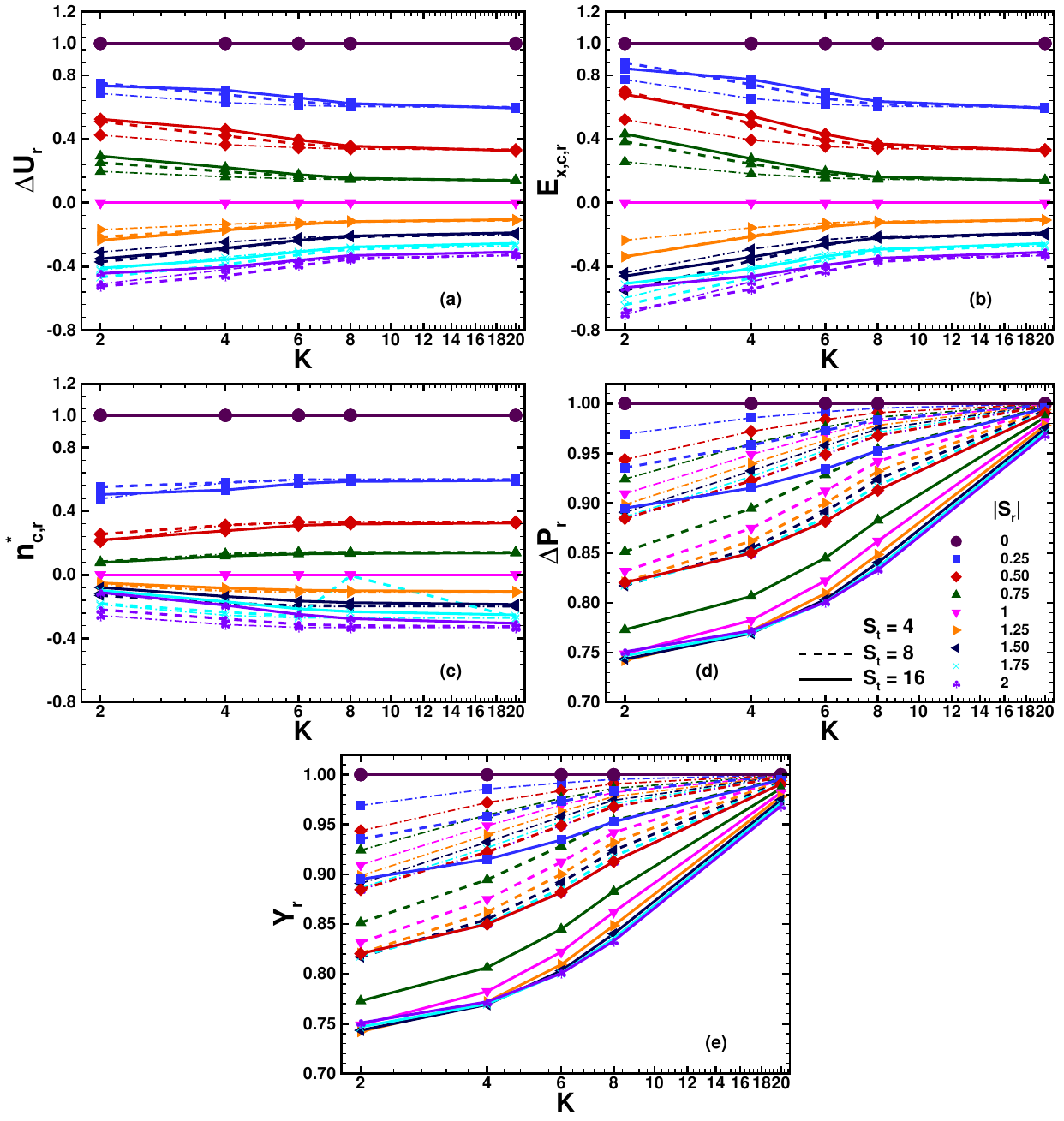}
\caption{Normalized electrokinetic flow quantities ($\theta_\text{r}$) as a function of dimensionless parameters ($K$, $S_\text{t}$, $|S_\text{r}|$).}
\label{fig:4a}
\end{figure} 
%
\subsection{Comparative analysis of electrokinetic flow characteristics in the similar and opposite asymmetrically charged micro-slits}
\begin{figure}[t!]
	\centering\includegraphics[width=1\linewidth]{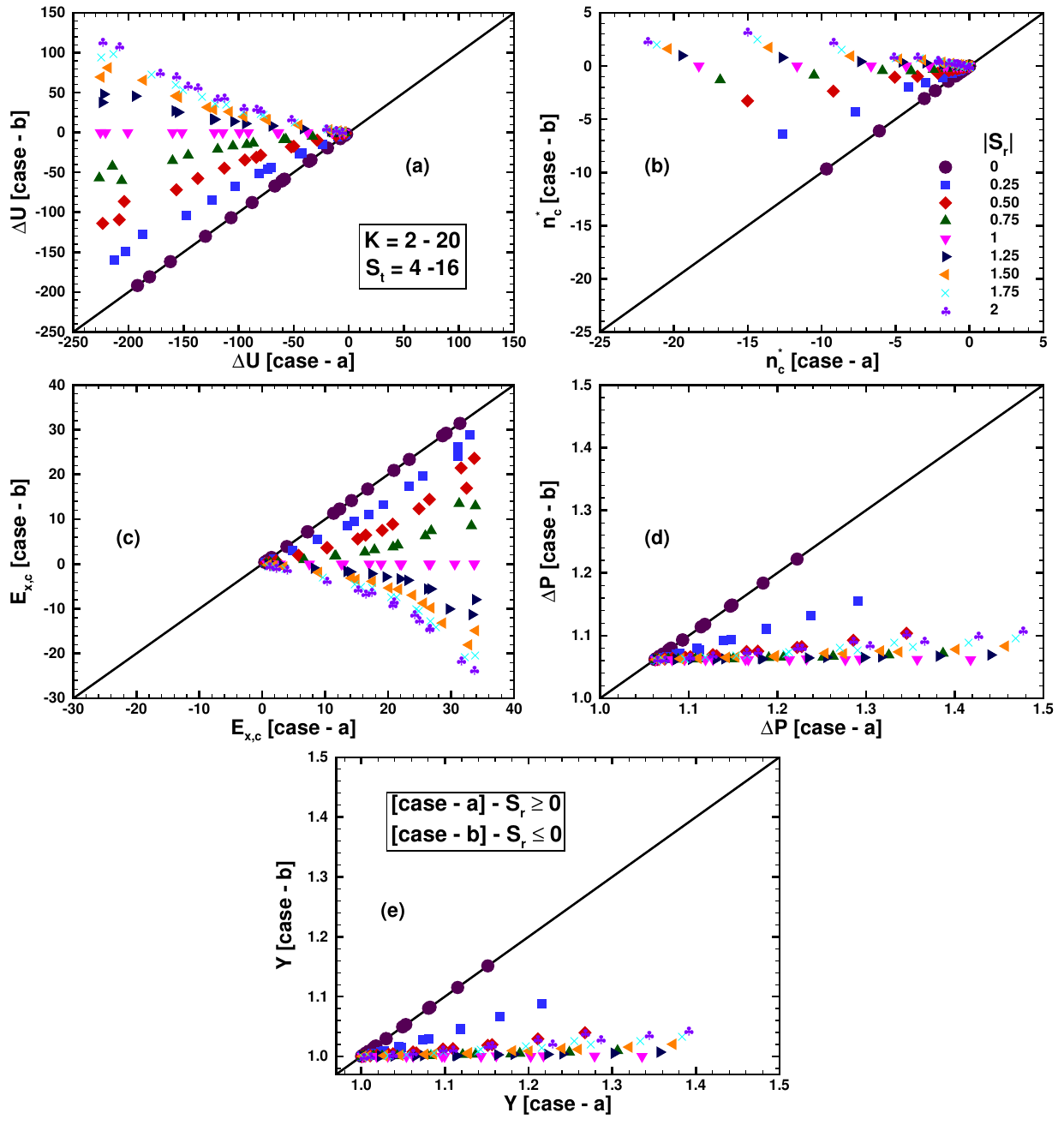}
	\caption{Parity chart of electrokinetic quantities ([\textbf{case - a}] vs [\textbf{case - b}]) in the charged micro slit as a function of dimensionless parameters (\mm{\sut,K,|\sur|}).}
	\label{fig:4b}
\end{figure}

%
The present study investigates the influence of opposite-charge asymmetry ($-2 \leq S_r \leq 0$, Case B) on the electroviscous (EV) flow of liquid electrolyte through a contraction-expansion micro-slit. In contrast, the effects of similar-charge asymmetry ($0 \leq S_r \leq 2$, Case A) in the same geometry, under otherwise identical conditions, have been analyzed in our recent work \citep{dhakar2023cfd}.  \figs\ref{fig:4a} and \ref{fig:4b} present a comparative analysis of electrokinetic flow quantities for both cases, where {Case A} corresponds to similar-charge asymmetry ($0 \leq S_r \leq 2$) and {Case B} to opposite-charge asymmetry ($-2 \leq S_r \leq 0$) over a broad range of dimensionless parameters (\tab\ref{tab:pm}). 

\noindent
The electrokinetic flow quantities ($\Delta U$, $n^\ast_\text{c}$, $E_\text{x,c}$, $\Delta P$, $Y$) are normalized ($\theta_\text{r}$, \eqn\ref{eq:MR}) to facilitate the comparative analysis of charge asymmetry (Case A and Case B).  \fig\ref{fig:4a} present a normalized electrokinetic flow quantities ($\theta_\text{r}$) over a broad range of dimensionless parameters (\tab\ref{tab:pm}).  At $S_\text{r} = 0$, the normalized quantities ($\theta_\text{r}$) equal unity, regardless of $K$ and $S_\text{t}$, since $S_\text{r} = 0$ represents the common condition shared by both Case A and Case B.

\noindent
\figs\ref{fig:4a}a and \ref{fig:4a}b show that both the normalized potential drop ($\Delta U_\text{r}$) and the normalized critical electric field strength ($E_\text{x,c,r}$) exhibit qualitatively similar variations with the dimensionless parameters. For instance, at $|S_\text{r}|=0$ and $|S_\text{r}|=1$, both $\Delta U_\text{r}$ and $E_\text{x,c,r}$ attain values of 1 and 0, respectively, regardless of $S_\text{t}$ and $K$. This occurs because the micro-slit becomes electrically neutral at $S_\text{r}=-1$ (i.e., $S_\text{b}=-S_\text{t}$), leading to $U$ and $E_\text{x}$ values close to zero. Both $\Delta U_\text{r}$ and $E_\text{x,c,r}$ increase (for $|S_\text{r}|<1$) and decrease (for $|S_\text{r}|>1$) as $K$ decreases from 20 to 2, independent of $S_\text{t}$. A similar trend is observed with increasing $S_\text{t}$, although the trend reverses at higher $S_\text{t}$. Furthermore, both quantities increase with decreasing $|S_\text{r}|$, regardless of $K$ and $S_\text{t}$. Consequently, the deviation of $\Delta U_\text{r}$ and $E_\text{x,c,r}$ between Case A and Case B grows with decreasing $S_\text{t}$ and increasing $|S_\text{r}|$. For $1 \leq |S_\text{r}|\leq 2$, the deviation increases, whereas for $0 \leq |S_\text{r}|\leq 1$, it decreases with decreasing $K$ (\figs\ref{fig:4a}a and \ref{fig:4a}b).

\noindent 
The normalized critical excess charge ($n^\ast_\text{c,r}$) as a function of the dimensionless parameters ($S_\text{t}, K, |S_\text{r}|$) is shown in \fig\ref{fig:4a}c. At $|S_\text{r}|=0$ and 1, $n^\ast_\text{c,r}$ equals 1 and 0, respectively, across the considered ranges of $S_\text{t}$ and $K$. At $S_\text{r}=-1$, the device becomes electrically neutral (equal and opposite wall charges), resulting in zero excess charge. When $K$ increases from 2 to 20 (i.e., thinning EDL), $n^\ast_\text{c,r}$ decreases for $|S_\text{r}|>1$ and increases for $|S_\text{r}|<1$, regardless of $S_\text{t}$. Similarly, $n^\ast_\text{c,r}$ decreases (for $|S_\text{r}|>1$) and increases (for $|S_\text{r}|<1$) with larger $S_\text{t}$, though the opposite trend appears at higher $S_\text{t}$, independent of $K$. In general, $n^\ast_\text{c,r}$ increases with decreasing $|S_\text{r}|$ for all other conditions. Consequently, the deviation in $n^\ast_\text{c,r}$ between case A and case B reduces with increasing $S_\text{t}$ and decreasing $|S_\text{r}|$. Correspondingly, an enhancement (for $0 \le |S_\text{r}| \le 1$) and a reduction (for $1 \le |S_\text{r}| \le 2$) in deviation is observed with decreasing $K$ (i.e., EDL thickening) (\fig\ref{fig:4a}c).

\noindent
The normalized pressure drop ($\Delta P_\text{r}$) as a function of dimensionless parameters ($S_\text{t},K,|S_\text{r}|$) is shown in \fig\ref{fig:4a}d. At $|S_\text{r}|=0$, $\Delta P_\text{r}=1$ for all other considered conditions. In general, $\Delta P_\text{r}$ increases with larger $K$ and smaller $S_\text{t}$, independent of $|S_\text{r}|$. With increasing $|S_\text{r}|$ (from 0 to 2), $\Delta P_\text{r}$ decreases, though the trend reverses at low $K$ and high $S_\text{t}$. Consequently, the deviation of $\Delta P_\text{r}$ between Case A and Case B becomes more pronounced with decreasing $K$ and increasing $S_\text{t}$ and $|S_\text{r}|$ (\fig\ref{fig:4a}

\noindent
Finally, \fig\ref{fig:4b} presents parity charts comparing the results of Case A (x--axis) and Case B (y--axis) for electrokinetic flow quantities, namely, (a) total potential drop ($\Delta U$), (b) critical excess charge ($n^\ast_\text{c}$), (c) critical induced electric field strength ($E_\text{x,c}$), (d) pressure drop ($\Delta P$), and (e) electroviscous correction factor ($Y$) over the considered ranges of parameters ($S_\text{t}, K, |S_\text{r}|$). The influence of charge asymmetry ($S_\text{r}$) is weaker in regions where Case A and Case B show close agreement. The deviations between Case A and Case B progressively diminish with decreasing $|S_\text{r}|$, decreasing $S_\text{t}$, and thinning of the electrical double layer (i.e., increasing $K$). Conversely, the discrepancies are more pronounced at larger $|S_\text{r}|$, higher $S_\text{t}$, and thicker electrical double layers (i.e., smaller $K$), highlighting the stronger role of charge asymmetry under these conditions (high $|S_\text{r}|$, high $S_\text{t}$, and low $K$). 
Overall, the influence of opposite charge asymmetry ($-2 \leq S_\text{r} \leq 0$, case B) on electroviscous flow (EVF) is found to be weaker than that of similar charge asymmetry ($0 \leq S_\text{r} \leq 2$, case A), for contraction-expansion micro slits under otherwise identical conditions,.
%
\section{Concluding remarks}
%
\noindent
This work numerically investigates the electroviscous flow (EVF) of symmetric ($1$:$1$) liquid electrolytes through oppositely charged contraction-expansion micro-slits. A coupled mathematical model (P-NP-NS), comprising Poisson’s, Nernst-Planck, and Navier-Stokes equations, is solved using the finite element method to compute the electrokinetic flow fields, including potential ($U$), velocity ($\mathbf{V}$), excess charge ($n^\ast$), induced electric field ($E_\text{x}$), pressure ($P$), and correction factor ($Y$), over the wide ranges of parameters ($K$, $S_\text{r}$, $S_\text{t}$, $Re$, $Sc$, $\beta$, $d_c$).
The results depict that the oppositely charged wall conditions significantly alter flow behavior in contraction-expansion geometries. The maximum increase in potential drop ($|\Delta U|$) and pressure drop ($|\Delta P|$) is 296.82\% (at $K=20$, $S_\text{t}=4$) and 14.57\% (at $S_\text{t}=16$, $S_\text{r}=0$) when $S_\text{r}$ decreases from $-1.25$ to $-2$ and $K$ decreases from 20 to 2, respectively. The electroviscous correction factor ($Y$) shows maximum enhancement of 14.02\% (at $S_\text{t}=16$, $K=2$), 11.81\% (at $K=2$, $S_\text{r}=0$), and 14.57\% (at $S_\text{t}=16$, $S_\text{r}=0$). Overall, the maximum increase in $Y$ reaches 15.13\% (at $S_\text{t}=16$, $K=2$, $S_\text{r}=0$) relative to the non-EVF case ($S_\text{k}=0$ or $K=\infty$), demonstrating the pronounced role of opposite charging in EVF through non-uniform slits. 
The flow quantities have been statistically correlated to provide the predictive correlations as a function of flow governing parameters.
Further, a predictive pseudo-analytical model is further developed to estimate pressure drop and the associated correction factor in EVF. The model evaluates the total pressure drop as the sum of Poiseuille contributions from uniform sections and losses due to sudden expansion-contraction via creeping flow through thin orifices. The analytical predictions agree with numerical results within $\pm3$\%, confirming the reliability of the simplified framework. The extended model is validated against numerical data, and parity charts confirm its predictive accuracy. Finally, a comparative analysis of similar (Case A) and opposite (Case B) charge asymmetries reveals that deviations between similar (Case A) and opposite (Case B) charge asymmetries diminish with decreasing $|S_\text{r}|$, lower $S_\text{t}$, and thinning of the electrical double layer (increasing $K$). Conversely, stronger asymmetry effects are observed at higher $S_\text{t}$, larger $|S_\text{r}|$, and smaller $K$. Overall, the influence of opposite charge asymmetry is found to be weaker than that of similar asymmetry under otherwise identical conditions. The correlations developed herein broaden the applicability of pseudo-analytical models for EVF, offering useful predictive tools for microchip and microfluidic device design.
%
\section*{Declaration of Competing Interest}
\noindent 
The authors declare that they have no known competing financial interests or personal relationships that could have appeared to influence the work reported in this paper.
%
%
\begin{spacing}{1.1}
\input{Nomenclature.tex}

\printnomenclature
\end{spacing}
\begin{spacing}{1.3}
\noindent \small
\bibliography{references}
\end{spacing}
%
%
%
%
%
%
%
%
%
%
%
%
\end{document}

%% file: nomenclature.tex
\fontsize{10}{10pt}\selectfont
 \nomenclature[g0]{\textit{Greek letters}}{}
 \nomenclature[d0]{\textit{Dimensionless groups}}{}
 \nomenclature[s0]{\textit{Subscripts and Superscripts}}{}
 \nomenclature[z0]{\textit{Abbreviations}}{}
%
\nomenclature[zcfd]{CFD}{computational fluid dynamics}
\nomenclature[zedl]{EDL}{electrical double layer}
\nomenclature[zeve]{EVE}{electroviscous effect}
\nomenclature[zevee]{EV}{electroviscous}
\nomenclature[zevf]{EVF}{electroviscous flow}
\nomenclature[zfem]{FEM}{finite element method}
\nomenclature[zpdps]{PDF}{pressure-driven flow}
%
\nomenclature[adc]{$d_{\text{c}}$}{contraction ratio ($=W_{\text{c}}/W$), --}
\nomenclature[aDj]{$\mathcal{D}_{j}$}{diffusivity of the ions of type j, assumed equal ($\mathcal{D}_{+}=\mathcal{D}_{-}=\mathcal{D}$), m$^2$/s}
\nomenclature[ae]{$e$}{elementary charge of a proton ($=1.602176634\times 10^{-19}$), C or A.s}
\nomenclature[aE]{$E_{\text{x}}$}{induced electric field strength ($=-\partial U/\partial x$), V/m or --}
\nomenclature[afj]{$\mathbf{f_\text{j}}$}{flux density of the ions of type j (\eqn\ref{eq:6}), 1/(m$^2$.s)}
\nomenclature[aIc]{$I_{\text{c}}$}{conduction current density (\eqn\ref{eq:2}), A/m$^2$ or --}
\nomenclature[aId]{$I_{\text{d}}$}{diffusion current density (\eqn\ref{eq:2}), A/m$^2$ or --}
\nomenclature[aIs]{$I_{\text{s}}$}{streaming current density (\eqn\ref{eq:2}), A/m$^2$ or --}
\nomenclature[akB]{$k_{\text{B}}$}{Boltzmann constant ($=1.380649\times 10^{-23}$), J/K}
\nomenclature[aLc]{$L_{\text{c}}$}{contraction section length, m or --}
\nomenclature[aLd]{$L_{\text{d}}$}{downstream outlet section length, m or --}
\nomenclature[aLu]{$L_{\text{u}}$}{upstream inlet section length, m or --}
\nomenclature[an+]{$n_{+}$}{local number density of positive ions (\eqn\ref{eq:6}), 1/m$^3$ or --}
\nomenclature[an-]{$n_{-}$}{local number density of negative ions (\eqn\ref{eq:6}), 1/m$^3$ or --}
\nomenclature[an0]{$n_{0}$}{bulk density of the ions of type j, 1/m$^3$}
\nomenclature[anj]{$n_{j}$}{local number density of the ions of type j, 1/m$^3$}
\nomenclature[ans]{$n^*$}{excess charge ($=n_{+}-n_{-}$), 1/m$^3$ or --}
\nomenclature[aP]{$P$}{pressure, Pa or --}
\nomenclature[aT]{$T$}{temperature, K}
\nomenclature[aU]{$U$}{total electrical potential, V or --}
\nomenclature[aU]{$U_{\text{c}}$}{characteristic total electrical potential, V}
\nomenclature[aV]{$\mathbf{V}$}{velocity vector, m/s or --}
\nomenclature[aVa]{$\overline{V}$}{average velocity of the fluid at the inlet, m/s}
\nomenclature[aVx]{$V_x$}{x-component of the velocity, m/s or --}
\nomenclature[aVy]{$V_y$}{y-component of the velocity, m/s or --}
\nomenclature[aW]{$W$}{cross-sectional width of inlet and outlet sections, m}
\nomenclature[aWc]{$W_{\text{c}}$}{cross-sectional width of contraction section, m}
\nomenclature[ax]{$x$}{streamwise coordinate, --}
\nomenclature[ay]{$y$}{transverse coordinate, --}
\nomenclature[aY]{$Y$}{electroviscous correction factor (\eqns\ref{eq:27}, and \ref{eq:40}), --}
\nomenclature[azj]{$z_{j}$}{valency of the ions of type j, assumed equal ($z_{+}=\rev{-}z_{-}=z$), --}
%
%
\nomenclature[gdP]{$\Delta P$}{pressure drop (\eqn\ref{eq:38}), --}
\nomenclature[geps0]{$\varepsilon_{\text{0}}$}{permittivity of free space (i.e. vaccum), F/m or C/(V.m)}
\nomenclature[gepsr]{$\varepsilon_{\text{r}}$}{dielectric constant (absolute or relative permittivity) of the electrolyte liquid, --}
\nomenclature[glambdad]{$\lambda_{\text{D}}$}{Debye length ($=K^{-1}$), --}
\nomenclature[gmu]{$\mu$}{viscosity, Pa.s}
\nomenclature[gmueff]{$\mu_\text{eff}$}{effective or apparent viscosity, Pa.s}
\nomenclature[gpsi]{$\psi$}{EDL potential (\eqn\ref{eq:4a}), V or --}
\nomenclature[grho]{$\rho$}{density of fluid, kg/m$^3$}
\nomenclature[grhoe]{$\rho_{\text{e}}$}{charge density of liquid, C/m$^3$}
\nomenclature[gsigmab]{$\sigma_\text{k}$}{surface charge density, C/m$^2$}
%
%
\nomenclature[dbeta]{$\mathit{\beta}$}{liquid parameter, --}
\nomenclature[dK]{$\mathit{K}$}{inverse Debye length, --}
\nomenclature[dPe]{$Pe$}{Peclet number ($={Re}\times\mathit{Sc}$), --}
\nomenclature[dRe]{$Re$}{Reynolds number, --}
\nomenclature[dSa]{$\mathit{S_\text{k}}$}{surface charge density, --}
\nomenclature[dSc]{$\mathit{Sc}$}{Schmidt number, --}
\nomenclature[dSb]{$\mathit{S_\text{r}}$}{surface charge density ratio, --}
%
\nomenclature[sz]{$0$}{without electroviscous effects}
\nomenclature[sc]{$c$}{contraction}
\nomenclature[sd]{$d$}{downstream}
\nomenclature[se]{$e$}{extra or excess}
\nomenclature[sea]{$i$}{inflow}
\nomenclature[seb]{$o$}{outflow}
\nomenclature[sec]{$w$}{walls}
\nomenclature[sf]{$evac$}{electroviscous and asymmetric charge}
\nomenclature[sm]{$m$}{mathematical}
\nomenclature[ss]{$s$}{statistical}
\nomenclature[su]{$u$}{upstream}
%